\documentclass[12pt]{article}
\pdfoutput=1
\usepackage[colorlinks,linkcolor=Blue,citecolor=Blue,bookmarks,bookmarksnumbered]{hyperref}
\usepackage[scaled=0.85]{helvet}
\usepackage{amsmath,amssymb,accents,mathrsfs,XoohmE}
\usepackage{graphicx,color}
\usepackage{booktabs}
\usepackage{multirow}
\usepackage{placeins}
\usepackage{amsmath}
\usepackage{bbold}
\usepackage{amsmath}
\usepackage[vcentermath]{youngtab}
\usepackage{tabu}
\usepackage{empheq}
\usepackage{appendix}

\usepackage{young}
\usepackage{XoohmE}
\usepackage{mathrsfs}

\definecolor{Green}  {rgb}{0.10,0.70,0.10} 
\definecolor{Orange} {rgb}{1.00,0.50,0.15} 
\definecolor{Red}    {rgb}{0.90,0.00,0.12} 
\definecolor{Purple} {rgb}{0.50,0.25,0.55} 
\definecolor{Turque} {rgb}{0.00,0.65,0.85} 
\definecolor{Blue}   {rgb}{0.00,0.00,1.00} 
\definecolor{Magenta}{rgb}{1.00,0.00,1.00} 
\definecolor{Gold}   {rgb}{1.00,0.75,0.25} 
\definecolor{Seaweed}{rgb}{0.01,0.24,0.09} 
\definecolor{Brown}  {rgb}{0.43,0.26,0.32} 
\definecolor{grey1}  {rgb}{0.20,0.20,0.20} 
\definecolor{grey2}  {rgb}{0.40,0.40,0.40} 
\definecolor{grey3}  {rgb}{0.60,0.60,0.60} 
\definecolor{grey4}  {rgb}{0.80,0.80,0.80} 
\definecolor{grey5}  {rgb}{0.90,0.90,0.90} 
\def\C#1#2{{\ifcase#1\or
             \color{Green}\or \color{Orange}\or \color{Red}\or
              \color{Purple}\or \color{Turque}\or \color{Blue}\or
               \color{Magenta}\or \color{Gold}\or \color{Seaweed}\or
                \color{Brown}\or\color{grey1}\or\color{grey2}\or
                 \color{grey3}\else\color{grey4}\fi#2}}

\definecolor{Slate} {rgb}{0.00,0.45,0.55}

\definecolor{DarkViolet}{rgb}{0.35,0,0.35}

\def\rD{{\rm D}}
\def\rI{{\rm I}}
\def\rJ{{\rm J}}
\def\rK{{\rm K}}
\def\rL{{\rm L}}
\def\rM{{\rm M}}
\def\rN{{\rm N}}
\def\rR{{\rm R}}

\def\hi{{\hat\imath}}

\def\hk{{\hat{k}}}

\def\fracm#1#2{\hbox{\large{${\frac{{#1}}{{#2}}}$}}}

\def\be{\begin{equation}}
\def\ee{\end{equation}}
\newcommand{\bea}{\begin{eqnarray}}
\newcommand{\eea}{\end{eqnarray}}
\newcommand{\ena}{\end{eqnarray}}

\def\pp{{\mathchoice
          {
              \kern 1pt%
              \raise 1pt
              \vbox{\hrule width5pt height0.4pt depth0pt
                    \kern -2pt
                    \hbox{\kern 2.3pt
                          \vrule width0.4pt height6pt depth0pt
                          }
                    \kern -2pt
                    \hrule width5pt height0.4pt depth0pt}%
                    \kern 1pt
           }
            {
              \kern 1pt%
              \raise 1pt
              \vbox{\hrule width4.3pt height0.4pt depth0pt
                    \kern -1.8pt
                    \hbox{\kern 1.95pt
                          \vrule width0.4pt height5.4pt depth0pt
                          }
                    \kern -1.8pt
                    \hrule width4.3pt height0.4pt depth0pt}%
                    \kern 1pt
            }
            {
              \kern 0.5pt%
              \raise 1pt
              \vbox{\hrule width4.0pt height0.3pt depth0pt
                    \kern -1.9pt  
                    \hbox{\kern 1.85pt
                          \vrule width0.3pt height5.7pt depth0pt
                          }
                    \kern -1.9pt
                    \hrule width4.0pt height0.3pt depth0pt}%
                    \kern 0.5pt
            }
            {
              \kern 0.5pt%
              \raise 1pt
              \vbox{\hrule width3.6pt height0.3pt depth0pt
                    \kern -1.5pt
                    \hbox{\kern 1.65pt
                          \vrule width0.3pt height4.5pt depth0pt
                          }
                    \kern -1.5pt
                    \hrule width3.6pt height0.3pt depth0pt}%
                    \kern 0.5pt
            }
        }}

\def\mm{{\mathchoice
                       {
                             \kern 1pt
               \raise 1pt    \vbox{\hrule width5pt height0.4pt depth0pt
                                  \kern 2pt
                                  \hrule width5pt height0.4pt depth0pt}
                             \kern 1pt}
                       {
                            \kern 1pt
               \raise 1pt \vbox{\hrule width4.3pt height0.4pt depth0pt
                                  \kern 1.8pt
                                  \hrule width4.3pt height0.4pt depth0pt}
                             \kern 1pt}
                       {
                            \kern 0.5pt
               \raise 1pt
                            \vbox{\hrule width4.0pt height0.3pt depth0pt
                                  \kern 1.9pt
                                  \hrule width4.0pt height0.3pt depth0pt}
                            \kern 1pt}
                       {
                           \kern 0.5pt
             \raise 1pt  \vbox{\hrule width3.6pt height0.3pt depth0pt
                                  \kern 1.5pt
                                  \hrule width3.6pt height0.3pt depth0pt}
                           \kern 0.5pt}
                       }}

\def\ad{{\kern0.5pt
                   \alpha \kern-5.05pt \raise5.8pt\hbox{$\textstyle.$}\kern
0.5pt}}

\def\bd{{\kern0.5pt
                   \beta \kern-5.05pt \raise5.8pt\hbox{$\textstyle.$}\kern
0.5pt}}

\def\qd{{\kern0.5pt
                   q \kern-5.05pt \raise5.8pt\hbox{$\textstyle.$}\kern
0.5pt}}
\def\Dot#1{{\kern0.5pt
     {#1} \kern-5.05pt \raise5.8pt\hbox{$\textstyle.$}\kern
0.5pt}}

\catcode`@=11
\def\un#1{\relax\ifmmode\@@underline#1\else
        $\@@underline{\hbox{#1}}$\relax\fi}
\catcode`@=12

\def\a{\alpha}
\def\b{\beta}
\def\c{\chi}
\def\d{\delta}
\def\e{\epsilon}

\def\g{\gamma}

\def\i{\iota}
\def\j{\psi}
\def\k{\kappa}
\def\l{\lambda}
\def\m{\mu}
\def\n{\nu}
\def\v{\nu}
\def\o{\omega}

\def\r{\rho}
\def\s{\sigma}
\def\t{\tau}

\def\L{\Lambda}
\def\O{\Omega}
\def\P{\Pi}

\def\S{\Sigma}

\def\ca{{\cal A}}

\def\cf{{\cal F}}

\def\cv{{\cal V}}

\def\dslash{\not{\hbox{\kern-2pt $\partial$}}}
\def\Dslash{\not{\hbox{\kern-4pt $D$}}}
\def\pslash{\not{\hbox{\kern-2.3pt $p$}}}
 \newtoks\slashfraction
 \slashfraction={.13}
 \def\slash#1{\setbox0\hbox{$ #1 $}
 \setbox0\hbox to \the\slashfraction\wd0{\hss \box0}/\box0 }
 
\def\kcr{{\hbox{\ro \char'170}}}                
\def\ktl{{\hbox{\ro \char'170}}}        
\def\ktr{{\hbox{\ro \char'170}}}        
\def\kbl{{\hbox{\ro \char'170}}}        
\def\kbr{{\hbox{\ro \char'170}}}        

\def\plpl{\raise-2pt\hbox{$\raise3pt\hbox{$_+$}\hskip-6.67pt\raise0.0pt
\hbox{$^+$}\hskip 0.01pt$}}
\def\mimi{\raise-2pt\hbox{$\raise3pt\hbox{$_-$}\hskip-6.67pt\raise0.0pt
\hbox{$^-$}\hskip 0.01pt$}} 

\def\bo{{\raise.15ex\hbox{\large$\Box$}}}               
\def\pa{\partial}                                       
\def\TH{{\raise.2ex\hbox{$\displaystyle \bigodot$}\mskip-4.7mu \llap H \;}}
\def\face{{\raise.2ex\hbox{$\displaystyle \bigodot$}\mskip-2.2mu \llap {$\ddot
        \smile$}}}                                      

\def\dt#1{\on{\hbox{\bf .}}{#1}}                
\def\Dot#1{\dt{#1}}

\def\Tilde#1{\widetilde{#1}}                    
\def\Hat#1{\widehat{#1}}                        
\def\leftrightarrowfill{$\mathsurround=0pt \mathord\leftarrow \mkern-6mu
        \cleaders\hbox{$\mkern-2mu \mathord- \mkern-2mu$}\hfill
        \mkern-6mu \mathord\rightarrow$}
\def\dvec#1{\vbox{\ialign{##\crcr
        \leftrightarrowfill\crcr\noalign{\kern-1pt\nointerlineskip}
        $\hfil\displaystyle{#1}\hfil$\crcr}}}           
\def\dt#1{{\buildrel {\hbox{\LARGE .}} \over {#1}}}     

\def\fracm#1#2{\hbox{\large{${\frac{{#1}}{{#2}}}$}}}
\def\sfrac#1#2{{\vphantom1\smash{\lower.5ex\hbox{\small$#1$}}\over
        \vphantom1\smash{\raise.4ex\hbox{\small$#2$}}}} 
\def\bfrac#1#2{{\vphantom1\smash{\lower.5ex\hbox{$#1$}}\over
        \vphantom1\smash{\raise.3ex\hbox{$#2$}}}}       
\def\afrac#1#2{{\vphantom1\smash{\lower.5ex\hbox{$#1$}}\over#2}}    

\def\pa{\partial}

\def\ad{{\dot{\alpha}}}
\def\bd{{\dot{\beta}}}

 \font\rOpe=cmsy10                        
 \def\ktl{{\hbox{\rOpe\char'170}}}        
 \def\kbl{{\hbox{\rOpe\char'170}}}        
 \def\kcr{{\reflectbox{\rOpe\char'170}}}        
 \def\ktr{{\reflectbox{\rOpe\char'170}}}        
 \def\kbr{{\reflectbox{\rOpe\char'170}}}        
 \def\Border{\vbox{\hsize0pt
        \setlength{\unitlength}{1mm}
        \newcount\xco
        \newcount\yco
        \xco=-21
        \yco=12
        \begin{picture}(0,0)(-7.5,0)
        \put(\xco,\yco){$\ktl$}
        \advance\yco by-1
        {\loop
        \put(\xco,\yco){$\kcr$}
        \advance\yco by-2
        \ifnum\yco>-240
        \repeat
        \put(\xco,\yco){$\kbl$}}
        \xco=170
        \yco=12
        \put(\xco,\yco){$\ktr$}
        \advance\yco by-1
        {\loop
        \put(\xco,\yco){$\kcr$}
        \advance\yco by-2
        \ifnum\yco>-240
        \repeat
        \put(\xco,\yco){$\kbr$}}
        \put(-19.5,13){\scalebox{.6065}{%
         University of Maryland Center for String and Particle  Theory \&\ Physics Department%
        |University of Maryland Center for String and Particle  Theory \&\ Physics Department}}
        \put(-19.5,-241.5){\scalebox{.5835}{%
         ****University of Maryland * Center for String and
         Particle  Theory* Physics Department****University of Maryland *Center
        for String and Particle  Theory* Physics Department}}
        \end{picture}
        \par\vskip-8mm}}
\definecolor{UMred}{rgb}{.9,.05,.2}
\definecolor{HUblue}{rgb}{.0,.3,.7}

\definecolor{Red}    {rgb}{0.90,0.00,0.12} 
\definecolor{Blue}   {rgb}{0.00,0.00,1.00} 
\definecolor{Green}  {rgb}{0.10,0.70,0.10} 
\definecolor{Turque} {rgb}{0.00,0.65,0.85} 
\definecolor{Orange} {rgb}{1.00,0.50,0.15} 
\definecolor{Magenta}{rgb}{1.00,0.00,1.00} 
\definecolor{Gold}   {rgb}{1.00,0.75,0.25} 
\definecolor{Seaweed}{rgb}{0.01,0.24,0.09} 
\definecolor{Purple} {rgb}{0.50,0.25,0.55} 
\definecolor{Brown}  {rgb}{0.43,0.26,0.32} 
\definecolor{grey1}  {rgb}{0.20,0.20,0.20} 
\definecolor{grey2}  {rgb}{0.40,0.40,0.40} 
\definecolor{grey3}  {rgb}{0.60,0.60,0.60} 
\definecolor{grey4}  {rgb}{0.80,0.80,0.80} 
\definecolor{grey5}  {rgb}{0.90,0.90,0.90} 
\def\C#1#2{{\ifcase#1\or
             \color{Red}\or \color{Green}\or \color{Blue}\or\
              \color{Turque}\or \color{Orange}\or \color{Magenta}\or 
               \color{Gold}\or \color{Seaweed}\or \color{Purple}\or
                \color{Brown}\or\color{grey1}\or\color{grey2}\or
                 \color{grey3}\else\color{grey4}\fi#2}}

\definecolor{Slate} {rgb}{0.00,0.45,0.55}

\newdimen\parshift\parshift=\parindent
\catcode`@=11
 \long\def\@footnotetext#1{\insert\footins{\reset@font\footnotesize
           \interlinepenalty\interfootnotelinepenalty\splittopskip%
            \footnotesep\splitmaxdepth\dp\strutbox\floatingpenalty\@MM%
             \hsize\columnwidth\addtolength{\hsize}{-2\parindent}
              \@parboxrestore\protected@edef\@currentlabel%
              {\csname p@footnote\endcsname\@thefnmark}%
                \color@begingroup%
                 \@makefntext{\rule\z@\footnotesep\ignorespaces#1%
                  \@finalstrut\strutbox}%
                \color@endgroup}}
 \long\def\@makefntext#1{\hglue\parshift%
           \vbox{\noindent\baselineskip=11pt plus.5pt minus.5pt\hb@xt@0em{\hss\@makefnmark\kern1pt}#1}}
\catcode`@=12

\newskip\humongous \humongous=0pt plus 1000pt minus 1000pt
\def\caja{\mathsurround=0pt}
\def\eqalign#1{\,\vcenter{\openup2\jot \caja
        \ialign{\strut \hfil$\displaystyle{##}$&$
        \displaystyle{{}##}$\hfil\crcr#1\crcr}}\,}
\newif\ifdtup

\makeatletter
\def\section{\@startsection{section}{1}{\z@}
        {3ex plus-1ex minus-.2ex}{1pt plus1pt}{\large\sf\bfseries\boldmath}}
\def\subsection{\@startsection{subsection}{2}{\z@}
         {1.5ex plus-1ex minus-.2ex}{0.01pt plus1pt}{\sf\slshape}}
\def\subsubsection{\@startsection{subsubsection}{3}{\z@}
          {1.5ex plus-1ex minus-.2ex}{0.01pt plus0.2pt}{\sf\boldmath}}
\def\paragraph{\@startsection{paragraph}{4}{\z@}
           {.75ex \@plus.5ex \@minus.2ex}{-2mm}{\sf\bfseries\boldmath}}
\makeatother

\def\downup#1#2{_{#1}^{#2}}

\def\ddown#1#2{_{#1#2}}
\def\ddn#1#2{_{#1#2}}

\def\ddd#1#2#3{_{#1#2#3}}

\def\uup#1#2{^{#1#2}}

\def\sw#1#2{_{#1}^{\ #2}}

\def\uuu#1#2#3{^{#1#2#3}}

\def\up#1{^{#1}}

\def\dn#1{_{#1}}
\def\sig#1#2#3{(\s^{#1})^{#2#3}}

\def\duuu#1#2#3#4{_{#1}^{\ #2#3#4}}

\def\til#1{\Tilde{#1}}

\def\pd#1{\partial\dn#1}
\def\pu#1{\partial\up#1}
\def\gup#1{\g^#1}
\def\DDAB{[\, {\rm D}\downup ai ~,~ {\rm D}\downup bj \, ]}
\def\Fmn{F\ddn\m\n}
\def\gamatdown#1#2#3#4{(\g#2{#1})_{#3#4}}

\def\gupsw#1#2#3{(\gup#1)\sw#2#3}
\def\gupcommsw#1#2#3#4{(\g\uup[#1\g\uup#2])\sw #3#4}
\def\gupdown#1#2#3{(\gup#1)\ddn#2#3}
\def\gupcommdown#1#2#3#4{(\g\uup[#1\g\uup#2])\ddn#3#4}
\def\ggfiupsw#1#2#3{(\gup5\gup#1)\sw#2#3}
\def\ggfiupdown#1#2#3{(\gup5\gup#1)\ddn#2#3}

\def\tila{\til A}
\def\tilb{\til B}
\def\tilf{\til F}
\def\tilg{\til G}
\def\tilv{\til\varphi}
\def\tilbmn#1#2{\til B_{#1#2}}
\def\tilpsi{\til\Psi}

\def\holtil{[\, \til {\rm D}\downup ai ~,~\til {\rm D}\downup bj \,]}
\def\holcrosstil{[\, {\rm D}\downup ai ~,~ \til {\rm D}\downup bj]}

\def\algcrosstil{\{\, {\rm D}\downup ai ~,~\til {\rm D}\downup bj \, \}}
\def\tilh#1#2#3{\til H\ddd#1#2#3}
\def\lap{\square}
\def\gpart{\gupdown\m ab\pd\m}
\def\gupbrack#1#2{\g\uup[#1\g\uup#2]}
\def\gdnbrack#1#2{\g\ddn[#1\g\ddn#2]}

\def\zeeh#1{Y_{#1}^{ijkl}}
\def\tilzeeh#1{\til{Y}_{#1}^{ijkl}}
\def\xee#1{X_{#1}^{ijkl}}
\def\tilxee#1{\til{X}_{#1}^{ijkl}}
\def\xeeh#1{W_{#1}^{ijkl}}
\def\tilxeeh#1{\til{W}_{#1}^{ijkl}}
\def\vee#1{V_{#1}^{ijkl}}
\def\tilvee#1{\til{V}_{#1}^{ijkl}}

 \allowdisplaybreaks
 \seceq

\definecolor{R}{rgb}{1,0,0}

\begin{document}

\thispagestyle{empty}
%
\noindent{\small
$~~~~~~~~~~~~~~~~~~~~~~~~~~~~~~~~~~~~~~~~~~~~~~~~~~~~~~~~~~~~~~~~~$
$~~~~~~~~~~~~~~~~~~~~~~~~~~~~~~~~~~~~~~~~~~~~~~~~~~~~~~~~~~~~~~~~~$
{}
}
\vspace*{8mm}
\begin{center}
{\large \bf
Infinite-Dimensional Algebraic {$\mathfrak{Spin}$}($N$) Structure in \\[2.3pt] 
Extended/Higher Dimensional 
SUSY Holoraumy for \\[5.0pt] 
Valise and On Shell Supermultiplet Representations
}   \\   [12mm]
{\large {
S.\ James Gates, Jr.\footnote{\href{mailto:sylvester\_gates@brown.edu}{sylvester\_gates@brown.edu}}$^a$,
Gabriel Hannon\footnote{\href{mailto:gabriel\_hannon@alumni.brown.edu }{gabriel\_hannon@alumni.brown.edu }}$^a$, \\
Rui Xian Siew\footnote{\href{mailto:ruixian.siew@duke.edu}{ruixian.siew@duke.edu}}$^b$, 
and 
Kory Stiffler\footnote{\href{mailto:kory\_stiffler@brown.edu}{kory\_stiffler@brown.edu}}$^{a,c}$
}}
\\*[12mm]
\centering
{\it ${}^a$ Brown Theoretical Physics Center and Department of Physics}\\
{\it Brown University, Providence, RI 02912-1843, USA }
\\[0.2pt]
\vspace*{8mm}
\emph{$^b$Department of Physics, Duke University, \\[1pt]
Durham, NC 27710, U.S.A.
\\[1pt]
}
\vspace*{8mm}
{\it ${}^c$ Department of Physics and Astronomy}\\
{\it   The University of Iowa,   Iowa City, IA 52242, USA}
~\\[20mm]
{ ABSTRACT}\\[4mm]
\parbox{132mm}{\parindent=2pc\indent\baselineskip=14pt plus1pt
We explore the relationship between holoraumy and Hodge duality beyond four dimensions. We find this relationship to be ephemeral beyond six dimensions: it is not demanded by the structure of such supersymmetrical theories. In four dimensions for the case of the vector-tensor $\cal N$ = 4 multiplet, however, we show that such a linkage is present. 
Reduction to 1D theories presents evidence for a linkage from higher-dimensional supersymmetry to an
infinite-dimensional algebra extending {$\mathfrak{Spin}$}($N$).}
 \end{center}
\vfill
\vfill
\clearpage

\section{Introduction}
\label{sec:INTRO}

It is well known the anticommutator of two supercharge generators closes on the 
generator of translations, as the supercharges are contained in a super Lie
algebra.  Some time ago \cite{Gates:2012xb,HoLoR1,HoLoR2}, it was noted in one dimensional 
theories, there exists supersymmetric representations where the {\em {commutator}}
of two supercharge generators on the fields also and simultaneously closes by
defining a new generator that was given the name of ``holoraumy,'' but the new 
generator involves the inclusion of an additional temporal derivative.  

Thus, on these representations the supercharges augmented by the holoraumy generator 
have the potential to form a genuine algebra, and not just a super Lie algebra.  
The one dimensional representations for which this is true are characterized by 
a distinguishing feature...the engineering dimensions of all the bosons in the 
representation are identical and the engineering dimensions of all the fermions 
in the representation are identical, but distinct from that of the bosons.  Such 
representations are called one dimensional ``valise'' supermultiplets.  The 
algebra of the supercharges closes on these representations.

Subsequently, it was demonstrated \cite{HoLoR3,HoLoR4} that such operators exist for
4D, $\cal N$ = 1
representations, as well as 4D, $\cal N$ = 2 representations \cite{HoLoR5}, i.\  e.\
on manifolds with more than one dimension.  However, these higher dimensional
valise representations only exist for on-shell (i.\ e.\ in the presence of equations
of motion) theories.  The condition of being on-shell is necessary for the higher 
dimensional theories to satisfy the same conditions that are required on the 
engineering dimensions of field variables in the one dimensional valise representations.

The fact that valise representations exist in both off-shell one dimensional supersymmetrical 
theories and on-shell higher dimensional supersymmetrical theories is the central
pillar for the concept of ``SUSY holography \cite{ENUF},'' i.\ e.\ the possibility 
that the kinematic structure of higher dimensional SUSY theories can be holographically 
embedded \cite{HoLoR3,HoLoR4,HoLoR5,ENUF,AdnkMAC} into one dimensional SUSY theories.

Most recently it has been noted \cite{UBQ1} that the ``holoraumy' involves both 
electromagnetic duality transformations and Hodge duality transformations in a 
number of ``on-shell'' supermultiplet representations of 4D, $\cal N$ = 1 supersymmetry.  
From this observed behavior, it was conjectured that more generally the commutator 
of two supercharges for higher dimensional and extended supersymmetrical representations 
was likely to possess the same property.  It is the purpose of this current work to 
provide calculational exploration of this conjecture.  The current
work will also explore these concepts in the context of higher dimensional 
supersymmetrical theories.

\section{Examples of Holoraumy in Higher Dimensions}
\label{sec:ChapH}

 \subsection{Lagrangian and Transformation Laws in 10D, 6D, and 4D}
The Lagrangian for the abelian vector supermultiplet takes a unified form in 10D, 6D, and 4D 
theories where explicitly one finds
\begin{equation}
\mathcal{L} = - \tfrac{1}{4} F_{\m\n} F^{\m\n} + i \tfrac{1}{2} \lambda^a (\sigma^\m)_{ab} 
\partial_{\m} \lambda^b  ~~~,
\label{eq:Lag}
\end{equation}
with $F_{\m\n} = \partial_{[\m} A_{\n]}$ and the spinor $\lambda^a$ is a real (i.e. Majorana)
fermionic field. The transformation laws in the 10D, 6D, and 4D theories are all of the exact
same form,
\begin{align}
{\rm D}{}_{a} A_\m ~=&~ (\sigma_\m)_{ab} \l^b  ~~~,  \nonumber \\
{\rm D}{}_a \l^b ~=&~ i (\s^{\m\n})_a{}^b \partial_\m A_\n  ~~~,
\label{D-ruls}	
\end{align}
where the ranges of the vector indices (i.e. $\mu$, $\nu$, etc.) and the spinor indices
(i.e. $a$, $b$, etc.) depend on the spacetime dimension of the bosonic manifold according to:
\begin{align}\label{e:10Dindices}
	\m,\n,\dots =&~ 0,1,2,\dots ,\, 9~~~,~~~a,b,c,\dots = 1,2,3,\dots,16~\text{in 10D} ~~~, \\
	\label{e:6Dindices}
	\m,\n,\dots =&~ 0,1,2,\dots , \, 5~~~,~~~a,b,c,\dots = 1,2,3,\dots,8~\text{in 6D}~~~,  \\
	\label{e:4Dindices}
	\m,\n,\dots =&~ 0,1,2,3~~~~~~~~~,~~~a,b,c, \dots = 1,2,3,4~\text{in 4D} ~~~.
	\end{align}
The explicit forms of the $\sigma$-matrices are given in Appendix~\ref{appen::sigma} for each respective 
manifold. These are a reordering and rearrangement of those used in~\cite{Gates:2019dyk}. We choose
these new conventions as this leads to a simple dimensional reduction by taking the upper left block 
of the $\sigma$-matrices in this work.
\subsection{10D Algebra and Holoraumy}
The 10D SUSY algebra (expressed in terms of the $\rm D$-operators) is
\begin{align}\label{e:10DBosonicAlgebra}
\{ \, {\rm D}{}_a ~,~ {\rm D}{}_b \,   \} A_\m=& i \, 2  (\sigma^\n)_{ab} \partial_\n A_\m - \partial_\m \Lambda_{ab} ~~~, \\
\label{e:10DFermionicAlgebra}
\{ \, {\rm D}{}_a ~,~ {\rm D}{}_b \,   \}  \l^c =& i \, 2   (\s^\m)_{ab}  \partial_\m \lambda^c + i \Sigma_{ab}{}^{cd} (\s^\m)_{de} \partial_\m \lambda^e
~~~,		
\end{align}
with the gauge term (the right most term in (\ref{e:10DBosonicAlgebra})) for the boson and 
the term proportional to the equation of motion term (the rightmost term in (\ref{e:10DFermionicAlgebra})).  These
take the respective explicit forms
\begin{align}
	\Lambda_{ab} =& i\, 2  (\sigma^\m)_{ab} A_\m  ~~~, \\
	\S_{ab}{}^{cd} =& - \tfrac{7}{8} (\s^\m)_{ab}(\s_\m)^{cd} + \tfrac{1}{1,920} (\s^{[5]})_{ab}(\s_{[5]})^{cd} ~~~.
\end{align}
In calculating the algebra for the fermions, the following Fierz identity is useful:
\begin{align}
    (\sigma^{\m\n})_{(a}{}^c (\sigma_\n)_{b)d}=& 2 (\sigma^\m)_{ab} \delta_d{}^c - \tfrac{7}{8} (\sigma^\n)_{ab} (\s_\n)^{cf} (\s^\m)_{fd} + \tfrac{1}{1,920} (\s^{[5]})_{ab}(\s_{[5]})^{cf} (\s^\m)_{fd} ~~~.
    \label{eq:HLRvm}
\end{align}
It can be seen from the expression in (\ref{eq:HLRvm}) that the final term is curiously similar in
its algebraic structure to the auxiliary field that is required to embed the 
lowest order open-string correction into the supergeometry of a 10D space construction \cite{FX1}.

The holoraumy is
\begin{align}\label{e:10BosonicHoloraumy}
	[ \, {\rm D}{}_a ~,~ {\rm D}{}_b \,   ] A_\m=& i (\sigma_{\m\n\r})_{ab} F^{\n\r}  ~=~ i 2 \, (\sigma_{\m}{}^{\n \r})_{ab}\, \pa{}_{\n} A{}_{\r}   ~~~, \\
	\label{e:10DFermionicHoloraumy}
	[ \, {\rm D}{}_a ~,~ {\rm D}{}_b \,   ]  \l^c =& -i (\sigma^{\m\n})_{[a}{}^c (\sigma_\n)_{b]d} \partial_\m \lambda^d ~~~.
\end{align}
It is readily apparent that the holoraumy calculation in (\ref{e:10BosonicHoloraumy}) does 
{\it {not}} involve the dual tensor of the 10D Maxwell field strength, at least not with the smallest number of products of linearly independent $\s$-matrices as written.\footnote{One can of course define the dual field strength as $\tilde{F}_{\mu_1\mu_2 \dots \mu_8} \sim \epsilon_{\mu_1 \mu_2 \dots \mu_8 [2]} F^{[2]}$ and write the bosonic holoraumy as proportional to $\s^{[7]}\tilde{F}_{\mu[7]}$, however, this reduces owing to identity $\sigma_{\mu\nu\alpha} = -\tfrac{1}{7!}\epsilon_{\mu\nu\alpha[7]}\sigma^{[7]}$.} Instead the Maxwell
tensor itself appears.  Thus the ubiquitous 
nature of the relation between holoraumy and electromagnetic duality noted in the work 
of \cite{UBQ1} in the context of 4D, $\cal N$ = 1 theories does {\it {not}} apply in the
case of the 10D, $\cal N$ = 1 super Maxwell theory.

\subsection{Reduction to 6D}
We reduce to 6D by setting to zero the `last' four components of the 10D gauge field $A{}_{\m}$ and
'last' eight components of the spinor field $ \l^a$ according to
\begin{align}
A_6 = A_7 = A_8= A_9 = 0~~~,~~~\l^a = 0~~~\text{for}~a = 9,10,\dots 16 ~~~
\end{align}
We consider the remaining field components to depend only on the 6D coordinates. The bosonic algebra reduces to that in Eq.~(\ref{e:10DBosonicAlgebra}) with indices ranging over the 6D values in Eq.~\eqref{e:6Dindices}. The bosonic holoraumy reduces to
\begin{align}
[\, {\rm D}{}_a ~,~ {\rm D}{}_b \,  ] A_\m =& \tfrac{1}{3} i (\sigma^{[3]})_{ab} \tilde{F}_{\m[3]} ~~~,
\label{e:6dHolor}
\end{align}
with the four-form dual Maxwell tensor equal to
\begin{align}
	\tilde{F}_{\k\l\m\n} = \tfrac{1}{2} \epsilon_{\k\l\m\n[2]}F^{[2]} ~~~, 
\end{align}
and with $\epsilon_{\k\l\m\n\r\s}$ the completely antisymmetric Levi-Civita tensor in 6D. In the above calculations we have made use of the following relationship, valid in 6D
\begin{align}\label{e:sigma6dual}
	\sigma_{\m\n\l} =& \tfrac{1}{6} \epsilon_{\m\n\l[3]}\sigma^{[3]}  ~~~.
\end{align}
Thus, although the duality interpretation of the holoraumy was {\it {not}} present for the theory in ten dimensions,
here we have found that it ``reappears'' for the six dimensional theory. This can obviously be seen as a function of the number of indices versus the dimension of the space-time: 6D is the critical dimension where the dual field strength would be able to appear in the holoraumy with linearly-independent $\s$-matrices, owing to the duality relationship in Eq.~\eqref{e:sigma6dual}, and the fact that the dual field strength has four indices in 6D.
\subsection{Reduction to 4D}
We continue and reduce from 4D to 6D by setting to zero
\begin{align}
	A_4 = A_5 = 0~~~,~~~ \l_5 = \l_6 = \l_7 = \l_8 = 0  ~~~,
\end{align}
and imposing the additional restriction on all remaining field components to depend only on the 4D coordinates. The algebra reduces to that in Eq.~(\ref{e:10DBosonicAlgebra}) with indices ranging over the 4D values in Eq.~\eqref{e:4Dindices}. The bosonic holoraumy reduces to
\begin{align}
[\, {\rm D}{}_a ~,~ {\rm D}{}_b \,  ] A_\m =& 2 (\gamma^5 \gamma^\n)_{ab} \tilde{F}_{\m\n} ~~~, 
\label{e:4dHolor}	
\end{align}
with the dual two-form equal to
\begin{align}
	\tilde{F}_{\m\n} = \tfrac{1}{2} \epsilon_{\m\n\k\l}F^{\k\l} ~~~.
\end{align}
In the above calculations we have used
\begin{align}
	(\sigma_{\m\n\l})_{ab} = i \epsilon_{\m\n\l\k} (\gamma^5 \sigma^\k)_{ab}  ~~~.
\end{align}

The results in (\ref{e:6dHolor}) and (\ref{e:4dHolor}) demonstrate an interesting pattern.
Apparently the relationship between the holoraumy of the gauge field and the electromagnetic
duality is present in the 6D and 4D theory, but this relationship ``evaporates,'' and is
not valid (\ref{e:10DFermionicHoloraumy}) in the 10D theory.

The fact of the ``evaporation'' raises an interesting question.  Ordinarily, and most
often one regards 4D, $\cal N$ = 4 theories as being the result of a dimensional reduction
from a 10D theory.  So are there \emph{no} on-shell 4D, $\cal N$ = 4 multiplets
that have relationships between holoraumy and electromagnetic duality?  The most obvious
context in which to explore this question is within 4D, $\cal N$ = 4 supergravity
theories.  However, clearly such an exploration is a substantial undertaking simply
due to the complexity of such supermultiplets.  There is an intermediate theory
which does not possess such a high degree of complication.  There exists a ``variant'' 
form \cite{SSW} of the 4D, $\cal N$ = 4 Maxwell supermultiplet where one of the spin-0 
fields is replaced by a skew second rank tensor.  In the following, we will study the
issue of a possible relationship between the holoraumy and electromagnetic duality
within this supermultiplet.

\section{On Shell Holoraumy Results}
\label{sec:HR}
The basis upon which we undertake the investigation is provided by two
separate 4D, $\cal N$ = 2 supermultiplets.  One consists of the 4D, $\cal N$ = 2
vector supermultiplet consisting of component fields, ($A$, $B$, $F$, $G$, $A
{}_{\mu}$, $d$, $\Psi{}_{a}^i$) and the second supermultiplet is the 4D, $\cal 
N$ = 2 tensor supermultiplet consisting of component fields, (${\Tilde A}$, ${\Tilde B}$, 
${\Tilde F}$, ${\Tilde G}$, ${\Tilde \varphi}$,  $B{}_{\mu \, \nu}$, ${\Tilde \Psi}
{}_{a}^i$).  The Latin indices $i$, $j$, $\dots$ here take on values of 1 and 2.  
There are four supercovariant derivative operators ${\rm D}\downup ai $, and ${\Tilde {\rm D}}
\downup ai $ and their realizations on the component fields were given in the work of 
\cite{N4VTM}.  We have included these, but they are relegated to Appendix \ref{appen:Transformation}. An 
advantage of this formulation is that the ${\rm D}\downup ai $ and ${\Tilde {\rm D}}\downup ai $ operators are each individually off-shell, 
i.e. close without the use of equations of motion and central charges.  This is 
not the case for the cross terms in the algebra between ${\rm D}\downup ai $ and ${\Tilde {\rm D}}\downup ai $.

The only totally on-shell fields from the first 4D, $\cal N$ = 2 supermultiplet 
are ($A$, $B$, $A{}_{\mu}$, $\Psi{}_{a}^i$) and for the second supermultiplet is the 4D, $\cal 
N$ = 2 supermultiplet consisting of component fields, (${\Tilde A}$, ${\Tilde B}$, 
${\Tilde \varphi}$,  $B{}_{\mu \, \nu}$, ${\Tilde \Psi}{}_{a}^i$).  Since
our goal in this work is only to consider the purely on-shell holoraumy 
we include the partial off-shell starting point results in an appendix as this contains possible results 
that will be important for future work and explorations. We perform the standard reduction to the on-shell system, imposing the equations of motion which has the effect of removing all auxiliary fields from the transformation laws. on-shell algebra results are explained in appendix~\ref{a:onshell}.

In order to realize an $\cal N $ = 4 on-shell SUSY system, we will require {\it {two}} independent
superspace derivative operators denoted by ${\rm D}{}_a^i$ and ${\Tilde {\rm D}}{}_a^i$ where
the ``isospin'' label on each take the values $i$ = 1, and 2. In the remainder of this section and the next two, we show the on-shell holoraumy results. These are derived from the fully on-shell transformation laws arrived at by taking the transformation laws in Appendix~\ref{appen:Transformation} and setting all auxiliary fields to zero ($F = G = d = {\Tilde F} = {\Tilde G}  =0)$. The fermionic holoraumies are shown upon enforcing the Dirac equation $(\g^\mu)_a{}^c \pa_\mu \Psi^k_c = (\g^\mu)_a{}^c \pa_\mu \widetilde{\Psi}^k_c = 0$. The terms involving the Dirac equation can be found in the associated \emph{Mathematica} code that is freely available online through \emph{GitHub} at \href{https://hepthools.github.io/Data/4DN4Holo/}{https://hepthools.github.io/Data/4DN4Holo/}.
\subsection{Vector Multiplet \texorpdfstring{${\rm D}$-${\rm D}$}{D-D} Holoraumy}
\vspace*{-24 pt}
\begin{align}
    \DDAB A&= -2\d\uup ij\ggfiupdown\m ab\pd\m B+\fracm{1}{2}\sig2ij\gupcommdown\m\n ab F\ddn\m\n ~~~, \nonumber\\
    \DDAB B &= 2\d\uup ij\ggfiupdown\m ab\pd\m A + i \, \fracm{1}{2}\sig2ij(\gup5\gupbrack\m\n)\ddn ab F\ddn\m\n ~~~, \nonumber\\
    \DDAB A\dn\m &=\sig2ij\, [ \,(\gdnbrack\m\n)\ddn ab\pu\n A+i(\gup5\gdnbrack\m\n)\ddn ab\pu\n B\, ]
    -\d\uup ij\e\duuu\m\n\a\b\ggfiupdown\n abF\ddn\a\b  ~~~, \nonumber\\
    \DDAB\Psi\downup ck &= i\, 2\sig2ij\sig2kl\gpart\Psi\downup cl
    \cr
    &~~~ +i\, 2\, [ \,\sig1ij\sig1kl+\sig3ij\sig3kl\, ]\ggfiupdown\m ab\gupsw5cd\pd\m\Psi\downup dl  ~~~.
\end{align}

\subsection{Vector Multiplet \texorpdfstring{${\rm D}$-${\Tilde {\rm D}}$}{D-D} Holoraumy}
\vspace*{-24 pt}
\begin{align}
    \holcrosstil{}A&=i\, 2 \sig2ij\ggfiupdown\m ab\pd\m\tilb - \fracm{2}{3}\d\uup ij\e\duuu\m\n\a\b\ggfiupdown\m ab\tilh\n\a\b
        \nonumber\\
    \holcrosstil{}B&=-2\sig1ij\ggfiupdown\m ab\pd\m\tila + 2\sig3ij\ggfiupdown\m ab\pd\m\tilv  \nonumber\\
    \holcrosstil{}A\dn\m&= 2\d\uup ij\gupdown5ab\pd\m\tilb + i(\gdnbrack\m\l)\ddn ab\pu\l\{\sig3ij\tila+\sig1ij\tilv\}
    +i \fracm{2}{3}\sig2ij\e\duuu\m\n\a\b\gupdown5ab\tilh\n\a\b  
    \nonumber \\
\holcrosstil{}\Psi\downup ck &= i\, 2\vee1\gpart\tilpsi\downup cl +2i\vee3\gupcommdown\m\n ab(\g\dn\n)\sw cd\pd\m\tilpsi\downup dl
 +i \,2\vee5\ggfiupdown\m ab\gupsw5cd\pd\m\tilpsi\downup dl 
\end{align}
where $\vee1$, $\vee3$, and $\vee5$ are defined in subsection (\ref{VMcrossfermion}).

\subsection{Vector Multiplet \texorpdfstring{${\Tilde {\rm D}}$-${\Tilde {\rm D}}$}{D-D} Holoraumy}
\vspace*{-24 pt}
\begin{align}
    \holtil{}A&=-2\d\uup ij\ggfiupdown\m ab\pd\m B-\fracm{1}{2}\sig2ij\gupcommdown\m\n ab\Fmn ~~~, \nonumber \\
    \holtil{}B&=2\d\uup ij\ggfiupdown\m ab\pd\m A - i \fracm{1}{2}\sig2ij(\gup5\gupbrack\m\n)\ddn ab\Fmn ~~~,  \nonumber\\
    \holtil{}A\dn\m &=-\sig2ij\, [ \, (\gdnbrack\m\n)\ddn ab\pu\n A+i(\gup5\gdnbrack\m\n)\ddn ab\pu\n B\, ]  
    -\d\uup ij\e\duuu\m\n\a\b(\gup5\g\dn\n)\ddn ab F\ddn \a\b  ~~~, \nonumber\\
    \holtil{}\Psi\downup ck &= -i\sig2ij\sig2kl\gupcommdown\m\n ab(\g\dn\m)\sw cd\pd\n\Psi\downup dl  
    \cr
    &~~~+2i\d\uup ij\d\uup kl\ggfiupdown\m ab\gupsw5cd\pd\m\Psi\downup dl  ~~~.
\end{align}

\subsection{Tensor Multiplet \texorpdfstring{${\rm D}$-${\rm D}$}{D-D} Holoraumy}
\vspace*{-24 pt}
\begin{align}
    \DDAB\tila &= -2\sig3ij\ggfiupdown\m ab\pd\m\tilb+2\sig2ij\gpart\tilv{} 
    +\fracm{2}{3}\sig1ij\e\duuu\m\n\a\l\ggfiupdown\m ab\tilh\n\a\b  ~~~, \nonumber\\
    \DDAB\tilb &= 2\ggfiupdown\m ab\, [ \, \sig3ij\pd\m\tila+\sig1ij\pd\m\tilv \, ] 
    -\fracm{2}{3}\sig2ij\e\duuu\m\n\a\b(\g\up\m)\ddn ab\tilh\n\a\b  ~~~, \nonumber\\
    \DDAB\tilv&=-2\sig1ij\ggfiupdown\m ab\pd\m\tilb - 2\sig2ij\gpart\tila 
    -\fracm{2}{3}\sig3ij\e\duuu\m\n\a\b\ggfiupdown\m ab\tilh\n\a\b  ~~~,  \nonumber\\
    \DDAB\tilbmn\m\n&= \sig2ij\e_{\m\n}^{\ \ \d\l}(\g\dn\d)\ddn ab\pd\l\tilb +\e_{\m\n}^{\ \ \d\l}(\gup5\g\dn\d)\ddn ab\, [ \,-\sig1ij\pd\l\tila+\sig3ij\pd\l\tilv\, ]\nonumber\\
    &{~~~\,} +\fracm{1}{3}\d\uup ij\e^{\l\a\b}_{\ \ \ [\m}(\gup5\g\ddn\n])\ddn ab\tilh\l\a\b ~~~, \nonumber\\
    \DDAB\tilpsi\downup ck &= -i\sig2ij\sig2kl\gupcommdown\m\n ab(\g\dn\m)\sw cd\pd\n\tilpsi\downup dl
     +i \,2\d\uup ij\d\uup kl\ggfiupdown\m ab\gupsw5cd\pd\m\tilpsi\downup dl ~~~.
\end{align}

\subsection{Tensor Multiplet \texorpdfstring{${\rm D}$-${\Tilde {\rm D}}$}{D-D} Holoraumy}
\vspace*{-24 pt}
\begin{align}
    \holcrosstil{}\tila &= 2\sig1ij\ggfiupdown\m ab\pd\m B+\fracm{i}{2}\sig3ij\gupcommdown\m\n ab\Fmn
    ~~~, \nonumber\\
    \holcrosstil{}\tilb&=-i\, 2\sig2ij\ggfiupdown\m ab\pd\m A  ~~~, \nonumber\\
    \holcrosstil{}\tilv &= -2\sig3ij\ggfiupdown\m ab\pd\m B+i\, \fracm{1}{2}\sig1ij\gupcommdown\m\n ab\Fmn~~~, \nonumber\\
    \holcrosstil{}\tilbmn\m\n&= \d\uup ij\e_{\m\n}^{\ \ \d\l}(\gup5\g\dn\d)\ddn ab\pd\l A+\d\uup ij(\gup5\g\ddn[\m)\ddn ab\partial\ddn\n]B  
-C_{ab}\sig2ij F_{\m\n} 
    + i\, \fracm{1}{2}\sig2ij\e_{\m\n}^{ \ \ \ \r\s}(\gup5)\ddn ab F_{\r\s} ~~~,  \nonumber \\
\holcrosstil{}\tilpsi\downup ck &= i\, 2 \tilvee1\gpart\Psi\downup cl +2i\tilvee3\gupcommdown\m\n ab(\g\dn\n)\sw cd\pd\m\Psi\downup dl\nonumber\\
&{~~~\,} +2i\tilvee5\ggfiupdown\m ab\gupsw5cd\pd\m\Psi\downup dl ~~~.
\end{align}
where $\tilvee1$, $\tilvee3$, and $\tilvee5$are defined in subsection (\ref{TMcrossfermionApp}).

\subsection{Tensor Multiplet \texorpdfstring{${\Tilde {\rm D}}$-${\Tilde {\rm D}}$}{D-D} Holoraumy}
\vspace*{-24 pt}
\begin{align}
    \holtil\tila&= 2\sig3ij\ggfiupdown\m ab\pd\m\tilb+2\sig2ij\gpart\tilv{}
    +\fracm{2}{3}\sig1ij\e\duuu\m\n\a\b\ggfiupdown\m ab\tilh\n\a\b  ~~~, \nonumber\\
    \holtil\tilb&=-2\ggfiupdown\m ab\pd\m\{\sig1ij\tilv{}+\sig3ij\tila\}
    +\fracm{2}{3}\sig2ij\e\duuu\m\n\a\b\gupdown\m ab\tilh\n\a\b ~~~, \nonumber\\
    \holtil{}\tilv{}&= 2\sig1ij\ggfiupdown\m ab\pd\m\tilb-2\sig2ij\gpart\tila
    -\fracm{2}{3}\sig3ij\e\duuu\m\n\a\b\ggfiupdown\m ab\tilh\n\a\b ~~~, \nonumber\\
    \holtil\tilbmn\m\n&=-\sig2ij\e_{\m\n}^{\ \ \d\l}(\g\dn\d)\ddn ab\pd\l\tilb +\e_{\m\n}^{\ \ \d\l}(\gup5\g\dn\d)\ddn ab\, \partial_\lambda[\, -\sig1ij\tila+\sig3ij\tilv \,] \nonumber\\
    &{~~~\,} +\fracm{1}{3}\d\uup ij\e^{\k\a\b}_{\ \ \ [\m}(\gup5\g\ddn\n])\ddn ab\tilh\k\a\b  ~~~, \nonumber\\
    \holtil\tilpsi\downup ck&= i\, 2 \sig2ij\sig2kl\gpart\Psi\downup cl
    \cr
    &~~~ -i \, 2 [\sig1ij\sig1kl+\sig3ij\sig3kl\, ] \ggfiupdown\m ab\gupsw5cd\pd\m\Psi\downup dl ~~~.
\end{align}

\clearpage
\section{Exploring Electromagnetic Rotations on Bosons in the On Shell Results}
\label{sec:EER}
\subsection{Vector Holoraumy}
\label{sec:EER1}
Starting with vector transformations, we can write
\begin{align}
     \DDAB{}A &= -2\d\uup ij(\gup5\gup\m)\ddn ab\pd\m B +\sig2ij(\g\uup[\m\g\uup\n])\ddn ab\pd\m A\dn\n ~~~, \nonumber \\
    \DDAB{}B &= 2\d\uup ij(\gup5\gup\m)\ddn ab\pd\m A + i \sig2ij(\gup5\g\uup[\m\g\uup\n])\ddn ab\pd\m A\dn\n ~~~,\nonumber  \\
      \DDAB A\dn\m &= \sig2ij\, [ \, ([\g\dn\m,\g\up\n])\ddn ab\pd\n A + i(\gup5[\g\dn\m,\g\up\n])\ddn ab\pd\n B\, ]
    - 2\d\uup ij\e\duuu\m\n\a\b(\gup5\g\dn\n)\ddn ab\pd\a A\dn\b ~~~.
\label{eq:H0LR1}    
\end{align}
These allowing us to construct a 6-by-6 matrix, which we will implicitly represent as a 3-by-3 matrix by introducing a set of row and column indices I, J which run from 0-5, with $\m$ and $\n$ representing the 0-3 indices. We also introduce a column vector, $\Phi\dn J$ to contain all the bosons as follows:
\begin{align}
     (\mathcal{B}^{ij(VM)}_{ab})\sw IJ &= 
 \begin{pmatrix}- 2\d\uup ij\e\duuu\m\r\l\n(\gup5\g\dn\r)\ddn ab\pd\l  & \sig2ij([\g\dn\m,\g\up\l])\ddn ab\pd\l & i\sig2ij(\gup5[\g\dn\m,\g\up\l])\ddn ab\pd\l 
 \\\sig2ij(\g\uup[\l\g\uup\n])\ddn ab\pd\l& 0 & -2\d\uup ij\ggfiupdown\l ab\pd\l \\
i\sig2ij(\gup5\g\uup[\l\g\uup\n])\ddn ab\pd\l &2\d\uup ij\ggfiupdown\l ab\pd\l &0 \end{pmatrix}
\label{eq:H0LR2}
\end{align}
\begin{align}
\Phi\dn J &= \begin{pmatrix}A\dn0\\A\dn1\\A\dn2\\A\dn3\\ A\\ B\end{pmatrix}
\label{eq:H0LR3}
\end{align}
\begin{align}
\DDAB{}\begin{pmatrix}A\dn\m \\A\\B
\end{pmatrix} &= (\mathcal{B}^{ij(VM)}_{ab})\sw IJ\begin{pmatrix}A\dn\n \\A\\B
\end{pmatrix} 
\label{eq:H0LR4}
\end{align}

Using this same type of framework, we can translate the tilded laws into a similar form
\begin{align}
    \holtil{}A&=-2\d\uup ij\ggfiupdown\m ab\pd\m B-\sig2ij\gupcommdown\m\n ab\pd\m A\dn\n ~~~, \nonumber \\
    \holtil{}B&=2\d\uup ij\ggfiupdown\m ab\pd\m A -i\sig2ij(\gup5\gupbrack\m\n)\ddn ab\pd\m A\dn\n ~~~, \nonumber  \\
    \holtil{}A\dn\m &=-\sig2ij\{(\gdnbrack\m\n)\ddn ab\pu\n A+i(\gup5\gdnbrack\m\n)\ddn ab\pu\n B\}\nonumber\\
    &{~~~\,}-2\d\uup ij\e\duuu\m\n\a\b(\gup5\g\dn\n)\ddn ab \pd\a A\dn\b ~~~.
\label{eq:H0LR5}
\end{align}
Exchanging some dummy indices and raising and lowering some space time indices, this leads to
\begin{align}
     (\til{\mathcal{B}}^{ij(VM)}_{ab})\sw IJ &= 
 \begin{pmatrix}- 2\d\uup ij\e\duuu\m\r\l\n(\gup5\g\dn\r)\ddn ab\pd\l  & -\sig2ij([\g\dn\m,\g\up\l])\ddn ab\pd\l & -i\sig2ij(\gup5[\g\dn\m,\g\up\l])\ddn ab\pd\l 
 \\-\sig2ij(\g\uup[\l\g\uup\n])\ddn ab\pd\l& 0 & -2\d\uup ij\ggfiupdown\l ab\pd\l \\
-i\sig2ij(\gup5\g\uup[\l\g\uup\n])\ddn ab\pd\l &2\d\uup ij\ggfiupdown\l ab\pd\l &0 \end{pmatrix}
\label{eq:H0LR6}
\end{align}
So we see that this is the same matrix as the untilded transformation law except with a sign change in all terms associated with the crossing between gauge terms and chiral fields.

For the vector multiplet ${\rm D}$-${\Tilde {\rm D}}$ holoraumy, the vector multiplets are transformed into the tensor multiplets. We have 

\begin{align}
    \holcrosstil{}A\dn\m&= 2\d\uup ij\gupdown5ab\pd\m\tilb + i(\gdnbrack\m\l)\ddn ab\pu\l\{\sig3ij\tila+\sig1ij\tilv\}
    +i 2\sig2ij\e\duuu\m\n\a\b\gupdown5ab \pd\n \til B_{\a\b} ~~~,\nonumber\\
    \holcrosstil{}A&=i\, 2 \sig2ij\ggfiupdown\m ab\pd\m\tilb -2\d\uup ij\e\duuu\m\n\a\b\ggfiupdown\m ab \pd\n \til B_{\a\b}
    ~~~,\nonumber\\
    \holcrosstil{}B&=-2\sig1ij\ggfiupdown\m ab\pd\m\tila + 2\sig3ij\ggfiupdown\m ab\pd\m\tilv ~~~.
\label{eq:H0LR7}
\end{align}
This can be succinctly written as
\begin{align}
 \holcrosstil \begin{pmatrix}A\dn\m \\A\\B
\end{pmatrix} &= (\mathcal{A}^{ij(VM)}_{ab})\sw IJ \begin{pmatrix}\til{B}\ddn\a\b \\\til{A} \\\til{B}\\\til{\varphi}
\end{pmatrix}
\label{eq:H0LR8}
\end{align}

\begin{align}
 &(\mathcal{A}^{ij(VM)}_{ab})\sw IJ =\cr
&\begin{pmatrix}
i 2\sig2ij\e\duuu\m\l\a\b\gupdown5ab \pd\l  & i\sig3ij(\gdnbrack\m\l)\ddn ab\pu\l & 2\d\uup ij\gupdown5ab\pd\m &  i\sig1ij(\gdnbrack\m\l)\ddn ab\pu\l\\
-2\d\uup ij\e\duuu\k\l\a\b\ggfiupdown\k ab \pd\l & 0 & i\, 2 \sig2ij\ggfiupdown\l ab\pd\l & 0 \\
0 & -2\sig1ij\ggfiupdown\l ab\pd\l  & 0 & 2\sig3ij\ggfiupdown\l ab\pd\l 
\end{pmatrix} 
\label{eq:H0LR9}
\end{align}
where the index $J := \alpha\beta$ for $J=0,\cdots,5$. 

\subsection{Tensor Holoraumy}

The steps for the three sectors of the bosonic holoraumy on the tensor supermultiplet
follows the same steps as used for the vector supermultiplet and yields the results
we report through the end of this chapter.
\begin{align}
    \DDAB\tila &= -2\sig3ij\ggfiupdown\m ab\pd\m\tilb+2\sig2ij\gpart\tilv{}
    +2\sig1ij\e\duuu\m\n\a\l\ggfiupdown\m ab\pd\n\tilbmn\a\l ~~~, \nonumber \\
    \DDAB\tilb &= 2\ggfiupdown\m ab\{\sig3ij\pd\m\tila+\sig1ij\pd\m\tilv\}
    -2\sig2ij\e\duuu\m\n\a\b(\g\up\m)\ddn ab\pd\n\tilbmn\a\b ~~~, \nonumber \\
    \DDAB\tilv&=-2\sig1ij\ggfiupdown\m ab\pd\m\tilb - 2\sig2ij\gpart\tila
    -2\sig3ij\e\duuu\m\n\a\b\ggfiupdown\m ab\pd\n\tilbmn\a\b ~~~,  \nonumber \\
    \DDAB\tilbmn\m\n&= \sig2ij\e_{\m\n}^{\ \ \d\l}(\g\dn\d)\ddn ab\pd\l\tilb +\e_{\m\n}{}^{ \d\l}(\gup5\g\dn\d)\ddn ab\{-\sig1ij\pd\l\tila+\sig3ij\pd\l\tilv\}\nonumber\\
    &{~~~\,} +\d\uup ij\e^{\l\a\b}{}_{ [\m}(\gup5\g\ddn\n])\ddn ab\pd\lambda\tilbmn\a\b
\label{eq:H0LR10}
\end{align}
\begin{align}
   &  (\mathcal{B}^{ij(TM)}_{ab})\sw IJ =  \cr
&
\text{\footnotesize 
$\begin{pmatrix}
\d\uup ij\e^{\l\a\b}{}_{ [\m}(\gup5\g\ddn\n])\ddn ab\pd\l & -\sig1ij\e_{\m\n\s}{}^{\l}\ggfiupdown\s ab\pd\l & \sig2ij\e_{\m\n\d}{}^{\l}\gupsw\d ab\pd\l & \sig3ij\e_{\m\n\s}{}^{ \l}\ggfiupdown\s ab\pd\l \\
2\sig1ij\e\duuu\k\l\a\b\ggfiupdown\k ab\pd\l & 0 & -2\sig3ij\ggfiupdown\l ab\pd\l & 2\sig2ij\gupdown\l ab\pd\l \\
-2\sig2ij\e\duuu\k\l\a\b\gupdown\k ab\pd\l & 2\sig3ij\ggfiupdown\l ab\pd\l  & 0 & 2\sig1ij\ggfiupdown\l ab\pd\l\\
-2\sig3ij\e\duuu\k\l\a\b\ggfiupdown\k ab\pd\l& -2\sig2ij\gupdown\l ab\pd\l  & -2\sig1ij\ggfiupdown\l ab\pd\l & 0\end{pmatrix}
$
}
\label{eq:H0LR11}
\end{align}

\begin{align}
\DDAB{}\begin{pmatrix}\til{B}\ddn\m\n \\\til{A} \\\til{B}\\\til{\varphi}
\end{pmatrix} &= (\mathcal{B}^{ij(TM)}_{ab})\sw IJ\begin{pmatrix}\til{B}\ddn\a\b \\\til{A} \\\til{B}\\\til{\varphi}
\end{pmatrix} = (\mathcal{B}^{ij(TM)}_{ab})\sw IJ\Phi\dn J
\label{eq:H0LR13}
\end{align}
\begin{align}
\Phi\dn J= \begin{pmatrix}
\tilbmn 01\\\tilbmn02\\\tilbmn03\\\tilbmn12\\\tilbmn13\\\tilbmn23\\\tila\\\tilb\\\tilv
\end{pmatrix}
\label{eq:H0LR12}
\end{align}
Here, the index $I  := \mu\nu$ and $J := \a\b$ for $I,J = 0,\dots 5$.
Then for the tilde transformations
\begin{align}
    \holtil\tila=& 2\sig3ij\ggfiupdown\m ab\pd\m\tilb+2\sig2ij\gpart\tilv{}
    +2\sig1ij\e\duuu\m\n\a\b\ggfiupdown\m ab\pd\n\tilbmn\a\b   \nonumber \\
    \holtil\tilb=&-2\ggfiupdown\m ab\pd\m\{\sig1ij\tilv{}+\sig3ij\tila\}
    +2\sig2ij\e\duuu\m\n\a\b\gupdown\m ab\pd\n\tilbmn\a\b   \nonumber \\
    \holtil{}\tilv{}=& 2\sig1ij\ggfiupdown\m ab\pd\m\tilb-2\sig2ij\gpart\tila
    -2\sig3ij\e\duuu\m\n\a\b\ggfiupdown\m ab\pd\n\tilbmn\a\b  \nonumber \\
    \holtil\tilbmn\m\n=&-\sig2ij\e_{\m\n}^{\ \ \d\l}(\g\dn\d)\ddn ab\pd\l\tilb +\e_{\m\n}^{\ \ \d\l}(\gup5\g\dn\d)\ddn ab\pd\l\{-\sig1ij\tila+\sig3ij\tilv\}\nonumber\\
    &{~~~\, }+\d\uup ij\e^{\k\a\b}_{\ \ \ [\m}(\gup5\g\ddn\n])\ddn ab\pd\k\tilbmn\a\b
\label{eq:H0LR14}
\end{align}
\begin{align}
     &(\til{\mathcal{B}}^{ij(TM)}_{ab})\sw IJ = \nonumber\\
&
\text{\footnotesize 
$
\begin{pmatrix}
\d\uup ij\e^{\l\a\b}{}_{ [\m}(\gup5\g\ddn\n])\ddn ab\pd\l & -\sig1ij\e_{\m\n\s}{}^{\l}\ggfiupdown\s ab\pd\l & -\sig2ij\e_{\m\n\d}{}^{ \l}\gupdown\d ab\pd\l & \sig3ij\e_{\m\n\s}{}^{\l}\ggfiupdown\s ab\pd\l \\
2\sig1ij\e\duuu\k\l\a\b\ggfiupdown\k ab\pd\l & 0 & 2\sig3ij\ggfiupdown\l ab\pd\l & 2\sig2ij\gupdown\l ab\pd\l \\
2\sig2ij\e\duuu\k\l\a\b\gupdown\k ab\pd\l & -2\sig3ij\ggfiupdown\l ab\pd\l  & 0 & -2\sig1ij\ggfiupdown\l ab\pd\l\\
-2\sig3ij\e\duuu\k\l\a\b\ggfiupdown\k ab\pd\l& -2\sig2ij\gupdown\l ab\pd\l  & 2\sig1ij\ggfiupdown\l ab\pd\l & 0\end{pmatrix}
$
}
\label{eq:H0LR15}
\end{align}

For the tensor multiplet ${\rm D}$-${\Tilde {\rm D}}$ holoraumy, the tensor multiplets are transformed into the vector multiplets. We have

\begin{align}
    \holcrosstil{}\tilbmn\m\n=& \d\uup ij\e_{\m\n}{}^{\d\l}(\gup5\g\dn\d)\ddn ab\pd\l A+\d\uup ij(\gup5\g\ddn[\m)\ddn ab\partial\ddn\n]B+C_{ab}\sig2ij \d_{[\m}{}^\r \partial_{\n]}  A_{\r} \cr
    &+ i\,\sig2ij\e_{\m\n}{}^{\l\r}(\gup5)\ddn ab \pd\l A_\r ~~~,\nonumber\\
    \holcrosstil{}\tila =& 2\sig1ij\ggfiupdown\m ab\pd\m B+i\sig3ij\gupcommdown\m\n ab\pd\m A_\n
    ~~~, \nonumber\\
    \holcrosstil{}\tilb =& -i\, 2\sig2ij\ggfiupdown\m ab\pd\m A  ~~~, \nonumber\\
    \holcrosstil{}\tilv =& -2\sig3ij\ggfiupdown\m ab\pd\m B+i\, \sig1ij\gupcommdown\m\n ab \pd\m A_\n~~~.
\label{eq:H0LR16}
\end{align}
\begin{align}
 \holcrosstil \begin{pmatrix}\til{B}\ddn\m\n \\\til{A} \\\til{B}\\\til{\varphi}
\end{pmatrix} &= (\mathcal{A}^{ij(TM)}_{ab})\sw IJ\begin{pmatrix}A\dn\a \\A\\B
\end{pmatrix}
\label{eq:H0LR17}
\end{align}

\newpage
\begin{align}
&(\mathcal{A}^{ij(TM)}_{ab})\sw IJ = \cr
&
\begin{pmatrix}
\sig2ij[i\e_{\m\n}{}^{ \l\a}(\gup5)\ddn ab\pd\l+C_{ab}\d_{[\m}{}^\a\partial_{\n]}] & \d\uup ij\e_{\m\n\k}{}^{\l}\ggfiupdown\k ab\pd\l & \d\uup ij(\g^5\g_{[\m})_{ab}\partial_{\n]}\\
-i\sig3ij\gupcommdown\a\l ab\pd\l & 0 & 2\sig1ij\ggfiupdown\l ab\pd\l \\
0& -2\sig2ij\ggfiupdown\l ab\pd\l  & 0 \\
-i\sig1ij\gupcommdown\a\l ab\pd\l & 0  & -2\sig3ij\ggfiupdown\l ab\pd\l 
\end{pmatrix}
\label{eq:H0LR18}
\end{align}
where here $I := \mu\nu$ for $I= 0, \dots, 5$ and $J := \alpha$ for $J = 0,1,2,3$.

\clearpage
\section{Exploring Electromagnetic Rotations on Fermions in the On-shell Results}
\label{sec:EERf}
The results in the last section refer to the evaluation of the holoraumy
calculations on the bosonic fields on-shell.  There are equivalent ways to reach
results on fermions.  One way to obtain these is by application of the ${\rm D}{}_a^i$ 
and ${\Tilde {\rm D}}{}_a^i$ operators to both sides of the equations in the 
previous section.  Alternately, one can directly obtain them after some
algebra in the off-shell formulation so we arrive at the results that follow. Below are the on-shell fermionic holoraumies upon enforcing the Dirac equation $(\g^\mu)_a{}^c \pa_\mu \Psi^k_c = (\g^\mu)_a{}^c \pa_\mu \widetilde{\Psi}^k_c = 0$. The terms involving the Dirac equation can be found in the associated \emph{Mathematica} code that is freely available online through \emph{GitHub} at \href{https://hepthools.github.io/Data/4DN4Holo/}{https://hepthools.github.io/Data/4DN4Holo/}.

\subsection{Vector Multiplet \texorpdfstring{${\rm D}$-${\rm D}$}{D-D} Fermionic On-Shell Holoraumy}
\vspace*{-24 pt}

\begin{align}
\DDAB{}\Psi\downup ck &=
i 2\, [ \, {\cal V}{}_1{}^{i j k l }\, \gupdown\m ab
+ {\cal V}{}_2{}^{i j k l } \,  ( \gup5\gup\m)\ddn ab (\gup5)\sw cd \, ] \,  \pd\m  \Psi\downup dl
\label{eq:VMFHddZed}    
\end{align}
\subsection{Vector Multiplet \texorpdfstring{${\rm D}$-${\Tilde {\rm D}}$}{D-D} On-Shell  Fermionic Holoraumy}
\vspace*{-24 pt}
\begin{align}
\holcrosstil{}\Psi\downup ck &= i 2 \, \  \vee1\gpart\tilpsi\downup cl 
+i 2 \, [ \, \vee{2}\gupcommdown\m\n ab(\g\dn\n)\sw cd\ 
+ \,\vee{3}\ggfiupdown\m ab\gupsw5cd \, ] \, \pd\m \tilpsi\downup dl 
\label{eq:VMFHddBZed}    
\end{align}
\subsection{Vector Multiplet \texorpdfstring{${\Tilde {\rm D}}$-${\Tilde {\rm D}}$}{D-D} On-Shell Fermionic Holoraumy}
\vspace*{-24 pt}
\begin{align}
\holtil{}\Psi\downup ck &= i \, [ \,  2 \,  {\cal V}{}_1{}^{i j k l } \,\ggfiupdown\m ab\gupsw5cd
+  {\cal V}{}_2{}^{i j k l }  \, \gupcommdown\m\n ab(\g\dn\n)\sw cd \, ] \, \pd\m  \Psi\downup dl 
\label{eq:VMFHdBdBZed}    
\end{align}

\subsection{Tensor Multiplet \texorpdfstring{${\rm D}$-${\rm D}$}{D-D} On-Shell Fermionic Holoraumy}
\vspace*{-24 pt}
\begin{align}
\DDAB{}\til\Psi\downup ck &= i 2 \, {\cal V}_1{}^{i j k l} \, \ggfiupdown\m ab\gupsw5cd\pd\m\tilpsi\downup   dl + i \,
{\cal V}_2{}^{i j k l} \, (\g{}^{[\m} \g{}^{\l]} )_{ab} \,( \g\dn\l ){}_{c}{}^{d}  \pd\m  \, \tilpsi\downup   dl 
\label{eq:TMFHddZed}    
\end{align}
\subsection{\label{TMcrossfermion}Tensor Multiplet \texorpdfstring{${\rm D}$-${\Tilde {\rm D}}$}{D-D} On-Shell Fermionic Holoraumy}
\vspace*{-24 pt}
\begin{align}
\holcrosstil{}\tilpsi\downup ck &= i \, 2\tilvee1\gpart\Psi\downup cl   + i \,2 \left[ \tilvee3\gupcommdown\m\n ab(\g\dn\n)\sw cd
 + \tilvee5\ggfiupdown\m ab\gupsw5cd \right] \partial_\mu \Psi\downup dl
\end{align}
\subsection{Tensor Multiplet \texorpdfstring{${\Tilde {\rm D}}$-${\Tilde {\rm D}}$}{D-D} On-Shell Fermionic Holoraumy}
\vspace*{-24 pt}
\begin{align}
    \holtil{}\tilpsi\downup ck &=
    i 2 \,\{ \, {\cal V}_1{}^{i j k l} \, \gupdown\m ab \delta_c{}^d + {\cal V}_2{}^{i j k l} \,  \ggfiupdown\m ab\gupsw5cd  \, \} \, \pd\m\til\Psi\downup dl
\label{eq:TMFHdBdBZed}    
\end{align}    
The explicit data about the $\cal V$-type and $V$-type coefficient tensors
is found by carefully referring respectively to each type of holoraumy 
(i.e. ${ {\rm D}}$-${ {\rm D}}$, ${ {\rm D}}$-${\Tilde {\rm D}}$, or 
${\Tilde {\rm D}}$-${\Tilde {\rm D}}$) acting on the field variable
as given in the tables of \ref{appen:holoraumy}.

\newpage

\section{Holoraumy Points To An Infinite-Dimensional Algebra}
\label{sec:HP}

\subsection{A 2D, (4,0) Superspace Truncation}
Having found evidence that the commutator of the supercharge operator when acting
on valise supermultiplets (either in one dimension or in higher dimensions), 
indicates an additional algebraic structure, it is natural to study
this via examination of the commutator algebra of the holoraumy operator
acting on fermions.  As can be seen from the systems analyzed previously 
such calculations tend to become rather involved.  Accordingly, we will 
follow a path that avoids these by using an appeal to SUSY holography.

It has long been known \cite{NF}\footnote{The reader is
directed to the work in \cite{BB} to see a more recent demonstration of such
and approach.} that theories which realize one degree of extendedness or 
spatial dimension can be represented by superfields that manifestly realize 
a lower degree or extendedness or dimensions.  In the following, we will
use the work of \cite{AdnkKoR} to gain insight into the structures that
follow from the equations shown in (\ref{eq:H0LR1}) - (\ref{eq:H0LR18}).

The work in \cite{AdnkKoR} is focused upon superfields in 2D, $(4, \, 0)$
superspace.  The work provides a complete analysis of all such superfields
with the property that a set of propagating bosons reside ``lower'' in 
a $\theta$-expansion than a set of propagating fermions.  This ensures that 
a dimensional reduction of the results will have to ``land'' on one of the 
supermultiplets considered in this section.  Also in the following, we will 
use the notational conventions seen in \cite{AdnkKoR}.

The analysis in \cite{AdnkKoR} found there are four and only four distinct
supermultiplets we need to consider.  In this work they are given the names
SM-I, SM-II, SM-III and SM-IV so that we can introduce a ``representation
label'' $\cal R$ that takes on these four values.  For each value of the
``representation label'' there are four bosons and four fermions.
In order to use as compact  a notation as possible, we denote these
fields by $\Phi_{i}^{(\cal R)}$ (bosons with $i$ = 1, $\dots$, 4) and
$\Psi_{+\, \hat k}^{(\cal R)}$ (fermions with $\hat k$ = 1, $\dots$, 4).
However, after obtaining the results in the 2D, (4,0) superspace,
a reduction to a 1D superspace streamlines the results.  This amounts
to ``dropping'' all spin-helicity indices on operators (i.e. ${\rm D}{}_{+{}_{\rm I}} 
\, \to \, {\rm D}{}_{{}_{\rm I}}$, $\pa_{\pp} \, \to \, \pa_0$) and 
fields $\Psi_{+\, \hat k}^{(\cal R)} \, \to \, \Psi_{\hat k}^{(\cal R)}$.

Before we continue, it may be convenient here to discuss the significance
of the SM-I, SM-II, SM-III and SM-IV supermultiplets.  There are two way to
demonstrate this and the presentation to follow with discuss both.  One
has its origin in the initial discovery of twisted superpotentials \cite{Gtm}, 
``twisted chiral supermultiplets'' \cite{GHRtm}.  The other perspective
is a mathematical one of more recent vintage \cite{CorL1,Bristow:2020rdf}.

The works of \cite{Gtm,GHRtm}, provided the first in the physics literature of examples in extended 
SUSY theories, there can exist supermultiplets with identical spectra of component fields, 
but which nevertheless are inequivalent.  The inequivalence of such supermultiplets is manifested
in two ways,  First the SUSY transformation laws are inequivalent.  As shown in the works
of \cite{Gtm,GHRtm}, this can led to an unexpected result.  Some sets of dynamical equations
that are consistent with supersymmetry {\it {require}} the simultaneous presence of
inequivalent supermultiplets.   

A mathematical perspective on this was recently enunciated in the works of 
\cite{CorL1,Bristow:2020rdf}.  These works inaugurated the use of a mathematical
concept (sometimes called the ``permutahedron'') to show that the inequivalences
of such SUSY representations can easily be detected mapping the transformation
laws of the component fields of a SUSY representation onto elements of the
permutation group.  When this is done, the permutahedron, provides a well-defined
metric on the space of one dimensional supermultiplets which easily allows for
the identification of the inequivalence.

In Appendix G of this work, the SUSY transformation laws of the SM-I, SM-II, SM-III
and SM-IV supermultiplets are explicitly given.  These are specified by giving four
matrices for each supermultiplet.  Mapping these onto elements of the permutation
group is done by simply taking the absolute values of the entries in the matrices.
When this is done, the following relationships are obtained.

\begin{table}[h]
\vspace{0.2cm}
\begin{center}
\footnotesize
\begin{tabular}{|c|c|c|c|c|}
\hline 
$\rm {Supermultiplet}$  & $|  {\rm L}_1 |$ & $|  {\rm L}_2 | $ & $|  {\rm L}_3 | $ & $|  {\rm L}_4 |$  \\ \hline
${\rm {SM}-}{\rm {I}}$  & $~~(243)~~$ & $~~(123)~~$ & $~~(134)~~$ & $~~(142)~~$  \\ \hline
${\rm {SM}-}{\rm {II}}$  & $~~(1243)~~$ & $~~(23)~~$ & $~~(14)~~$ & $~~(1342)~~$  \\ \hline
${\rm {SM}-}{\rm {III}}$  & $~~(1243)~~$ & $~~(14)~~$ & $~~(23)~~$ & $~~(1342)~~$  \\ \hline
${\rm {SM}-}{\rm {IV}}$  & $~~(23)~~$ & $~~(1342)~~$ & $~~(1243)~~$ & $~~(14)~~$ \\ \hline
\end{tabular}
\caption{Supermultiplet Transformation Law/Permutation Elements \label{tab:Ix}}
\end{center}
\end{table} 
\noindent
In writing these results, we have utilized the cycle notation for the elements of the permutation
group of order four as in the works of \cite{CorL1,Bristow:2020rdf}.  The matrix sets of $\{\,  |  {\rm L}_1 |, \,
|  {\rm L}_2 |, \,  |  {\rm L}_3 |, \, |  {\rm L}_4 | \, \}$ given here apply to each supermultipet
as indicated.  For the purpose of the
physics vantage point the sets should be considered as unordered sets.  With this restriction
only the SM-I and SM-II paring will lead to the same type of dynamical properties as discovered
for the chiral vs. twisted chiral pair noted in the works published in 1984.

For each representation, the supercharges (${\rm D}{}_{{}_{\rm I}} $ with 
$\rm I$ = 1, $\dots$, 4) are realized by the transformations
\be \eqalign{
{\rm D}{}_{{}_{\rm I}} \Phi_{i}^{(\cal R)} ~=~ i \, \left( {\rm L}{}_{{}_{\rm I}}^{(\cal R)} \right) 
{}_{i \, {\hat k}}  \, \, \Psi_{\hat k}^{(\cal R)}  ~~~,~~~
{\rm D}{}_{{}_{\rm I}} \Psi_{\hat k}^{(\cal R)} ~=~ \left( {\rm R}{}_{{}_{\rm I}}^{(\cal R)} \right)
{}_{{\hat k} \, i}  \, \pa_{0} \, \Phi_{i}^{(\cal R)}  ~~~,
}  \label{eq:VH1}
\ee  
where $ {\rm L}{}_{{}_{\rm I}}^{(\cal R)}$ and $ {\rm R}{}_{{}_{\rm I}}^{(\cal R)}$ are matrices
whose explicit values depend on the representation under consideration.  These values are given in
appendix~\ref{appen:LRmatrices}.  These matrices also satisfy the equations.
\begin{align}\label{eq:ortho}
	{
	{\rm R}}_\rI^{(\cal R)} = ({  {\rm L}}_\rI^{(\cal R)})^{-1} = ({  {\rm L}}_\rI^{(\cal R)})^{T}  ~~~.
\end{align}
It is a direct calculation to show
\be  \eqalign{
[\, {\rm D}{}_{{}_{\rm I}}  ~,~ {\rm D}{}_{{}_{\rm J}} \, ] \, \Psi_{ \hat k}^{({\cal R})} 
~&=~  2\, \left[ \,  {\tilde{ \rm V}}^{({\cal R})}{}_{{}_{\rm I}}{}_{{}_{\rm J}} \, \right]  {}_{{\hat 
k} \, {\hat \ell}}  \,  \pa_{0} \, \Psi_{\hat \ell}^{({\cal R})}    ~~~,
}  \label{eq:VH2}
\ee
where
\be \eqalign{
\left[ \, {\tilde {\rm V}}{}_{\rI\rJ}^{({\cal R})}  \,\right] {}_\hi{\,}^\hk ~&=~ -i\,  \fracm12 \, 
\left[ \, (\,{\rm R}_\rI {}^{({\cal R})} \,)_\hi{}^j\>(\, {\rm L}_\rJ {}^{({\cal R})} \,
)_j{}^\hk ~-~ (\,{\rm R}_\rJ  {}^{({\cal R})} \, )_\hi{}^j\>(\,{\rm L}_\rI
{}^{({\cal R})} 
\,)_j{}^\hk  \, \right]
~~.
}  \label{eq:VH4}
\ee
It must be {\it {emphasized}} that the result shown in (\ref{eq:VH4}) is {\it {only}} valid for
valise supermultiplets, and is {\it {not}} valid for general supermultiplets.

Use of the explicit forms of the ``V-matrices'' from the final appendix shows that
\newline \indent
\be \eqalign{  {~~~~~~~}
\left[ \, {\tilde {\rm V}}{}_{\rI\rJ}^{({\cal R})}  ~,~ {\tilde {\rm V}}{}_{{\rm K} {\rm L}}^{({\cal R})}  \, \right] ~=~ 
- \, i \, 2 \, \left[ \, \delta{}_{{\rm J} {\rm K}} \, {\tilde {\rm V}}{}_{\rI\rL}^{({\cal R})} ~-~
\delta{}_{{\rm I} {\rm K}} \, {\tilde {\rm V}}{}_{\rJ\rL}^{({\cal R})}  ~+~
\delta{}_{{\rm I} {\rm L}} \, {\tilde {\rm V}}{}_{\rJ\rK}^{({\cal R})}  ~-~
\delta{}_{{\rm J} {\rm L}} \, {\tilde {\rm V}}{}_{\rI\rK}^{({\cal R})}  
\, \right] ~~~~,
}  \label{eq:VH5}
\ee
a result that is uniformly satisfied on all the representations. In fact, it was proven in~\cite{Mak:2018qsr} that all $\tilde{V}_{\rI \rJ}$ satisfy Eq.~(\ref{eq:VH5}), so long as the associated $ {\rm L}_{\rI}^{(\cal R)}$
 and ${\rm R}_{\rI}^{(\cal R)} $ satisfy the $ {\cal {GR}}({\rm d},\, {\cal N}$) ``garden algebra.'' The factor of two in Eq.~\eqref{eq:VH5} along with the fact that $\tilde{V}_{IJ}^{(\mathcal{R})}$ squares to the identity demonstrates that the holoraumy operator $\tilde{V}_{IJ}^{(\mathcal{R})}$ is a representation of {$\mathfrak{Spin}$}($N$).
 
 To demonstrate the $4D$, $\mathcal{N}=4$ VT multiplets relationship to the SM-i multiplets, we dimensionally reduce the transformation laws, choosing temporal gauge $A_0 = \tilde{B}_{12} = \tilde{B}_{23}  = \tilde{B}_{31} = 0$ and considering all other fields to depend only on time. We then define the $16 \times 16$ $\rL_\rI^{(VT)}$ and $\rR_\rI^{(VT)}$ matrices through Eq.~\eqref{eq:VH1} with $\rI = 1, 2, 3, \dots 16$, $i, j, \dots = 1,2,3, \dots 16$, and $\hat{k}, \hat{l},\dots = 1,2,3, \dots 16$ and the identifications
 \begin{align}\label{e:D16}
	\rD_\rI \equiv& \left (
	\rD_a^1,
	\rD_b^2,	-\widetilde{\rD}_c^1, -\widetilde{\rD}_d^2
	\right) 
	~~~,~~~
	i\Psi_{\hat{k}}^{(VT)} \equiv \left (
	\widetilde{\Psi}_a^1,
	\widetilde{\Psi}_b^2,
	\Psi_c^1 ,
	\Psi_d^2
	\right)
\end{align}
and
\begin{align}
    \Phi_i^{(VT)} =  \left(\widetilde{A}, \widetilde{B}, \int \widetilde{F}~dt, \int \widetilde{G}~dt, \widetilde{\varphi}, \widetilde{B}_{12}, \widetilde{B}_{23}, \widetilde{B}_{31}, A, B, \int F~dt, \int G~dt, A_1, A_2, A_3, \int d~dt  \right)
\end{align}
The explicit form of the resulting $\rL_\rI^{(VT)}$ matrices are given in appendix~\ref{a:L16}. The $\rR_\rI^{(VT)}$ matrices can be found through the orthogonality relationship~\ref{eq:ortho} for all $\rI = 1,2,3,\dots,16$. The above definitions of  $\Psi_{\hat{k}}^{(VT)}$ and $\Phi_i^{(VT)}$ are chosen to line up with those for $4D$, $\mathcal{N}=4$ vector-chiral ($VC$) multiplet of~\cite{Gates:2011zv,Calkins:2014exa} for the fields in common. The definition of $\rD_\rI$ in terms of the $VT$ supercharges is chosen to identify with the following definition in terms of the $VC$ supercharges 
\begin{align}\label{e:VCSupercharges}
    \rD_\rI = \left( \rD_a , \rD^{3}_b , \rD^{1}_c , \rD^{2}_d \right)~~~\text{for $VC$ supercharges}
\end{align}
to align the transformation laws of these two multiplets for the fields in common. The explicit form of the $\rL_\rI^{(VC)}$ matrices defined in~\cite{Calkins:2014exa} can also be found in appendix~\ref{a:L16}. The $\rR_\rI^{(VC)}$ matrices can be found through the orthogonality relationship~\ref{eq:ortho} for all $\rI = 1,2,3,\dots,16$.

Since the two multiplets $VT$ and $VC$ describe the same on-shell physics in $4D$ though clearly have different auxiliary fields, we compare their 1D reductions through the dot-product like \emph{gadget} to determine if this distinction is made there as well~\cite{Gates:2012xb}
\begin{align}
	\mathcal{G}(\mathcal{R},\mathcal{R}') = \frac{1}{\rd_{min}(N) N (N-1)}\sum_{IJ} \tilde{V}_{\rI\rJ}^{(\mathcal{R})}\tilde{V}_{\rI\rJ}^{(\mathcal{R'})}
\end{align}
where $\rd_{min}(16) = 128$ and $\rd_{min}(4) = 4$ and the $\tilde{V}_{\rI\rJ}^{(VT)}$ and $\tilde{V}_{\rI\rJ}^{(VC)}$ are calculated using Eq.~\eqref{eq:VH4} for $\rI, \rJ = 1,2,3,\dots, 16$. Their gadgets are
\begin{align}
    \mathcal{G}(VT,VC) = \frac{11}{240}~~~,~~~\mathcal{G}(VT,VT) = \frac{43}{480} ~~~,~~~\mathcal{G}(VC,VC) = \frac{1}{10} 
\end{align}
Thinking of the gadget as a kind of dot product, we define an ``angle'' $\theta(\mathcal{R},\mathcal{R}')$ between two representations as~\cite{Gates:2012xb}
\begin{align}
    \theta(\mathcal{R},\mathcal{R}') = \cos^{-1}\left( \frac{\mathcal{G}(\mathcal{R},\mathcal{R}')}{\sqrt{\mathcal{G}(\mathcal{R},\mathcal{R})\mathcal{G}(\mathcal{R}',\mathcal{R}')}}\right)
\end{align}
Any angle aside from zero indicates two representations are distinct in the sense of the gadget. We find for $VT$ and $VC$
\begin{align}
     \theta(VT,VC) = 61.04^{\circ}
\end{align}
to four significant figures.
This means the gadget can distinguish these two multiplets at the one-dimensional reduction, or adinkra, level.

The complete set of  $\tilde{V}^{(VT)}_{\rI \rJ}$ do not furnish a representation of {$\mathfrak{Spin}$}($16$), however they do furnish a representation of {$\mathfrak{Spin}$}($8$), i.e. satisfy Eq.~\eqref{eq:VH5},  for the subset  $\rI, \rJ, \rK, \rL  = 1, 2, \dots 8$ of $16 \times 16$ matrices as well as  for the subset $\rI, \rJ, \rK, \rL = 9, 10, \dots 16$ of $16 \times 16$ matrices.  The two {$\mathfrak{Spin}$}($8$) subalgebras satisfying~\eqref{eq:VH5} for $N =8$ can be understood as arising from the two off-shell 4D, $\mathcal{N}=2$ submultiplets. Thus it is not surprising that the $\tilde{V}^{(VC)}_{\rI \rJ}$  do not enjoy such a {$\mathfrak{Spin}$}($8$) substructure of $16 \times 16$ matrices as it is the dimensional reduction of a single 4D, $\mathcal{N}=2$ off-shell multiplet and a 4D, $\mathcal{N}=2$ on-shell multiplet as demonstrated explicitly in~\cite{Gates:2011zv}.   As neither the full 4D, $\mathcal{N}=4$ vector-tensor multiplet nor the full 4D, $\mathcal{N}=4$ vector-chiral multiplet close off-shell, it is expected~\cite{Bristow:2020rdf} that the non-closure terms for the complete sets $\tilde{V}^{(VT)}_{\rI \rJ}$ and $\tilde{V}^{(VC)}$ each take the form
\be \eqalign{  {~~~~~~~}
\left[ \, {\tilde {\rm V}}{}_{\rI\rJ}^{({\cal R})}  , {\tilde {\rm V}}{}_{{\rm K} {\rm L}}^{({\cal R})}  \, \right] = 
- i 2  \left[ \, \delta{}_{{\rm J} {\rm K}} \, {\tilde {\rm V}}{}_{\rI\rL}^{({\cal R})} -
\delta{}_{{\rm I} {\rm K}} \, {\tilde {\rm V}}{}_{\rJ\rL}^{({\cal R})}  +
\delta{}_{{\rm I} {\rm L}} \, {\tilde {\rm V}}{}_{\rJ\rK}^{({\cal R})}  -
\delta{}_{{\rm J} {\rm L}} \, {\tilde {\rm V}}{}_{\rI\rK}^{({\cal R})}  
\, \right] + \mathcal{N}_{\rI\rJ\rK\rL}{}^{\rM\rN(\mathcal{R})} \tilde{\rm V}^{(\mathcal{R})}_{\rM\rN}
}  \label{eq:VH5NonClosure}
\ee
for some non-closure coefficients $ \mathcal{N}_{\rI\rJ\rK\rL}{}^{\rM\rN(\mathcal{R})}$. Further analysis of the substructures within $\tilde{V}^{(VT)}_{\rI \rJ}$ and $\tilde{V}^{(VC)}_{\rI \rJ}$ can be found in the Mathematica code on \emph{GitHub} at
\newline
\href{https://hepthools.github.io/Data/4DN4Holo/}{https://hepthools.github.io/Data/4DN4Holo/}.

We conclude this section by focusing on representations that satisfy Eq.~\eqref{eq:VH4}, such as the four SM-i multiplets. We will uncover associated 1D infinite dimensional algebras with holoraumy matrices as essential building blocks. As a notational device, we can make a definition $ \Delta{}{}^{{[p]}}_{{}_{{\rm I} {\rm J}}}$
 $\equiv$ $\fracm 12 [\, {\rm D}{}_{{}_{\rm I}}  ~,~ {\rm D}{}_{{}_{\rm J}} \, ]\, (\pa_{0}){}^{p - 1}$. In terms of this notation, the equation in (\ref{eq:VH2}) can be cast into the form
\be  \eqalign{
\Delta{}{}^{{[1]}}_{{}_{{\rm I} {\rm J}}} \, \Psi_{ \hat k}^{({\cal R})} 
~&=~   \left[ \,  {\tilde{ \rm V}}^{({\cal R})}{}_{{}_{\rm I}}{}_{{}_{\rm J}} \, \right]  {}_{{\hat 
k} \, {\hat \ell}}  \,  \pa_{0} \, \Psi_{\hat \ell}^{({\cal R})}    ~~~.
}  \label{eq:VHx1}
\ee
Applying the operator $ \pa_{0}^{[-1]}$ to both sides of this yields
\be  \eqalign{
\Delta{}{}^{{[0]}}_{{}_{{\rm I} {\rm J}}} \, \Psi_{ \hat k}^{({\cal R})} 
~&=~  \left[ \,  {\tilde{ \rm V}}^{({\cal R})}{}_{{}_{\rm I}}{}_{{}_{\rm J}} \, \right]  {}_{{\hat 
k} \, {\hat \ell}} \,  \Psi_{\hat \ell}^{({\cal R})}    ~~~.
}  \label{eq:VHx1U}
\ee

The equation in (\ref{eq:VHx1}) additionally implies
\be \eqalign{ {~~~~~~~~~~~~~}
\Delta{}{}^{{[1]}}_{{}_{{\rm I} {\rm J}}} \,
\Delta{}{}^{{[1]}}_{{}_{{\rm K} {\rm L}}} \,
\Psi_{ \hat k}^{({\cal R})} ~&=~  
\left[ \,  {\tilde{ \rm V}}^{({\cal R})}{}_{{}_{\rm K}}{}_{{}_{\rm L}} \, \right]  
{}_{{\hat k} \, {\hat \ell}}  \,  \pa_{0} \,\left( 
\Delta{}{}^{{[1]}}_{{}_{{\rm I} {\rm J}}} \,
\Psi_{ \hat \ell}^{({\cal R})}  \right)   ~~~,  \cr
 ~&=~  
\left[ \,  {\tilde{ \rm V}}^{({\cal R})}{}_{{}_{\rm K}}{}_{{}_{\rm L}} \, \right]{}_{{\hat k} \, {\hat \ell}} 
\, \left[ \,  {\tilde{ \rm V}}^{({\cal R})}{}_{{}_{\rm I}}{}_{{}_{\rm J}} \,  
\, \right] {}_{{\hat \ell} \, {\hat h}}  \, \left( \pa_{0}  \, \pa_{0}
\Psi_{{ \hat h} }^{({\cal R})}  \right)   ~~~,  {~~~}
} \label{eq:VH6}
\ee
and 
\be \eqalign{ {~~~~~}
\Delta{}{}^{{[1]}}_{{}_{{\rm I} {\rm J}}} \,
\Delta{}{}^{{[1]}}_{{}_{{\rm K} {\rm L}}} \,
\Delta{}{}^{{[1]}}_{{}_{{\rm M} {\rm N}}} \,
\Psi_{ \hat k}^{({\cal R})} ~&=~ 
\left[ \,  {\tilde{ \rm V}}^{({\cal R})}{}_{{}_{\rm M}}{}_{{}_{\rm N}} \, \right]{}_{{\hat k} \, {\hat \ell}} 
\, \left[ \,  {\tilde{ \rm V}}^{({\cal R})}{}_{{}_{\rm K}}{}_{{}_{\rm L}} \,  
\, \right] {}_{{\hat \ell} \, {\hat h}}
\,  \pa_{0}  \, \pa_{0} \,\left( 
\Delta{}{}^{{[1]}}_{{}_{{\rm I} {\rm J}}} \,
\Psi_{ \hat h}^{({\cal R})}  \right)   ~~~,  \cr
 ~&=~
\left[ \,  {\tilde{ \rm V}}^{({\cal R})}{}_{{}_{\rm M}}{}_{{}_{\rm N}} \, \right]{}_{{\hat k} \, {\hat \ell}} 
\, \left[ \,  {\tilde{ \rm V}}^{({\cal R})}{}_{{}_{\rm K}}{}_{{}_{\rm L}} \,  
\, \right] {}_{{\hat \ell} \, {\hat h}} 
\, \left[ \,  {\tilde{ \rm V}}^{({\cal R})}{}_{{}_{\rm I}}{}_{{}_{\rm J}} \,  
\, \right] {}_{{\hat h} \, {\hat j}} 
\, \left( \pa_{0}  \,  \pa_{0}  \, \pa_{0}
\Psi_{{ \hat j} }^{({\cal R})}  \right)  
} \label{eq:VH6z}
\ee
so that furthermore the result in (\ref{eq:VH6}) implies
\be \eqalign{ {~~~~~~}
\left[ \, 
\Delta{}{}^{{[1]}}_{{}_{{\rm I} {\rm J}}} ~,~
\Delta{}{}^{{[1]}}_{{}_{{\rm K} {\rm L}}} \, \right]
\Psi_{ \hat k}^{({\cal R})} 
 ~=~ &- \, \left(
\left[ \,  {\tilde{ \rm V}}^{({\cal R})}{}_{{}_{\rm I}}{}_{{}_{\rm J}} ~,~
  {\tilde{ \rm V}}^{({\cal R})}{}_{{}_{\rm K}}{}_{{}_{\rm L}} \,  
\, \right] \right) {}_{{\hat k} \, {\hat h}}  \, \left( \pa_{0}  \, \pa_{0}
\Psi_{{ \hat h} }^{({\cal R})}  \right)      ~~~,  \cr
} \label{eq:VH7}
\ee
and next by use of (\ref{eq:VH5}) we obtain
\be \eqalign{ {~~~~~~}
\left[ \, 
\Delta{}{}^{{[1]}}_{{}_{{\rm I} {\rm J}}} ~,~
\Delta{}{}^{{[1]}}_{{}_{{\rm K} {\rm L}}} \, \right]
\Psi_{ \hat k}^{({\cal R})} 
 ~=~ & 2 i \, \left(
\delta{}_{{\rm J} {\rm K}} \, {\tilde {\rm V}}{}_{\rI\rL}^{({\cal R})} ~-~
\delta{}_{{\rm I} {\rm K}} \, {\tilde {\rm V}}{}_{\rJ\rL}^{({\cal R})} 
 \right) {}_{{\hat k} \, {\hat h}}  \, \left( \pa_{0}  \, \pa_{0}
\Psi_{{ \hat h} }^{({\cal R})}  \right)     \cr
~ &+~ 2i \,  \left(
\delta{}_{{\rm I} {\rm L}} \, {\tilde {\rm V}}{}_{\rJ\rK}^{({\cal R})}  ~-~
\delta{}_{{\rm J} {\rm L}} \, {\tilde {\rm V}}{}_{\rI\rK}^{({\cal R})}  
\right) {}_{{\hat k} \, {\hat h}}  \, \left( \pa_{0}  \, \pa_{0}
\Psi_{{ \hat h} }^{({\cal R})}  \right)   ~~~.  \cr
} \label{eq:VH7zz}
\ee
Using (\ref{eq:VHx1}), this can be rewritten in the form
\be \eqalign{ {~}
\left[ \, 
\Delta{}{}^{{[1]}}_{{}_{{\rm I} {\rm J}}} \,,\,
\Delta{}{}^{{[1]}}_{{}_{{\rm K} {\rm L}}} \, \right]
\Psi_{ \hat k}^{({\cal R})} 
 \,&=\,i\, 2 \, \left(
\delta{}_{{\rm J} {\rm K}} \,  \left( \Delta{}{}^{{[2]}}_{{}_{{\rm I} {\rm L}}} 
\Psi_{{ \hat k} }^{({\cal R})}  \right)\,-\,
\delta{}_{{\rm I} {\rm K}} \, \left( \Delta{}{}^{{[2]}}_{{}_{{\rm J} {\rm L}}}  
\Psi_{{ \hat k} }^{({\cal R})}  \right)
 \right)  
 {~~~~~~~~~~~~~~~}     \cr
~ &~~~~+~ i\, 2 \, \left(
\delta{}_{{\rm I} {\rm L}} \, \left( \Delta{}{}^{{[2]}}_{{}_{{\rm J} {\rm K}}} 
\Psi_{{ \hat k} }^{({\cal R})}  \right) \,-\,
\delta{}_{{\rm J} {\rm L}} \,  \left( \Delta{}{}^{{[2]}}_{{}_{{\rm I} {\rm K}}}
\Psi_{{ \hat k} }^{({\cal R})}  \right)
\right)      ~~~,  \cr
} \label{eq:VH8}
\ee
and to this equation, we can apply the differential operator $\pa_{0}^{[S + T - 2]} $
where $S$ $\ge$ 1 and $T$ $\ge $ 1.  This yields
\be \eqalign{ {~}
\left[ \, 
\Delta{}{}^{{[S]}}_{{}_{{\rm I} {\rm J}}} \,,\,
\Delta{}{}^{{[T]}}_{{}_{{\rm K} {\rm L}}} \, \right]
\Psi_{ \hat k}^{({\cal R})} 
 \,&=\,i\, 2 \, \left(
\delta{}_{{\rm J} {\rm K}} \,  \left( \Delta{}{}^{{[S + T]}}_{{}_{{\rm I} {\rm L}}}  
\Psi_{{ \hat k} }^{({\cal R})}  \right)\,-\,
\delta{}_{{\rm I} {\rm K}} \, \left( \Delta{}{}^{{[S + T]}}_{{}_{{\rm J} {\rm L}}}  
\Psi_{{ \hat k} }^{({\cal R})}  \right)
 \right)  {~~~~~~~~~~~~~~~}     \cr
~ &~~~~+~ i\, 2 \, \left(
\delta{}_{{\rm I} {\rm L}} \, \left( \Delta{}{}^{{[S + T]}}_{{}_{{\rm J} {\rm K}}}  
\Psi_{{ \hat k} }^{({\cal R})}  \right) \,-\,
\delta{}_{{\rm J} {\rm L}} \,  \left( \Delta{}{}^{{[S + T]}}_{{}_{{\rm I} {\rm K}}} 
\Psi_{{ \hat k} }^{({\cal R})}  \right)
\right)    ~~~.  \cr
} \label{eq:VH8dd}
\ee
This makes it apparent there exists a definition of a closure property of the 
collection of operators $\Delta{}{}^{{[S]}}_{{}_{{\rm I} {\rm K}}}$ acting 
on the valise supermultiplet as in this equation.

Next we apply the operator $\pa_{0}^{[S + T + U - 3]} $ where $S$ $\ge$ 1, $U$ $\ge$ 1, and $T$ $\ge $ 1
to the equation in (\ref{eq:VH6z}) to derive
\be \eqalign{ {~~~~~}
\Delta{}{}^{{[S]}}_{{}_{{\rm I} {\rm J}}} \,
\Delta{}{}^{{[T]}}_{{}_{{\rm K} {\rm L}}} \,
\Delta{}{}^{{[U]}}_{{}_{{\rm M} {\rm N}}} \,
\Psi_{ \hat k}^{({\cal R})} ~&=~  \, 
\left[ \,  {\tilde{ \rm V}}^{({\cal R})}{}_{{}_{\rm M}}{}_{{}_{\rm N}} \, \right]{}_{{\hat k} \, {\hat \ell}} 
\, \left[ \,  {\tilde{ \rm V}}^{({\cal R})}{}_{{}_{\rm K}}{}_{{}_{\rm L}} \,  
\, \right] {}_{{\hat \ell} \, {\hat h}} 
\, \left[ \,  {\tilde{ \rm V}}^{({\cal R})}{}_{{}_{\rm I}}{}_{{}_{\rm J}} \,  
\, \right] {}_{{\hat h} \, {\hat j}} 
\, \left( \pa_{0}^{[S + T + U ]}
\Psi_{{ \hat j} }^{({\cal R})}  \right)   ~~~,  {~~~}
} \label{eq:VH6zX}
\ee
and due to the form of this equation, it immediately follows that
\be \eqalign{ 
( \,
& [ \, \Delta{}{}^{{[S]}}_{{}_{{\rm I} {\rm J}}} \,  , \, [ \,
\Delta{}{}^{{[T]}}_{{}_{{\rm K} {\rm L}}} \, , \, 
\Delta{}{}^{{[U]}}_{{}_{{\rm M} {\rm N}}} \, ]  \, ]  \, +\, 
[ \, \Delta{}{}^{{[T]}}_{{}_{{\rm K} {\rm L}}} \,   , \, [ \,
\Delta{}{}^{{[U]}}_{{}_{{\rm M} {\rm N}}} \, , \,
\Delta{}{}^{{[S]}}_{{}_{{\rm I} {\rm J}}} \, ]  \, ]   \, +\,  \cr
&[ \, \Delta{}{}^{{[U]}}_{{}_{{\rm M} {\rm N}}} \,   , \, [ \,
\Delta{}{}^{{[S]}}_{{}_{{\rm I} {\rm J}}} \,  , \, 
\Delta{}{}^{{[T]}}_{{}_{{\rm K} {\rm L}}} \,  ]  \,  ] 
 \, ) \, \Psi_{ \hat k}^{({\cal R})} ~=~ 0   ~~~.  {~~~}
} \label{eq:VH6zY}
\ee
The results in (\ref{eq:VHx1}) - (\ref{eq:VH6zY}) indicate the set of operators $\Delta{}{}^{{[R]}}_{{}_{{\rm I} {\rm J}}}$
(with $R$ $\ge $ 0), form an infinite dimensional algebra when acting on a valise supermultiplet representation. Thus the holoraumy operator $\tilde{V}_{IJ}^{(\mathcal{R})}$ being derived from one of the higher dimensional SUSY representations $\mathcal{R}$ = SM-I, SM-II, SM-III, or SM-IV provides the linkage from higher dimensional SUSY to a infinite-dimensional extension of {$\mathfrak{Spin}$}($N$).

While in past works we have used the \emph{fermionic} holoraumies in explanations of their algaebraic significance to identify different 4D supermultiplets~\cite{Gates:2012xb,HoLoR2,HoLoR3,HoLoR4,HoLoR5,AdnkKoR}, the significance of the \emph{bosonic} holoraumies remains unclear. This is a question currently under study.
\newpage
\section{Conclusion}
\label{sec:ConCL}

In this work, we have explored (to a greater extent than previously) the range of validity of the
interconnection between Hodge duality, noted in the work of \cite{UBQ1}, and the concept of
holoraumy.  We find that up to six dimensions such a relation holds in supersymmetrical
Maxwell theories.  However,
beyond this, in the case of Maxwell theory, the interconnection vanishes.  This would
suggest that in four dimensional theories, it could appear that the connection
cannot hold beyond 4D theories with $\cal N$ = 2 supersymmetry.  However, by
explicit calculations within the context of the 4D, $\cal N$ = 4 supersymmetry
such connections are present for the 4D, $\cal N$ = 4 vector-tensor supermultiplet.
Finally, by reduction to 1D SUSY QM theories, evidence was given that the 
holoraumy operator is both a representation of {$\mathfrak{Spin}$}($N$) 
\emph{and} a member of a set of an infinite number of generators
that are closed and satisfy a Jacobi identity.  Thus holoraumy
appears to be a part of an infinite-dimensional extension of {$\mathfrak{Spin}$}($N$).

Now let us discuss the important distinctions between the standard formulation of the abelian (VC) multiplet versus the (VT) multiplet. For this comparison, we look at the absolute values of the $\rL_\rI^{(VC)}$ and $\rL_\rI^{(VT)}$ matrices listed in appendix~\ref{a:L16}.
\begin{align}
    |\rL_\rI^{(VC)}| =& \mathscr{V}^{1}_{(1)}\otimes |\rL_\rI^{(SM-I)}| + \mathscr{V}^{1}_{(2)}\otimes |\rL_\rI^{(SM-I)}|+\mathscr{V}^{1}_{(3)}\otimes |\rL_\rI^{(SM-I)}|+\mathscr{V}^{1}_{(4)}\otimes |\rL_\rI^{(SM-II)}| \\
    |\rL_{\rI+4}^{(VC)}| =& \mathscr{V}^{2}_{(1)}\otimes |\rL_\rI^{(SM-I)}| + \mathscr{V}^{2}_{(2)}\otimes |\rL_\rI^{(SM-I)}|+\mathscr{V}^{2}_{(3)}\otimes  |\rL_\rI^{(SM-I)}|+\mathscr{V}^{2}_{(4)}\otimes |\rL_\rI^{(SM-II)}| \\
    |\rL_{\rI+8}^{(VC)}| =& \mathscr{V}^{4}_{(1)}\otimes |\rL_\rI^{(SM-I)}| + \mathscr{V}^{4}_{(2)}\otimes  |\rL_\rI^{(SM-I)}|  +\mathscr{V}^{4}_{(3)}\otimes  |\rL_\rI^{(SM-I)}|+\mathscr{V}^{4}_{(4)}\otimes |\rL_\rI^{(SM-II)}| \\
    |\rL_{\rI+12}^{(VC)}| =& \mathscr{V}^{3}_{(1)}\otimes |\rL_\rI^{(SM-I)}|+ \mathscr{V}^{3}_{(2)}\otimes |\rL_\rI^{(SM-I)}|  +\mathscr{V}^{3}_{(3)}\otimes  |\rL_\rI^{(SM-I)}|+\mathscr{V}^{3}_{(4)}\otimes  |\rL_\rI^{(SM-II)}| \\
    |\rL_\rI^{(VT)}| =& \mathscr{V}^{1}_{(1)}\otimes |\rL_\rI^{(SM-I)}| + \mathscr{V}^{1}_{(2)}\otimes |\rL_\rI^{(TM)}|+\mathscr{V}^{1}_{(3)}\otimes |\rL_\rI^{(SM-I)}|+\mathscr{V}^{1}_{(4)}\otimes |\rL_\rI^{(SM-II)}| \\
    |\rL_{\rI+4}^{(VT)}| =& \mathscr{V}^{2}_{(1)}\otimes |\rL_\rI^{(SM-I)}| + \mathscr{V}^{2}_{(2)}\otimes  |\rL_\rI^{(TM)}| +\mathscr{V}^{2}_{(3)}\otimes  |\rL_\rI^{(SM-I)}|+\mathscr{V}^{2}_{(4)}\otimes  |\rL_\rI^{(SM-II)}| \\
    |\rL_{\rI+8}^{(VT)}| =& \mathscr{V}^{4}_{(1)}\otimes \rL_\rI^{(SM-I)}| + \mathscr{V}^{4}_{(2)}\otimes |\rL_\rI^{(TM)}|  +\mathscr{V}^{4}_{(3)}\otimes  |\rL_\rI^{(SM-I)}|+\mathscr{V}^{4}_{(4)}\otimes  |\rL_\rI^{(SM-II)}|\\
    |\rL_{\rI+12}^{(VT)}| =& \mathscr{V}^{3}_{(1)}\otimes |\rL_\rI^{(SM-I)}| + \mathscr{V}^{3}_{(2)}\otimes |\rL_\rI^{(TM)}|  +\mathscr{V}^{3}_{(3)}\otimes  |\rL_\rI^{(SM-I)}|+\mathscr{V}^{3}_{(4)}\otimes  |\rL_\rI^{(SM-II)}|
\end{align}
Considering the absolute values allows us to map these matrices into elements associated with the permutahedron~\cite{CorL1,Bristow:2020rdf}. At this permutahedron level, it is clearly seen that the difference between the two supermultiplets corresponds to switching the $TM$ to $SM-I$ (called $TS$ and $CS$ in~\cite{CorL1,Bristow:2020rdf}, respectively).

 \vspace{.05in}
 \begin{center}
\parbox{4in}{{\it ``Aim at high things but not presumptuously. \\ ${~}$
Endeavor to succeed--expect not to succeed.''
 \\ ${~}$ 
\\ ${~}$ }\,\,-\,\, Michael Faraday}
 \parbox{4in}{
 $~~$}  
 \end{center}
 
  \noindent
{\bf Acknowledgments}\\[.1in] \indent
The research of S.J.G., Jr. and K.\ Stiffler is supported in part by the endowment of 
the Ford Foundation Professorship of Physics at Brown University and they gratefully 
acknowledge the support of the Brown Theoretical Physics Center (BTPC).
Additional acknowledgment is given by Gabriel Hannon, and Rui Xian Siew for their 
participation in the 2019 SSTPRS (Student Summer Theoretical Physics Research Session) 
program at Brown University. We express gratitude for the referee's comments which contributed to increased transparency in our final work.

\clearpage
\appendix


\clearpage
\section{Useful Gamma Matrix Identities}
\label{appen:Gmtrx}
\subsection{Definitions And Conventions}
\label{appen:Gmtrix}
Note that here and throughout, we use the ``unweighted brackets" to indicate symmetrization or antisymmetrization of two indices, regardless of whether they are spinor or spacetime indices.
We use the following representation for the gamma matrices in 4D:
\begin{align}
    (\g^0)_{\a}{}^{\b}=i \s^3 \otimes \s^2~~,~~(\g^1)_{\a}{}^{\b}=I_2 \otimes \s^1~~,~~(\g^2)_{\a}{}^{\b}=\s^2 \otimes \s^2~~,~~(\g^3)_{\a}{}^{\b}=I_2 \otimes \s^3~~,
\end{align}
where $I_2$ is the $2\times 2$ identity matrix, and $\s^1,\s^2,\s^3$ are the Pauli matrices.
Also in 4D we have the $\gamma^5$ matrix:
\begin{align}
	(\gamma^5)_\a{}^\b =i\g^0\g^1\g^2\g^3= - \sigma^2 \otimes I_2~~.
\end{align}
The totally antisymmetric Levi-Civita tensor $\e^{\m\n\r\s}$ is defined by $\e^{0123}=-1$.
The spacetime indices of the gamma matrices and the Levi-Civita tensor are lowered and raised using the mostly plus Minkowski metric $\eta_{\m\n}$ and its inverse $\eta^{\m\n}$ respectively:
\begin{align}
    \g^\m \eta_{\m\n}=\g_\n~~,~~\g_\m \eta^{\m\n}=\g^\n~~,~~\eta^{\m\r}\eta_{\r\n}=\d_{\n}^\m~~.
\end{align}
The spinor indices are lowered and raised using the spinor metric $C_{ab}$ and its inverse $C^{ab}$ respectively:
\begin{align}
    (\g^\m)_a{}^c C_{cb}=(\g^\m)_{ab}~~,~~C^{ac}(\g^\m)_c{}^b=(\g^\m)^{ab}~~,
\end{align}
where the spinor metric and its inverse are defined by
\begin{align}
    C_{ab}=-i \s^3 \otimes \s^2~~,~~C_{ab}C^{cb}=\d_a{}^c~~.
\end{align}
\subsection{Gamma Matrix Identities}
\begin{align}
    (\g^{(\m}\g^{\n)})_a{}^b&=2\eta^{\m\n}\d_a{}^b\\
    \gup\m\gup5 &= -\gup5\gup\m\\
    \g\up\m\g\up\n\g\up\r &= \eta\uup\m\n\g\up\r + \eta\uup\n\r\g\up\m -\eta\uup\m\r\g\up\n - i\e_{\s}{}^{ \m\n\r}\g\up5\g\up\s\\
    \g\up\m\g\dn\n\g\up\r &= \d\downup\m\n\g\up\r + \d\downup\v\r\g\up\m - \eta\uup\m\r\g\dn\n + i\e_{\s\n}{}^{ \m\r}\g\up5\g\up\s\\
    \g\up5\g\uup[\a\g\uup\b] &= -\fracm{i}{2}\e^{\a\b\m\n}\g\ddn[\m\g\ddn\n]\\
    \g\up\m\g\uup[\a\g\uup\b] &= 2\eta^{\m[\a}\g\uup\b] + i2\e^{\a\b\m\n}\g\up5\g\dn\n\\
    \g\uup[\a\g\uup\b]\g\up\m &= -2\eta^{\m[\a}\g\uup\b] + i2\e^{\a\b\m\n}\g\up5\g\dn\n\\
    \g\up5\g\up\a\g\up\b &= \fracm1{2} \, {\g\up5} \,(\g\uup[\a\g\uup\b] + \g\uup(\a\g\uup\b)) = -\fracm{i}{4}\e^{\a\b\m\n}\g\ddn[\m\g\ddn\n] +\eta\uup\a\b \g\up5\\
    \gup5\gup\m\g\uup[\a\g\uup\b] &= -2\eta\uuu\m[\a\g\uup\b]\gup5 + 2i \e^{\a\b\m\l}\g\dn\l\\ 
   ( \gup\m\g\dn\m )_a{}^b &= 4 \d_a{}^b\\
    \gup\m\gup\n\g\dn\m &= -2\gup\n\\
   ( \gup\m\gup\n\gup\r\g\dn\m )_a{}^b&= 4\eta\uup\n\r\d_a{}^b\\
    (\gamma^{\mu} \gamma^{\nu} \gamma^{\rho} \gamma^{\sigma} \gamma^{5})_a{}^a&=4 i \epsilon^{\mu \nu \rho \sigma}
\end{align}

\subsection{Spinor Index Symmetries}
\begin{align}
    C\ddn ab &= -C\ddn ba\\
    (\g\dn\m)\ddn ab &= (\g\dn\m)\ddn ba\\
    (\g\up5)\ddn ab &= -(\g\up5)\ddn ba\\
    (\g\up\m\g\up5)\ddn ab &= -  (\g\up\m\g\up5)\ddn ba\\
    (\g\up\m\g\up\n)\ddn ab &= - (\g\up\n\g\up\m)\ddn ba\\
    (\g\uup[\m\g\uup\n])\ddn ab &= (\g\uup[\m\g\uup\n])\ddn ba\\
    (\g\uup(\m\g\uup\n))\ddn ab &= -(\g\uup(\m\g\uup\n))\ddn ba  
\end{align}
$~$
\newline
\section{Useful Higher D Sigma Matrix Identities} 
\label{appen::sigma}

The defining relationship of the sigma matrices is
\begin{align}
	(\sigma_{(\m})_{ac} (\sigma_{\n)})^{cb} = 2 \eta_{\m\n} \delta_{a}{}^b~,
\end{align}
with $\eta_{\m\n}$ mostly plus
\begin{align}
	\eta_{\m\n}= diag(-1,+1,+1, \dots, +1).
\end{align}
The three form matrix is
\begin{align}
(	\sigma_{\l\m\n})_{ab} = \tfrac{1}{3!} (\sigma_{[\l} \sigma_\m \sigma_{\n]})_{ab}
\end{align}
Denoting $n$-anytisymmetric indices as  $[n]$, the $n$-form matrix is
\begin{align}
	\sigma_{[n]} = \sigma_{a_1 a_2 a_3 \dots a_n} = \tfrac{1}{n!} \sigma_{[a_1} \sigma_{a_2} \sigma_{a_3} \dots \sigma_{a_n]}
\end{align}
where the matrix indices either both down or both up for odd $n$ and one down and one up for even $n$.

 We use the following representation for the sigma matrices in 10D, a reordering and rearrangement of those used in~\cite{Gates:2019dyk}. 
\begin{align}
	(\sigma^\m)_{ab} =& \left\{
	\begin{array}{c}
		I_2 \otimes I_2 \otimes I_2 \otimes I_2\\
		I_2 \otimes I_2 \otimes I_2 \otimes \sigma^1 \\
		\sigma^3 \otimes I_2 \otimes \sigma^2 \otimes \sigma^2 \\
		I_2 \otimes I_2 \otimes I_2 \otimes \sigma^3 \\
		I_2 \otimes \sigma^2 \otimes \sigma^1 \otimes \sigma^2 \\
		I_2 \otimes \sigma^2 \otimes \sigma^3 \otimes \sigma^2 \\
		\sigma^1 \otimes I_2 \otimes \sigma^2 \otimes \sigma^2 \\
		\sigma^2 \otimes \sigma^3 \otimes I_2 \otimes \sigma^2 \\
		\sigma^2 \otimes \sigma^1 \otimes I_2 \otimes \sigma^2 \\
		\sigma^2 \otimes \sigma^2 \otimes \sigma^2 \otimes \sigma^2 
	\end{array}
	\right.~~~,~~~	(\sigma^\m)^{ab} = \left\{
	\begin{array}{c}
		-I_2 \otimes I_2 \otimes I_2 \otimes I_2\\
		I_2 \otimes I_2 \otimes I_2 \otimes \sigma^1 \\
		\sigma^3 \otimes I_2 \otimes \sigma^2 \otimes \sigma^2 \\
		I_2 \otimes I_2 \otimes I_2 \otimes \sigma^3 \\
		I_2 \otimes \sigma^2 \otimes \sigma^1 \otimes \sigma^2 \\
		I_2 \otimes \sigma^2 \otimes \sigma^3 \otimes \sigma^2 \\
		\sigma^1 \otimes I_2 \otimes \sigma^2 \otimes \sigma^2 \\
		\sigma^2 \otimes \sigma^3 \otimes I_2 \otimes \sigma^2 \\
		\sigma^2 \otimes \sigma^1 \otimes I_2 \otimes \sigma^2 \\
		\sigma^2 \otimes \sigma^2 \otimes \sigma^2 \otimes \sigma^2 
	\end{array}
	\right.
\end{align}
The above representation allows for a simple reduction to 6D and 4D by taking the upper left block of the first six and four $\sigma$-matrices, respectively. Specifically, we have in 6D
\begin{align}
	(\sigma^\m)_{ab} =& \left\{
	\begin{array}{c}
		I_2 \otimes I_2 \otimes I_2\\
		I_2 \otimes I_2 \otimes \sigma^1 \\
		I_2 \otimes \sigma^2 \otimes \sigma^2 \\
		I_2 \otimes I_2 \otimes \sigma^3 \\
		\sigma^2 \otimes \sigma^1 \otimes \sigma^2 \\
		\sigma^2 \otimes \sigma^3 \otimes \sigma^2 \\
	\end{array}
	\right.~~~,~~~	(\sigma^\m)^{ab} = \left\{
	\begin{array}{c}
		-I_2 \otimes I_2 \otimes I_2\\
		I_2 \otimes I_2 \otimes \sigma^1 \\
		I_2 \otimes \sigma^2 \otimes \sigma^2 \\
		I_2 \otimes I_2 \otimes \sigma^3 \\
	    \sigma^2 \otimes \sigma^1 \otimes \sigma^2 \\
		\sigma^2 \otimes \sigma^3 \otimes \sigma^2 \\
	\end{array}
	\right.
\end{align}
and in 4D
\begin{align}
	(\sigma^\m)_{ab} =& \left\{
	\begin{array}{c}
		I_2 \otimes I_2\\
		I_2 \otimes \sigma^1 \\
		\sigma^2 \otimes \sigma^2 \\
		I_2 \otimes \sigma^3 
	\end{array}
	\right.~~~,~~~	(\sigma^\m)^{ab} = \left\{
	\begin{array}{c}
		- I_2 \otimes I_2\\
		 I_2 \otimes \sigma^1 \\
		\sigma^2 \otimes \sigma^2 \\
		 I_2 \otimes \sigma^3 
	\end{array}
	\right.
\end{align}


$~$\newline
\section{SUSY Transformation Laws}
\label{appen:Transformation}
The 4D, $\cal N$ = 4 Abelian Vector-Tensor Supermultiplet can be formulated in terms of four sets of transformation laws
implemented on the two distinct 4D, $\cal N$ = 2 supermultiplets as presented in the work of \cite{N4VTM}. The first two transformation laws are the standard 
4D, $\cal N$ = 2 transformation laws acting solely on the component fields {\it {within}} the vector supermultiplet 
and separately acting solely on the component fields {\it {within}} the tensor supermultiplet.
The remaining two describe supersymmetry variations on the component fields {\it {between}} the vector and the tensor
supermultiplets.  To distinguish these two types of SUSY charges, we use the symbols ${\rm D}\downup ai$ for the first type
and ${\Tilde {\rm D}}\downup ai$ for the second type.
\subsection{D-Type SUSY VM Realization}
The notations $(A, \, B,\, A\dn\m, \, \Psi\downup aj , \, F,\, G, \, d)$ denote the
component fields of the 4D, $\cal N$ = 2 vector multiplet.  These transform as 
under the operator ${\rm D}\downup ai$ as,
\begin{align}
    {\rm D}\downup ai A &= \d\uup ij\Psi\downup aj  ~~~, \nonumber \\ 
    {\rm D}\downup ai B &= i\d\uup ij(\gup5)\sw ab\Psi\downup bj ~~~,  \nonumber \\
    {\rm D}\downup ai F &= \d\uup ij\gupsw\m ab\pd\m\Psi\downup bj ~~~,  \nonumber \\ 
    {\rm D}\downup ai G &= i\sig3ij(\gup5\gup\m)\sw ab\pd\m\Psi\downup bj ~~~,  \nonumber \\
    {\rm D}\downup ai A\dn\m &= i\sig2ij(\g\dn\m)\sw ab\Psi\downup bj ~~~, \nonumber  \\
    {\rm D}\downup ai d &= i\sig1ij(\gup5\gup\m)\sw ab\pd\m\Psi\downup bj ~~~,  \nonumber  \\
    {\rm D}\downup ai \Psi\downup bj &= \d\uup ij\{i(\gup\m)\ddn ab\pd\m A - (\gup5\gup\m)\ddn ab\pd\m B - iC\ddn ab F\}\nonumber\\
    &{~~\,~} + (\gup5)\ddn ab\{\sig3ij G+ \sig1ijd\}+ \fracm{1}{4}\sig2ij(\g\uup[\m\g\uup\n])\ddn ab F\ddn\m\n ~~~,
\end{align}
under the action of ${{\rm D}}\downup ai$.
\subsection{\texorpdfstring{${\rm D}$}{D}-Type SUSY TM Realization}
The notations $({\Tilde A}, \, {\Tilde B},\, {\Tilde \varphi},\, {\Tilde B}\dn{\m \n}, \, {\Tilde \Psi}\downup aj , \, {\Tilde F},\, {\Tilde G})$ denote the
component fields of the 4D, $\cal N$ = 2 tensor multiplet.  These transform as 
under the operator ${\rm D}\downup ai$ as,
\begin{align}
    {\rm D}\downup ai \til A &= \sig3ij\til\Psi\downup aj ~~~,  \nonumber \\
    {\rm D}\downup ai\til B &= i\d\uup ij\gupsw5ab\til\Psi\downup bj ~~~,  \nonumber \\
    {\rm D}\downup ai\til F &= \d\uup ij\gupsw\m ab\partial\dn\m\til\Psi\downup bj  ~~~, \nonumber \\  
    {\rm D}\downup ai\til G &= i\d\uup ij\ggfiupsw\m ab\partial\dn\m\til\Psi\downup bi  ~~~, \nonumber \\
    {\rm D}\downup ai\til\varphi&= \sig1ij\til\Psi\downup aj  ~~~, \nonumber \\
    {\rm D}\downup ai\til B\ddn\m\n &= -\fracm{i}{4}\sig2ij(\g\ddn[\m\g\ddn\n])\sw ab\til\Psi\downup bj  ~~~, \nonumber \\
    {\rm D}\downup ai \til\Psi\downup bj &= \d\uup ij\{-\ggfiupdown\m ab\partial\dn\m\til B - iC\ddown ab\til F + \gupdown5ab\til G\}
    \nonumber\\
    &{~~\,~}+ i\gupdown\m ab\partial\dn\m\{\sig3ij\til A + \sig1ij\til\varphi\}- i\sig2ij\e\duuu\m\n\a\b\ggfiupdown\m ab\partial\dn\n\til B\ddn\a\b ~~~.
    \end{align}
\subsection{\texorpdfstring{$\Tilde {\rm D}$}{D}-Type SUSY VM Realization}
The transformation of the VM fields under the operator ${\Tilde {\rm D}}\downup ai$ look as,
\begin{align}
\til {\rm D}\downup ai A &= i\sig2ij\tilpsi{}\downup aj\\
\til {\rm D}\downup ai B &= -\sig2ij\gupsw5ab\tilpsi{}\downup bj\\
\til {\rm D}\downup ai F &= i\sig2ij\gupsw\m ab\pd\m \tilpsi{}\downup bj\\
\til {\rm D}\downup ai G &= i\sig1ij\ggfiupsw\m ab\pd\m\tilpsi\downup bj\\
\til {\rm D}\downup ai d &= -i\sig3ij\ggfiupsw\m ab\pd\m\tilpsi\downup bj\\
\til {\rm D}\downup ai A\dn\m &= 
\d\uup ij(\g\dn\m)\sw
ab\tilpsi\downup bj\\
\til {\rm D}\downup ai \Psi\downup bj &= i\sig2ij\{-\ggfiupdown\m ab \pd\m \tilb-iC\ddn ab\tilf+\gupdown5ab\tilg\}\nonumber\\
&{~~\,~}+i\gupdown\m ab\pd\m\{-\sig1ij\tila+\sig3ij\til\varphi\}-\d\uup ij\e\duuu\m\n\a\b\ggfiupdown\m ab\pd\n\tilbmn\a\b
\end{align}
\subsection{\texorpdfstring{$\Tilde {\rm D}$}{D}-Type SUSY TM Realization}
The transformation of the TM fields under the operator ${\Tilde {\rm D}}\downup ai$ look as,
\begin{align}
    \til {\rm D}\downup ai\tila &= -\sig1ij\Psi\downup aj\\
     \til {\rm D}\downup ai\tilb &=-\sig2ij\gupsw5ab\Psi\downup bj\\
     \til {\rm D}\downup ai\tilf &= i\sig2ij\gupsw\m ab\pd\m\Psi\downup bj\\
      \til {\rm D}\downup ai\tilg &= - \sig2ij\ggfiupsw\m ab\pd\m\Psi\downup bj\\
       \til {\rm D}\downup ai\til\varphi &= \sig3ij\Psi\downup aj\\
        \til {\rm D}\downup ai\tilbmn\m\n &= -\fracm{1}{4}\d\uup ij([\g_\m,\g_\n])_{ a}^{~b}\Psi\downup bj\\
     \til {\rm D}\downup ai\tilpsi\downup bj &= i\sig2ij\{i\gupdown\m ab\pd\m A - \ggfiupdown\m ab\pd\m B - iC\ddn abF\}\nonumber\\
     &{~~\,~}+\gupdown5ab\{\sig1ijG-\sig3ijd\}-\fracm{i}{4}\d\uup ij\gupcommdown\m\n abF\ddn\m\n
\end{align}
A complete discussion of the algebra associated with these transformations can be found the 
work given in \cite{N4VTM}.

\section{Algebra}\label{a:onshell}
The algebra of the full transformation laws shown in appendix~\ref{appen:Transformation} is as in~\cite{N4VTM}. The on-shell transformation laws are found by imposing the equations of motion, effectively setting all auxiliary fields to zero. The on-shell bosonic algebra is equivalent to the bosonic algebra in~\cite{N4VTM} upon setting all auxiliary fields to zero.  Some of the results of the on-shell algebra for the fermions are shown below, the rest follow a similar structure and can be found explicitly in the \emph{Mathematica} code on \href{https://hepthools.github.io/Data/4DN4Holo/}{\emph{GitHub}}. We have
\begin{align}
	\{ \rD_a^i , \rD_b^j \} \Psi_c^k=& 
	2 i \delta^{ij} (\g^\mu)_{ab}  \pa_\mu\Psi_c^k  + \left(Z^{ijkl}\right)_{abc}{}^{d}(\g^\mu)_{d}{}^{e} \pa_\mu \Psi_e^l \cr
	\{ \rD_a^i , \rD_b^j \} \widetilde{\Psi}_c^k=& 
	2 i \delta^{ij} (\g^\mu)_{ab}  \pa_\mu \widetilde{\Psi}_c^k  + \left(\widetilde{Z}^{ijkl}\right)_{abc}{}^{d}(\g^\mu)_{d}{}^{e} \pa_\mu \widetilde{\Psi}_e^l \cr 
	\{ \widetilde{\rD}_a^i , \widetilde{\rD}_b^j \} \Psi_c^k=& 
	2 i \delta^{ij} (\g^\mu)_{ab}  \pa_\mu\Psi_c^k  + \left(\mathcal{Z}^{ijkl}\right)_{abc}{}^{d}(\g^\mu)_{d}{}^{e} \pa_\mu \Psi_e^l \cr
	\{ \widetilde{\rD}_a^i , \widetilde{\rD}_b^j \} \widetilde{\Psi}_c^k=& 
	2 i \delta^{ij} (\g^\mu)_{ab}  \pa_\mu \widetilde{\Psi}_c^k  + \left(\widetilde{\mathcal{Z}}^{ijkl}\right)_{abc}{}^{d}(\g^\mu)_{d}{}^{e} \pa_\mu \widetilde{\Psi}_e^l 
\end{align}
with
\begin{align}
	\left(Z^{ijkl}\right)_{abd}{}^d = & - i\tfrac{3}{4}  \delta^{ij}\delta^{kl} \left(\gamma^\alpha\right)_{ab}(\g_\alpha)_c{}^d - i\tfrac{1}{32} \delta^{ij}\delta^{kl} ([\g^{\alpha} , \g^{\nu} ])_{ab}([\g_{\alpha} , \g_{\nu} ])_{c}{}^d  \cr
	&- i \tfrac{1}{4} (\s^1)^{ij} (\s^1)^{kl} (\g^\alpha)_{ab}(\g_\alpha)_c{}^d + i \tfrac{1}{32} (\s^1)^{ij} (\s^1)^{kl} ([\g^\alpha, \g^\nu])_{ab} ([\g_\alpha, \g_\nu])_c{}^d \cr
	&+ i \tfrac{1}{32} (\s^3)^{ij} (\s^3)^{kl} ([\g^\alpha, \g^\nu])_{ab} ([\g_\alpha, \g_\nu])_c{}^d - i \tfrac{1}{4} (\s^3)^{ij} (\s^3)^{kl} (\g^\alpha)_{ab} (\g_\alpha)_c{}^d \cr
	& - i \tfrac{1}{4} (\s^2)^{ij} (\s^2)^{kl} (\g^5 \g^{\alpha})_{ab} (\g^5\g_{\alpha})_c{}^d - i \tfrac{3}{4} (\s^2)^{ij} (\s^2)^{kl} C_{ab} \delta_c{}^d  \cr
	&- i \tfrac{3}{4} (\s^2)^{ij} (\s^2)^{kl} (\g^5)_{ab} (\g^5)_{c}{}^d
\end{align}
and the terms of the form $\widetilde{Z}$, $\mathcal{Z}$, and $\widetilde{\mathcal{Z}}$ found explicitly in the \emph{Mathematica} code on \href{https://hepthools.github.io/Data/4DN4Holo/}{\emph{Git hub}}. 

For the cross terms we have
\begin{align}
	\{ \rD_a^i , \widetilde{\rD}_b^j \} \Psi_c^k=& 2 i \mathcal{X}_1^{ijkl} (\g^\mu)_{ab} \pa_\mu \widetilde{\Psi}^l_c + 2 i \mathcal{X}_2^{ijkl} ([\g^\mu, \g^\nu])_{ab} (\g_\nu)_{c}{}^d \pa_\mu \widetilde{\Psi}^l_d \cr
	&+2i \mathcal{X}_3^{ijkl} (\g_{\nu})_{ab} (\g^\n \g^\m)_c{}^d \pa_\mu \widetilde{\Psi}^l_d + 2 i \mathcal{X}_4^{ijkl} (\g^5\g^\m)_{ab} (\g^5)_{c}{}^d \pa_\mu \widetilde{\Psi}^l_d\cr
	&+2 i \mathcal{X}_5^{ijkl} (\g^5\g_\n)_{ab} (\g^5\g^\n\g^\m)_c{}^d \pa_\mu \widetilde{\Psi}^l_d + 2 i \mathcal{X}_6^{ijkl} (\g^5)_{ab} (\g^5\g^\m)_c{}^d \pa_\mu \widetilde{\Psi}^l_d \cr
	&+2 i \mathcal{X}_7^{ijkl} C_{ab} (\g^\m)_{c}{}^d \pa_\mu \widetilde{\Psi}^l_d + 2 i \mathcal{X}_8^{ijkl} ([\g^\m,\g^\n])_{ab} ([\g_\m , \g_\n] \g^\lambda)_c{}^d \pa_\lambda \widetilde{\Psi}^l_d
\end{align}
\begin{align}
	\{ \rD_a^i , \widetilde{\rD}_b^j \} \widetilde{\Psi}_c^k=& 2 i \widetilde{\mathcal{X}}_1^{ijkl} (\g^\mu)_{ab} \pa_\mu \Psi^l_c + 2 i \widetilde{\mathcal{X}}_2^{ijkl} ([\g^\mu, \g^\nu])_{ab} (\g_\nu)_{c}{}^d \pa_\mu \Psi^l_d \cr
	&+2i \widetilde{\mathcal{X}}_3^{ijkl} (\g_{\nu})_{ab} (\g^\n \g^\m)_c{}^d \pa_\mu \Psi^l_d + 2 i \widetilde{\mathcal{X}}_4^{ijkl} (\g^5\g^\m)_{ab} (\g^5)_{c}{}^d \pa_\mu \Psi^l_d\cr
	&+2 i \widetilde{\mathcal{X}}_5^{ijkl} (\g^5\g_\n)_{ab} (\g^5\g^\n\g^\m)_c{}^d \pa_\mu \Psi^l_d + 2 i \widetilde{\mathcal{X}}_6^{ijkl} (\g^5)_{ab} (\g^5\g^\m)_c{}^d \pa_\mu \Psi^l_d \cr
	&+2 i \widetilde{\mathcal{X}}_7^{ijkl} C_{ab} (\g^\m)_{c}{}^d \pa_\mu \Psi^l_d + 2 i \widetilde{\mathcal{X}}_8^{ijkl} ([\g^\m,\g^\n])_{ab} ([\g_\m , \g_\n] \g^\lambda)_c{}^d \pa_\lambda \Psi^l_d
\end{align}
\begin{align}
\mathcal{X}_x^{ijkl}=& \a_x\d^{ij}\sig2kl+\b_x\sig1ij\sig3kl+\d_x\sig3ij\sig1kl+\k_x\sig2ij\d^{kl} \cr
\widetilde{\mathcal{X}}_x^{ijkl}=& \a_x\d^{ij}\sig2kl+\b_x\sig1ij\sig3kl+\d_x\sig3ij\sig1kl+\k_x\sig2ij\d^{kl}
\end{align}

\begin{table}[!h]
\centering
\begin{tabular}{|c|c|c|c|c|}
\hline
 & $\a$ & $\b$ & $\d$ & $\k$ \\ \hline
$\mathcal{X}_1^{ijkl}$ & $i$/4 & -3/4 & 3/4 & -$i$/4 \\ \hline
$\mathcal{X}_2^{ijkl}$ &-$i$/8 &-1/8  & 1/8  & $i$/8 \\ \hline
$\mathcal{X}_3^{ijkl}$ & -$i$/16 & 3/16 & -3/16 & 5$i$/16 \\ \hline
$\mathcal{X}_4^{ijkl}$ & -$i$/4 & -1/4 & 1/4 & $i$/4 \\ \hline
$\mathcal{X}_5^{ijkl}$ & 3$i$/16 & -1/16 & 1/16 & $i$/16 \\ \hline
$\mathcal{X}_6^{ijkl}$ & -9$i$/16 & -1/16 & 1/16 & $i$/16  \\ \hline
$\mathcal{X}_7^{ijkl}$ & 3$i$/16 & -5/16 & 5/16 & 5$i$/16 \\ \hline
$\mathcal{X}_8^{ijkl}$ & -$i$/128 & -1/128 & 1/128 & $i$/128 \\  \hline
\end{tabular}
\quad
\begin{tabular}{|c|c|c|c|c|}
\hline
 & $\a$ & $\b$ & $\d$ & $\k$ \\ \hline
$\widetilde{\mathcal{X}}_1^{ijkl}$ & -$i$/4 & -3/4 &3/4 & -$i$/4 \\ \hline
$\widetilde{\mathcal{X}}_2^{ijkl}$ &$i$/8 &-1/8  & 1/8 & $i$/8 \\ \hline
$\widetilde{\mathcal{X}}_3^{ijkl}$ & $i$/16 & 3/16 & -3/16 & 5$i$/16 \\ \hline
$\widetilde{\mathcal{X}}_4^{ijkl}$ & -$i$/4 & 1/4 & -1/4 & -$i$/4 \\ \hline
$\widetilde{\mathcal{X}}_5^{ijkl}$ & 3$i$/16 & 1/16 & -1/16 & -$i$/16 \\ \hline
$\widetilde{\mathcal{X}}_6^{ijkl}$ & -9$i$/16 & 1/16 & -1/16 & -$i$/16  \\ \hline
$\widetilde{\mathcal{X}}_7^{ijkl}$ & 3 $i$/16 & 5/16 & -5/16 & -5$i$/16 \\ \hline
$\widetilde{\mathcal{X}}_8^{ijkl}$ & $i$/128 & -1/128 & 1/128 & $i$/128 \\  \hline
\end{tabular}
\end{table}
The above demonstrates the closure of the on-shell algebras $\{\rD_a^i , \rD_b^j\}$ and $\{\widetilde{\rD}_a^i , \widetilde{\rD}_b^j\}$ upon enforcing the equations of motion $(\g^\mu)_{d}{}^{e} \pa_\mu \Psi_e^l = (\g^\mu)_{d}{}^{e} \pa_\mu \widetilde{\Psi}_e^l  = 0$ as expected.
\newpage
\section{Holoraumy}
\label{appen:holoraumy}

In the results of this section, we present only new results by deriving the holoraumies for these 
supermultiplets under the action of the four operators ${\rm D}\downup ai $ and ${\Tilde {\rm D}}\downup ai 
$.  This will be undertaken in three sectors, the ${\rm D}$-${\rm D}$ sector, 
${\Tilde {\rm D}}$-${\Tilde {\rm D}}$ sector, and the ${\rm D}$-${\Tilde {\rm D}}$ sector. In this section and the next, calculations are from the full transformation laws in appendix~\ref{appen:Transformation} and we will sometimes refer to these as ``off-shell" in the sense that the underlying $\mathcal{N}=2$ tensor and vector multiplets close off-shell although of course the composite $\mathcal{N}=4$ vector-tensor multiplet does not.

\subsection{Vector Multiplet \texorpdfstring{${\rm D}$-${\rm D}$}{D-D} Bosonic Holoraumy}

In the following equations, the ${\rm D}$-${\rm D}$ subsector of the holoraumy is
presented on the bosonic fields of the 4D, $\cal N$ = 2 vector supermultiplet.
We find
\begin{align}
    \DDAB{}A &= -\, 2 \, \d\uup ij\, [ \,  (\gup5\gup\m)\ddn ab\pd\m B + iC\ddn ab F \, ]
    + 2\, \gupdown5ab \, [ \, \sig3ijG + \sig1ijd \, ] \nonumber  \\ 
    &{~~~\,} +\fracm{1}{2}\sig2ij(\g\uup[\m\g\uup\n])\ddn abF\ddn\m\n ~~~, 
\nonumber \\
    \DDAB{}B &= 2\, \d\uup ij\, [ \, (\gup5\gup\m)\ddn ab\pd\m A + \gupdown5ab F\, ] 
    + i\, 2 \, C\ddn ab [ \, \sig3ij G+ \sig1ij d\, ]  \nonumber  \\ 
    &{~~~\,}+ i \fracm{1}{2} \sig2ij(\gup5\g\uup[\m\g\uup\n])\ddn abF\ddn\m\n ~~~,  
\nonumber\ \\
    \DDAB{}F &= 2\, \d\uup ij\, [ \,-iC\ddn ab\square A + \gupdown5ab\square B\, ] 
-2 \, (\gup5\gup\m)\ddn ab\pd\m \, [ \, \sig3ij G + \sig1ij d \, ]  \nonumber  \\ 
    &{~~~\,}- 2\sig2ij\gupdown\m ab\pu\n F\ddn\m\n ~~~,
\nonumber\\
    \DDAB{}G &= 2\, \sig3ij\, [ \,\gupdown5ab\square A + i C\ddn ab\square B + 
    (\gup5\gup\m)\ddn ab\pd\m F\, ] + 2\, \sig2ij(\gup\m)\pd\m d  
     \nonumber  \\ 
    &{~~~\,}-2\sig1ij(\gup5\gup\m)\ddn ab\pu\n F\ddn\m\n ~~~,
\nonumber\\
    \DDAB{}d &= 2\, \sig1ij \, [ \, \gupdown5ab\square A + iC\ddn ab\square B +(\gup5\gup\m)\ddn ab\pd\m F\, ]  
    -2\, \sig2ij\gupdown\m ab\pd\m G \nonumber  \\ 
    &{~~~\,}+ 2\sig3ij(\gup5\gup\m)\ddn ab\pu\n\Fmn{} ~~~,
    \nonumber\\
    \DDAB A\dn\m &= \sig2ij \, [ \, ([\g_\mu , \g^\n  ])\ddn ab\pd\n A + i(\gup5[\g_\mu , \g^\n  ])\ddn ab\pd\n B + 2(\g\dn\m)\ddn ab F\, ] \nonumber\\
    &{~~\,~} +2(\gup5\g\dn\m)\ddn ab\, [ \, \sig1ij G- \sig3ij d \, ] - \d\uup ij\e\duuu\m\n\a\b(\gup5\g\dn\n)\ddn abF\ddn\a\b ~~~,
\end{align}
for the fields of the vector supermultiplet holoraumy of this type.

\subsection{Vector Multiplet \texorpdfstring{${\rm D}$-${\Tilde {\rm D}}$}{D-D} Bosonic Holoraumy}
\vspace*{-24 pt}
\begin{align}
    \holcrosstil{}A &= 2\sig2ij[ \,i\ggfiupdown\m ab\pd\m\tilb - C\ddn ab\tilf-i\gupdown5abG\, ]
    -\fracm{2}{3}\d\uup ij\e\duuu\m\n\a\b\ggfiupdown\m ab\tilh\n\a\b  ~~~, \nonumber \\
    \holcrosstil{}B&= 2\sig2ij[ \,-i\gupdown5ab\tilf+C\ddn ab\tilg\, ]
    +2\ggfiupdown\m ab\pd\m[ \,-\sig1ij\tila+\sig3ij\til\varphi\, ] ~~~, \nonumber \\
    \holcrosstil{}F&= i \, 2 \sig2ij[ \,-\gupdown5ab\lap\tilb+\ggfiupdown\m ab\pd\m\tilg\, ]
    + i \, 2 C\ddn ab[ \,\sig1ij\lap\tila-\sig3ij\lap\til\varphi\, ] ~~~, \nonumber \\
    \holcrosstil{}G&= i \, 2 \sig1ij\gpart\tilg+2\d\uup ij\gupdown5ab\lap\til\varphi ~~~,\nonumber  \\
    \holcrosstil{}d&=-i \,2\sig3ij\gpart\tilg-2\d\uup ij\gupdown5ab\lap\tila ~~~,  \nonumber \\
    \holcrosstil{}A\dn\m &= 2\d\uup ij[ \,\gupdown5ab\pd\m\tilb-(\gup5\g\dn\m)\ddn ab\tilg\, ]
    +i\, (\gdnbrack\m\n)_{ab}\pu\n[ \,\sig3ij\tila+\sig1ij\til\varphi\, ]\nonumber\\
    &{~~~\,} + i \, \fracm{2}{3}\sig2ij\e\duuu\m\n\a\b\gupdown5ab\tilh\n\a\b ~~~,
\end{align}
where $\tilh\m\a\b=\pd\m\til B_{\a\b} +\pd\a\til B_{\b\m} +\pd\b\til B_{\m\a} $.
\subsection{Vector Multiplet \texorpdfstring{${\Tilde {\rm D}}$-${\Tilde {\rm D}}$}{D-D} Bosonic Holoraumy}
\vspace*{-24 pt}
\begin{align}
    \holtil{}A &=-2\d\uup ij [ \, 2\ggfiupdown\m ab\pd\m B+2iC\ddn abF\, ]
    +2\gupdown5ab [ \, \sig3ijG+\sig1ijd\, ]\nonumber\\
    &{~~~\,} -\fracm{1}{2}\sig2ij(\gupbrack\m\n)\ddn abF\ddn\m\n   ~~~, \nonumber \\
    \holtil{}B &= 2\d\uup ij [ \, \ggfiupdown\m ab\pd\m A+\gupdown5abF\, ]
    +2iC\ddn ab [ \, \sig3ijG+\sig1ijd\, ]\nonumber\\
    &{~~~\,}-\fracm{i}{2}\sig2ij(\gup5\gupbrack\m\n)\ddn abF\ddn\m\n  ~~~, \nonumber \\
    \holtil{}F&= 2\d\uup ij [ \, -i C\ddn ab\lap A+\gupdown5ab\lap B\, ]
    -2\ggfiupdown\m ab \partial_\mu [ \, \sig3ijG+\sig1ijd\, ]\nonumber\\
    &{~~~\,} +2\sig2ij\gupdown\m ab  \pu\n F\ddn\m\n  ~~~,  \nonumber \\
    \holtil G&= 2\sig3ij [ \, \gupdown5ab\lap A+iC\ddn ab\lap B+\ggfiupdown\m ab\pd\m F\, ]
    +2\sig2ij\gpart{}d \nonumber\\
    &{~~~\,} + 2\sig1ij\ggfiupdown\m ab\pu\n F\ddn \m\n  ~~~, \nonumber \\
    \holtil d&= 2\sig1ij [ \, \gupdown5ab\lap A+iC\ddn ab\lap B+\ggfiupdown\m ab\pd\m F\, ]\nonumber\\
    &{~~~\,} -2\sig2ij\gpart{}G- 2\sig3ij\ggfiupdown\m ab\pu\n F\ddn \m\n  ~~~, \nonumber \\
    \holtil A\dn\m &=-\sig2ij [ \, (\gdnbrack\m\n)\ddn ab\pu\n A+i(\gup5\gdnbrack\m\n)\ddn ab\pu\n B+2(\g\dn\m)\ddn ab F\, ]\nonumber\\
    &{~~~\,} -2(\gup5\g\dn\m)\ddn ab [ \, \sig1ijG-\sig3ijd\, ]
    -\d\uup ij\e\duuu\m\l\k\d(\gup5\g\dn\d)\ddn abF\ddn\l\k ~~~.
\end{align}

\subsection{Vector Multiplet \texorpdfstring{${\rm D}$-${\rm D}$}{D-D} Fermionic Holoraumy}
Following on in a similar fashion, the ${\rm D}$-${\rm D}$ subsector of the holoraumy is
presented on the fermionic fields of the 4D, $\cal N$ = 2 vector supermultiplet next.
\begin{align}
    \DDAB{}\Psi\downup ck &=
    i 2\, [ \, {\cal V}{}_1{}^{i j k l }\, (\g^\m)_{ab} \delta_c{}^d
    + {\cal V}{}_2{}^{i j k l } \,  ( \gup5\gup\m)\ddn ab (\gup5)\sw cd \, ] \,  \pd\m  \Psi\downup dl
    \nonumber \\
    &{~~\,~}+i {\cal V}{}_3{}^{i j k l } \, [ \, C\ddn ab(\gup\l)\sw cd +
   \gupdown5ab(\gup5\gup\l)\sw cd +
   (\gup5\gup\n)\ddn ab(\gup5\g\dn\n\gup\l)\sw cd \, ] \pd\l\Psi\downup dl  \nonumber \\
    &{~~\,~}+i {\cal V}{}_4{}^{i j k l } \,
    [ \, C\ddn ab (\gup\l)\sw cd +  (\gup5)\ddn ab(\gup5\gup\l)\sw cd - (\gup5 \gup\n)\ddn ab(\gup5\g\dn\n\gup\l)\sw cd    \, ] \pd\l  \Psi\downup dl  \nonumber \\
    &{~~\,~}+i\, {\cal V}{}_5{}^{i j k l } \, [ \, \gupdown\m ab(\g\dn\m\gup\l)\sw cd +  
    \fracm{1}{8} \, (\g\uup[\m\g\uup\n])\ddn ab (\g\ddn[\m\g\ddn\n]\gup\l)\sw cd  \, ] \,\pd\l\Psi\downup dl  ~~~,
\label{eq:VMFHdd}    
\end{align}
where the factors of $ {\cal V}{}_x{}^{i j k l }$ are defined by the expression
\begin{align}\label{e:cV}
{\cal V}{}_x{}^{i j k l } 
 &=     {\Hat \a}_x \sig2ij\, \sig2kl+ {\Hat \b}_x \, [ \, \sig1ij\sig1kl + \sig3ij\sig3kl  \, ] \, +
{\Hat \k}_x \d{}^{ij}\d{}^{kl}   ~~~,
\end{align}
and the corresponding coefficients ${\Hat \a}{}_x$, ${\Hat \b}{}_x$, and ${\Hat \k}{}_x$ are given in the following table.
\begin{table}[h]
\centering
\begin{tabular}{|c|c|c|c|}
\hline
 & $\Hat \a$ & $\Hat \b$  & $\Hat \k$ \\ \hline
${\cal V}{}_1$ & 1 & 0 & 0 \\ \hline
${\cal V}{}_2$ & 0 & 1 & 0 \\ \hline
${\cal V}{}_3$ & 0 & 0 & -1 \\ \hline
${\cal V}{}_4$ & 0 & 1 & 0 \\ \hline
${\cal V}{}_5$ & -1 & 0 & 0 \\ \hline
\end{tabular}
\end{table}

\subsection{\label{VMcrossfermion}Vector Multiplet \texorpdfstring{${\rm D}$-${\Tilde {\rm D}}$}{D-D} Fermionic Holoraumy}
The calculation yields
\begin{align}
\holcrosstil{}\Psi\downup ck &= i 2 \, \  \vee1\gpart\tilpsi\downup cl 
+i 2 \, [ \, \vee{2}\gupcommdown\m\n ab(\g\dn\n)\sw cd\ 
+ \,\vee{3}\ggfiupdown\m ab\gupsw5cd \, ] \, \pd\m \tilpsi\downup dl  \nonumber \\
&{~~~\,} + i 2 \, [ \, \vee{4}(\g\dn\n)\ddn ab(\gup\n\gup\l)\sw cd
+\vee{5}\gupcommdown\m\n ab(\g{}_{[\m}\g{}_{\n]} \gup\l)\sw cd \, ] \, \pd\l\tilpsi\downup dl
\nonumber \\
&{~~~\,} + i 2 \, [ \,  \vee6(\gup5\g\dn\n)\ddn ab(\gup5\gup\n\gup\l)\sw cd
+\vee7\gupdown5ab\ggfiupsw\l cd \, ] \, \pd\l\tilpsi\downup dl  \nonumber \\
&{~~~\,} +i 2 \,\vee8C\ddn ab\gupsw\l cd\pd\l\tilpsi\downup dl  ~~~,
\end{align}
and in these equations, we have introduced $\vee x$ by use of the
definition $\vee x \,=\,  i {\cal V}{}_x{}^{i r k l } \sig2rj $, with $\mathcal{V}_x^{irkl}$ defined in Eq.~\eqref{e:cV},
where the corresponding coefficient ${\Hat \a}{}_x$, ${\Hat \b}{}_x$, and ${\Hat \k}{}_x$ are given in the following table.
\begin{table}[h]
\centering
\begin{tabular}{|c|c|c|c|}
\hline
 & $\Hat \a$ & $\Hat \b$  & $\Hat \k$ \\ \hline
$\vee1$ & 1/4 & 1/4  & -1/4 \\ \hline
$\vee{2}$ &-1/8 &-1/8    &1/8 \\ \hline
$\vee{3}$ &3/4 & -1/4  & -3/4\\ \hline
$\vee{4}$ &-1/4 &-1/4   & 0 \\ \hline
$\vee{5}$ &-1/32 &-1/32    & 0\\ \hline
$\vee6$ &-1/4 &1/4    &1/2 \\ \hline
$\vee7$ &-1/4 & 1/4   & -1/2\\ \hline
$\vee8$ &1/2 &0  & -1/4 \\ \hline
\end{tabular}
\end{table}
\subsection{Vector Multiplet \texorpdfstring{${\Tilde {\rm D}}$-${\Tilde {\rm D}}$}{D-D} Fermionic Holoraumy}
The calculation yields
\begin{align}
    \holtil{}\Psi\downup ck &= i \, [ \, \tfrac{1}{2} \,  {\cal V}{}_1{}^{i j k l } \,\ggfiupdown\m ab\gupsw5cd 
    + \tfrac{1}{4} {\cal V}{}_2{}^{i j k l }  \, \gupcommdown\m\n ab(\g\dn\n)\sw cd \, ] \, \pd\m  \Psi\downup dl  
    \nonumber\\
    &{~~\,} + i\, {\cal V}{}_3{}^{i j k l } \, 
    [\, \gupdown\m ab(\g\dn\m\gup\l)\sw cd+\fracm{1}{8}\gupcommdown\m\n ab(\gdnbrack\m\n\gup\l)\sw cd]\pd\l\Psi\downup dl   \nonumber\\
    &{~~\,}+i\,  {\cal V}{}_4{}^{i j k l } \, [\, \ggfiupdown\m ab(\gup5\g\dn\m\gup\l)\sw cd -C\ddn ab\gupsw\l cd +\gupdown5ab\ggfiupsw\l cd\,]\pd\l\Psi\downup dl \nonumber\\
    &{~~\,}+i\,  {\cal V}{}_5{}^{i j k l } \, [\,\ggfiupdown\m ab(\gup5\g\dn\m\gup\l)\sw cd+C\ddn ab\gupsw\l cd-\gupdown5ab\ggfiupsw\l cd\,]\pd\l\Psi\downup dl  ~~~, 
\end{align}
along with the corresponding coefficients ${\Hat \a}{}_x$, ${\Hat \b}{}_x$, and ${\Hat \k}{}_x$ given in the following table.
\begin{table}[h]
\centering
\begin{tabular}{|c|c|c|c|}
\hline
 & $\Hat \a$ & $\Hat \b$  & $\Hat \k$ \\ \hline
${\cal V}{}_1$ & 0 & 0 & 1 \\ \hline
${\cal V}{}_2$ & 1 & 0 & 0 \\ \hline
${\cal V}{}_3$ & 1 & 0 & 0 \\ \hline
${\cal V}{}_4$ & 0 & 0 & - 1 \\ \hline
${\cal V}{}_5$ & 0 & 1 & 0 \\ \hline
\end{tabular}
\end{table}  

The next series of calculations turn to the results for the holoraumy calculations for the 4D, $\cal N $ = 2 supermultiplet.

\subsection{Tensor Multiplet \texorpdfstring{${\rm D}$-${\rm D}$}{D-D} Bosonic Holoraumy}
\vspace*{-24 pt}
\begin{align}
\DDAB{}\til A &= 2\, \sig3ij \, [ \,-(\g\up5\g\up\m)\ddn ab\pd \m\til B - iC\ddn ab\til F + (\g\up5)\ddn ab\til G \, ] 
+2\, \sig2ij\gamatdown\m^ab\pd \m\til\varphi  \nonumber\\
&{~~~\,}  +\fracm{2}{3}\sig1ij\e_{\m}^{\ \v\a\b}(\g\up5\g\up\m)\ddn ab\til H\ddd\n\a\b   ~~~, 
\nonumber\\
\DDAB{}\til B &= 2\, \d\uup ij\, [ \,(\g\up5)\ddn ab \til F + i C\ddn ab\til G\, ] \, 
+ 2\, (\g\up5\g\up\m)\ddn ab\pd \m(\sig3ij\til A + \sig1ij\til\varphi)\nonumber\\
&{~~~\,} -\fracm{2}{3}\sig2ij\e_{\m}^{\ \n\a\b}\gamatdown\m^ab\til H\ddd\n\a\b  ~~~, \nonumber\\
\DDAB{}\til F &= 2\d\uup ij\, [ \,(\g\up5)\ddn ab\square \til B - (\g\up5\g\up\m)\ddn ab\pd \m \til G\, ] \,
-i\, 2C\ddn ab\square\, [ \,\sig3ij\til A + \sig1ij\til\varphi\, ] \,\nonumber\\
&{~~~\,}- i\fracm{1}{3}\sig2ij\e\duuu\l\v\a\b(\gup5\g\uup[\m\g\uup\l])\ddn ab\pd\m \til H\ddd\n\a\b ~~~, \nonumber\\
\DDAB{}\til G &= 2\d\uup ij\, [ \,iC\ddn ab\square\til B + (\g\up5\g\up\m)\ddn ab \partial \dn\m\til F\, ] \,
+ 2(\g\up5)\ddn ab\square\, [ \,\sig3ij \til A + \sig1ij \til\varphi\, ] \, \nonumber\\
&{~~~\,} +\fracm{1}{3}\sig2ij\e\duuu\l\n\a\b(\g\uup[\m\g\uup\l])\ddn ab\pd\m\til H\ddd\n\a\b ~~~,
\nonumber\\
\DDAB{}\til\varphi &= 2\sig1ij\, [ \,(-\g\up5\g\up\l)\ddn ab\partial\dn\l\til B - iC\ddn ab\til F + \gamatdown5^ab\til G\, ] \,
- 2\sig2ij\gamatdown\l^ab\partial\dn\l\til A
\nonumber\\
&{~~~\,}  - \tfrac{2}{3} \sig3ij\e_{\l}^{\ \v\a\b}(\g\up5\g\up\l)\ddn ab\til H\ddd\n\a\b
~~~, \nonumber\\
\DDAB{}\til B_{\m\n} &= \sig2ij \, [\, \e_{\m\n\d}^{\ \ \ \l}\, (\gup\d)\ddn ab\pd\l\til B -\fracm{1}{2}(\g\ddn[\m\g\ddn\n])\ddn ab\til F -\fracm{1}{2}i(\gup5\g\ddn[\m\g\ddn\n])\ddn ab\til G\, ] \, \nonumber\\
&{~~~\,} +\e_{\m\n\d}^{\ \ \ \l}(\gup5\gup\d)\ddn ab\pd\l\, [ \,-\sig1ij\til A+\sig3ij\til\varphi\, ] \,
+\fracm{1}{3}\d\uup ij \e^{\k\a\b}_{\ \ \ \ [\m}(\gup5\g\ddn\n])\ddn ab\til H\ddd\k\a\b ~~~.
\end{align}

\subsection{Tensor Multiplet \texorpdfstring{${\rm  D}$-${\Tilde {\rm D}}$}{D-D} Bosonic Holoraumy}
\vspace*{-24 pt}
\begin{align}
    \holcrosstil{}\tila &= 2\sig1ij\, [ \,\ggfiupdown\m ab\pd\m B+iC\ddn abF\, ] \,
    -2\d\uup ij\gupdown5abd+ i\fracm{1}{2}\sig3ij\gupcommdown\m\n ab F\ddn\m\n   \nonumber\\
    \holcrosstil{}\tilb&= -i\, 2\sig2ij\, [ \,\ggfiupdown\m ab\pd\m A+\gupdown5abF\, ] \,  \nonumber \\
    \holcrosstil{}\tilf&= -2\sig2ij\, [ \,C\ddn ab\lap A+i\gupdown5ab\lap B\, ] \,  \nonumber  \\
    \holcrosstil{}\tilg &= 2\sig2ij \, [\, -i\gupdown5ab\lap A+C\ddn ab\lap B-i\ggfiupdown\m ab\pd\m F\, ] \,\nonumber\\
   &{~~~\,}  +i\, 2\gpart\, [ \,-\sig1ijG+\sig3ijd\, ] 
    -\fracm{1}{2}\d\uup ij(\gup5\gup\l\gupbrack\m\n)\ddn ab\pd\l F\ddn\m\n    \nonumber \\
    \holcrosstil{}\til\varphi &= -2\sig3ij\, [ \,\ggfiupdown\m ab\pd\m B+iC\ddn ab F\, ] \,
    +2\d\uup ij\gupdown5abG+\fracm{i}{2}(\sigma^1)^{ij}\gupcommdown\m\n abF\ddn\m\n   \nonumber \\
    \holcrosstil{}\tilbmn\m\n&= \d\uup ij\e_{\m\n}^{\ \ \d\l}(\gup5\g\dn\d)\ddn ab\pd\l A+\d\uup ij(\gup5\g\ddn[\m)\ddn ab\partial\ddn\n]B
    -C_{ab}\sig2ij F_{\m\n} \cr
    &~~~ +\fracm{i}{2}\sig2ij\e_{\m\n}^{ \ \ \ \r\s}(\gup5)\ddn ab F_{\r\s} 
\end{align}

\subsection{Tensor Multiplet \texorpdfstring{${\Tilde {\rm D}}$-${\Tilde {\rm D}}$}{D-D} Bosonic Holoraumy}
\vspace*{-24 pt}
\begin{align}
    \holtil\tila &= 2\sig3ij\{\ggfiupdown\m ab\pd\m\tilb+iC\ddn ab\tilf-\gupdown5ab\tilg\} \nonumber\\    
    &{~~\,~}+2\sig2ij\gpart\til\varphi +\fracm{2}{3}\sig1ij\e\duuu\m\n\a\b\ggfiupdown\m ab\tilh\n\a\b\\
    \holtil\tilb &= 2\d\uup ij\{\gupdown5ab\tilf+iC\ddn ab
    \tilg\} -2\ggfiupdown\m ab\pd\m\{\sig3ij\til A+\sig1ij\til\varphi\}\nonumber\\
    &{~~\,~}+\fracm{2}{3}\sig2ij\e\duuu\m\n\a\b\gupdown\m ab\tilh\n\a\b\\
    \holtil{}\tilf&= 2\d\uup ij\{\gupdown5ab\lap\tilb-\ggfiupdown\m ab\pd\m\tilg\}+ 2iC\ddn ab\{\sig3ij\lap\tila+\sig1ij\lap\til\varphi\}\nonumber\\
    &{~~\,~}+\fracm{i}{3}\sig2ij\e\duuu\m\n\a\b(\gup5\gupbrack\l\m)\ddn ab\pd\l\tilh\n\a\b\\
    \holtil{}\tilg &=2\d\uup ij\{iC\ddn ab\lap\tilb +\ggfiupdown\m ab\pd\m\tilf\}-2\gupdown5ab\{\sig3ij\lap\tila+\sig1ij\lap\til\varphi\}\nonumber\\
    &{~~\,~}-\fracm{1}{3}\sig2ij\e\duuu\m\n\a\b(\gupbrack\l\m)\ddn ab\pd\l\tilh\n\a\b\\
    \holtil\til\varphi &= 2\sig1ij\{\ggfiupdown\m ab\pd\m\tilb+iC\ddn ab\tilf-\gupdown5ab\tilg\}\nonumber\\
    &{~~\,~}-2\sig2ij\gpart\tila -\fracm{2}{3}\sig3ij\e\duuu\m\n\a\b\ggfiupdown\m ab\tilh\n\a\b\\
    \holtil\tilbmn\m\n &=\sig2ij\{\e_{\m\n}^{\ \ \l\d}(\g\dn\d)\ddn ab\pd\l\tilb +\fracm{1}{2}(\gdnbrack\m\n)\ddn ab\tilf +i\fracm{1}{2}(\gup5\gdnbrack\m\n)\ddn ab\tilg\}\nonumber\\
    &{~~\,~}+\e_{\m\n}^{\ \ \l\d}(\gup5\g\dn\d)\ddn ab\pd\l\{\sig1ij\tila-\sig3ij\til\varphi\}
    +\fracm{1}{3}\d\uup ij\e^{\k\a\b}_{\ \ \ \ [\m}(\gup5\g\ddn\n])\ddn ab\tilh\k\a\b
\end{align}

\subsection{Tensor Multiplet \texorpdfstring{${\rm D}$-${\rm D}$}{D-D} Fermionic Holoraumy}
\vspace*{-24 pt}
\begin{align}
\DDAB{}\til\Psi\downup ck &= i 2 \, {\cal V}_1{}^{i j k l} \, \ggfiupdown\m ab\gupsw5cd\pd\m\tilpsi\downup   dl + i \,
{\cal V}_2{}^{i j k l} \, (\g{}^{[\m} \g{}^{\l]} )_{ab} \,( \g\dn\l ){}_{c}{}^{d}  \pd\m  \, \tilpsi\downup   dl  \nonumber\\
&{~~~\,} + i \, {\cal V}_3{}^{i j k l} \, [ \,\gupdown\m ab(\g\dn\m\gup\l)\sw cd +\fracm{1}{8}\, (\g\uup[\m\g\uup\n])\ddn ab(\g\ddn[\m\g\ddn\n] \gup\l)\sw cd
\, ] \, \pd\l  \tilpsi\downup   dl\nonumber\\
&{~~~\,} +  i \, {\cal V}_4{}^{i j k l} \, [ \ggfiupdown\m ab(\gup5\g\dn\m\gup\l)\sw cd -C\ddn ab\gupsw\l cd + \gupdown5ab\ggfiupsw\l cd]\pd\l\tilpsi\downup   dl\nonumber\\
&{~~~\,} + i \, {\cal V}_5{}^{i j k l} \, [ \ggfiupdown\m ab(\gup5\g\dn\m\gup\l)\sw cd  + C\ddn ab\gupsw\l cd -\gupdown5ab\ggfiupsw\l cd]\pd\l\tilpsi\downup   dl ~~~,
\end{align}
along with the corresponding coefficients ${\Hat \a}{}_x$, ${\Hat \b}{}_x$, and ${\Hat \k}{}_x$ given in the following table.
\begin{table}[h]
\centering
\begin{tabular}{|c|c|c|c|}
\hline
 & $\Hat \a$ & $\Hat \b$  & $\Hat \k$ \\ \hline
${\cal V}{}_1$ & 0 & 0 & 1 \\ \hline
${\cal V}{}_2$ & 1 & 0 & 0 \\ \hline
${\cal V}{}_3$ & 1 & 0 & 0 \\ \hline
${\cal V}{}_4$ & 0 & 0 & -1 \\ \hline
${\cal V}{}_5$ & 0 & -1 & 0 \\ \hline
\end{tabular}
\end{table} 

\subsection{\label{TMcrossfermionApp}Tensor Multiplet \texorpdfstring{${\rm D}$-${\Tilde {\rm D}}$}{D-D} Fermionic Holoraumy}
The holoraumy calculation yields
\begin{align}
\holcrosstil{}\tilpsi\downup ck &= i \, 2\tilvee1\gpart\Psi\downup cl + i\, 2\tilvee2(\g\dn\n)\ddn ab(\gup\n\gup\m)\sw cd\pd\m\Psi\downup dl \nonumber \\
&{~~~\,} + i \,2\tilvee3\gupcommdown\m\n ab(\g\dn\n)\sw cd\pd\m\Psi\downup dl+ i\, 2\tilvee4\gupcommdown\m\n ab(\gamma_{[\mu}\gamma_{\nu]}\gup\l)\sw cd\pd\l\Psi\downup dl
\nonumber \\
&{~~~\,} + i\, 2\tilvee5\ggfiupdown\m ab\gupsw5cd \partial_\mu \Psi\downup dl+ 2i\tilvee6(\gup5\g\dn\n)\ddn ab(\gup5\gup\n\gup\m)\sw cd\pd\m\Psi\downup dl
\nonumber \\
&{~~~\,} + i \, 2\tilvee7\gupdown5ab\ggfiupsw\m cd\pd\m\Psi\downup dl+2i\tilvee8C\ddn ab\gupsw\m cd\pd\m\Psi\downup dl
\end{align}
In these equations, we have introduced $\tilvee x$ by use of the
definition $\tilvee x \,=\,  i {\cal V}{}_x{}^{i r k l } \sig2rj $ where the corresponding coefficient ${\Hat \a}{}_x$, ${\Hat \b}{}_x$, and ${\Hat \k}{}_x$ are given in the following table.
\begin{table}[!h]
\centering
\begin{tabular}{|c|c|c|c|}
\hline
 & $\Hat \a$ & $\Hat \b$  & $\Hat \k$ \\ \hline
$\tilvee1$ & 1/4 & -1/4  & 1/4 \\ \hline
$\tilvee2$ &-1/4 &1/4    &0 \\ \hline
$\tilvee3$ & -1/8 & 1/8   & -1/8 \\ \hline
$\tilvee4$ &-1/32 & 1/32  &0 \\ \hline
$\tilvee5$ &-3/4 &-1/4    & -3/4 \\ \hline
$\tilvee6$& 1/4 & 1/4   & 1/2 \\ \hline
$\tilvee7$& 1/4 &1/4    &-1/2 \\ \hline
$\tilvee8$& - 1/2 & 0  &-1/4 \\ \hline
\end{tabular}
\end{table}
\newpage

$ ~$\newline
\subsection{Tensor Multiplet \texorpdfstring{${\Tilde {\rm D}}$-${\Tilde {\rm D}}$}{D-D} Fermionic Holoraumy}
\vspace*{-24 pt}
\begin{align}
    \holtil{}\tilpsi\downup ck &=
    i 2 \,\{ \, {\cal V}_1{}^{i j k l} \, \gupdown\m ab + {\cal V}_2{}^{i j k l} \,  \ggfiupdown\m ab\gupsw5cd  \, \} \, \pd\m\til\Psi\downup dl             \nonumber\\
    &{~~\,~}+ i\, {\cal V}_3{}^{i j k l} \, [\,\gupdown\m ab(\g\dn\m\gup\l)\sw cd+\fracm{1}{8}(\g\uup[\m\g\uup\n])\ddn ab(\g\ddn[\m\g\ddn\n]\gup\l)\sw cd\,]\pd\l\til\Psi\downup dl\nonumber\\
    &{~~\,~}+i \, {\cal V}_4{}^{i j k l} \, [\,(\gup5\gup\m)\ddn ab(\gup5\g\dn\m\gup\l)\sw cd+C\ddn ab(\gup\l)\sw cd+\gupdown5ab(\gup5\gup\l)\sw cd\,]\pd\l\til\Psi\downup dl\nonumber\\
    &{~~\,~}+i\, {\cal V}_5{}^{i j k l}  \, [\,(\gup5\gup\m)\ddn ab(\gup5\g\dn\m\gup\l)\sw cd-C\ddn ab(\gup\l)\sw cd-\gupdown5ab(\gup5\gup\l)\sw cd\,]\pd\l\Psi\downup dl ~~~,
\end{align}
along with the corresponding coefficients ${\Hat \a}{}_x$, ${\Hat \b}{}_x$, and ${\Hat \k}{}_x$ given in the following table.
\begin{table}[h]
\centering
\begin{tabular}{|c|c|c|c|}
\hline
 & $\Hat \a$ & $\Hat \b$  & $\Hat \k$ \\ \hline
${\cal V}{}_1$ & 1 & 0 & 0 \\ \hline
${\cal V}{}_2$ & 0 & -1 & 0 \\ \hline
${\cal V}{}_3$ & -1 & 0 & 0 \\ \hline
${\cal V}{}_4$ & 0 & 0 & -1 \\ \hline
${\cal V}{}_5$ & 0 & 1 & 0 \\ \hline
\end{tabular}
\end{table}

\newpage
\section{Alternative Off-Shell \texorpdfstring{${\rm D}$-${\Tilde {\rm D}}$}{D-D} Fermionic Holoraumy and Anticommutator}
\label{appen:AOS}
The cross term fermion calculations can take several different forms. In section \ref{appen:holoraumy} we used the $V$ and $\cv$ coefficients, which correspond to two different pauli matrix bases which emphasize i-j symmetries, and a gamma basis which can most easily be used for imposing equations of motion in the on-shell case. Here, different variations are included, particularly with `Y' coefficients which are in the same basis as the results from \cite{N4VTM}. 

Calculations also completed in the improved Pauli basis with old gamma basis
\begin{align}
\holcrosstil{}\Psi\downup ck &= \xee1\gpart\tilpsi\downup cl
+\xee2(\g\dn\n)\ddn ab\gupcommsw\n\m cd\pd\m\tilpsi\downup dl\cr
&{~~\,~}+\xee3\gupcommdown\m\n ab(\g\dn\n)\sw cd\pd\m\tilpsi\downup dl+\xee4(\gup5\gupbrack\m\n)\ddn ab(\gup5\g\dn\n)\sw cd\pd\m\tilpsi\downup dl\cr
&{~~\,~}+\xee5\ggfiupdown\m ab\gupsw5cd\pd\m\tilpsi\downup dl+\xee6(\gup5\g\dn\n)\ddn ab(\gup5\gupbrack\n\m)\sw cd\pd\m\tilpsi\downup dl\cr
&{~~\,~}+\xee7\gupdown5ab\ggfiupsw\m cd\pd\m\tilpsi\downup dl+\xee8C\ddn ab\gupsw\m cd\pd\m\tilpsi\downup dl\\
\nonumber\\
\holcrosstil{}\tilpsi{}\downup ck &= \tilxee1\gpart\Psi\downup cl
+\tilxee2(\g\dn\n)\ddn ab\gupcommsw\n\m cd\pd\m\Psi\downup dl\cr
&{~~\,~}+\tilxee3\gupcommdown\m\n ab(\g\dn\n)\sw cd\pd\m\Psi\downup dl+\tilxee4(\gup5\gupbrack\m\n)\ddn ab(\gup5\g\dn\n)\sw cd\pd\m\Psi\downup dl\cr
&{~~\,~}+\tilxee5\ggfiupdown\m ab\gupsw5cd\pd\m\Psi\downup dl+\tilxee6(\gup5\g\dn\n)\ddn ab(\gup5\gupbrack\n\m)\sw cd\pd\m\til\Psi\downup dl\cr
&{~~\,~}+\tilxee7\gupdown5ab\ggfiupsw\m cd\pd\m\til\Psi\downup dl+\tilxee8C\ddn ab\gupsw\m cd\pd\m\Psi\downup dl
\end{align}
And we can write down the X coefficients as:
\begin{align}
\xee x= \a_x\d^{ij}\sig2kl+\b_x\sig1ij\sig3kl+\d_x\sig3ij\sig1kl+\k_x\sig2ij\d^{kl}
\end{align}
\begin{table}[!h]
\centering
\begin{tabular}{|c|c|c|c|c|}
\hline
 & $\a$ & $\b$ & $\d$ & $\k$ \\ \hline
$\xee1$ & 0 & 0 &0 & 1/2 \\ \hline
$\xee2$ &1/4 &(-1/4)$i$  & (1/4)$i$ & 0 \\ \hline
$\xee3$ & 0 & 0 & 0 & (-1/4) \\ \hline
$\xee4$ & (-1/4) & (1/4)$i$ & (-1/4)$i$ & 0 \\ \hline
$\xee5$ & -1 & 0 & 0 & (1/2) \\ \hline
$\xee6$ & (1/4) & (1/4) & (-1/4) &(-1/2)  \\ \hline
$\xee7$ & (1/2) & (1/2)$i$ & (-1/2)$i$ & 1 \\ \hline
$\xee8$ & -1 & 0 & 0 & (1/2) \\ \hline
\end{tabular}
\quad
\begin{tabular}{|c|c|c|c|c|}
\hline
 & $\a$ & $\b$ & $\d$ & $\k$ \\ \hline
$\tilxee1$ & 0 & 0 & 0 & (-1/2) \\ \hline
$\tilxee2$ & (1/4) & (1/4)$i$ & (-1/4)$i$ &0  \\ \hline
$\tilxee3$ & 0 & 0 & 0 & (1/4) \\ \hline
$\tilxee4$ & (-1/4) & (-1/4)$i$ & (1/4)$i$ & 0 \\ \hline
$\tilxee5$ & 1 & 0 & 0 & (1/2)\\ \hline
$\tilxee6$ & (-1/4) & (1/4)$i$ & (-1/4)$i$ & (-1/2) \\ \hline
$\tilxee7$ & (-1/2) & (1/2)$i$ & (-1/2)$i$ & 1 \\ \hline
$\tilxee8$ & 1 & 0 & 0 &(1/2)  \\ \hline
\end{tabular}
\end{table}

\newpage
Similarly for the anticommutator we have 
\begin{align}
\algcrosstil{}\Psi\downup ck &= \xeeh1\gpart\tilpsi\downup cl
+\xeeh2(\g\dn\n)\ddn ab\gupcommsw\n\m cd\pd\m\tilpsi\downup dl\cr
&{~~\,~}+\xeeh3\gupcommdown\m\n ab(\g\dn\n)\sw cd\pd\m\tilpsi\downup dl+\xeeh4(\gup5\gupbrack\m\n)\ddn ab(\gup5\g\dn\n)\sw cd\pd\m\tilpsi\downup dl\cr
&{~~\,~}+\xeeh5\ggfiupdown\m ab\gupsw5cd\pd\m\tilpsi\downup dl+\xeeh6(\gup5\g\dn\n)\ddn ab(\gup5\gupbrack\n\m)\sw cd\pd\m\tilpsi\downup dl\cr
&{~~\,~}+\xeeh7\gupdown5ab\ggfiupsw\m cd\pd\m\tilpsi\downup dl+\xeeh8C\ddn ab\gupsw\m cd\pd\m\tilpsi\downup dl\\
\nonumber\\
\algcrosstil{}\tilpsi{}\downup ck &= \tilxeeh1\gpart\Psi\downup cl
+\tilxeeh2(\g\dn\n)\ddn ab\gupcommsw\n\m cd\pd\m\Psi\downup dl\cr
&{~~\,~}+\tilxeeh3\gupcommdown\m\n ab(\g\dn\n)\sw cd\pd\m\Psi\downup dl+\tilxeeh4(\gup5\gupbrack\m\n)\ddn ab(\gup5\g\dn\n)\sw cd\pd\m\Psi\downup dl\cr
&{~~\,~}+\tilxeeh5\ggfiupdown\m ab\gupsw5cd\pd\m\Psi\downup dl+\tilxeeh6(\gup5\g\dn\n)\ddn ab(\gup5\gupbrack\n\m)\sw cd\pd\m\til\Psi\downup dl\cr
&{~~\,~}+\tilxeeh7\gupdown5ab\ggfiupsw\m cd\pd\m\til\Psi\downup dl+\tilxeeh8C\ddn ab\gupsw\m cd\pd\m\Psi\downup dl
\end{align}
And we can write down the $W$ coefficients as:
\begin{align}
W_x^{ijkl}= \a_x\d^{ij}\sig2kl+\b_x\sig1ij\sig3kl+\d_x\sig3ij\sig1kl+\k_x\sig2ij\d^{kl}
\end{align}
\begin{table}[!h]
\centering
\begin{tabular}{|c|c|c|c|c|}
\hline
 & $\a$ & $\b$ & $\d$ & $\k$ \\ \hline
$\xeeh1$ & (-1/2) & (-3/2)$i$ & (3/2)$i$ & 0 \\ \hline
$\xeeh2$ & 0 &0  & 0 & (-1/4) \\ \hline
$\xeeh3$ & 1/4 & (-1/4)$i$ & (1/4)$i$ & 0 \\ \hline
$\xeeh4$ & 0 & 0 & 0 & (1/4) \\ \hline
$\xeeh5$ & (-1/2) & (-1/2)$i$ & (1/2)$i$ & -1 \\ \hline
$\xeeh6$ & -(1/2) & 0 & 0 &(-1/4)  \\ \hline
$\xeeh7$ & 1 &0 & 0 & (-1/2) \\ \hline
$\xeeh8$ & (-1/2) & (-1/2)$i$ & (1/2)$i$ & -1 \\ \hline
\end{tabular}
\quad
\begin{tabular}{|c|c|c|c|c|}
\hline
 & $\a$ & $\b$ & $\d$ & $\k$ \\ \hline
$\tilxeeh1$ & (1/2) & (-3/2)$i$ & (3/2)$i$ & 0 \\ \hline
$\tilxeeh2$  & 0 &0  & 0 & (-1/4)  \\ \hline
$\tilxeeh3$ &(-1/4) & (-1/4)$i$ & (1/4)$i$ & 0 \\ \hline
$\tilxeeh4$ & 0 & 0 & 0 & (1/4)  \\ \hline
$\tilxeeh5$ &(-1/2) & (1/2)$i$ & (-1/2)$i$ & 1\\ \hline
$\tilxeeh6$ & -(1/2) & 0 & 0 &(1/4)  \\ \hline
$\tilxeeh7$ & 1 &0 & 0 & (1/2)  \\ \hline
$\tilxeeh8$ & (-1/2) & (1/2)$i$ & (-1/2)$i$ & 1   \\ \hline
\end{tabular}
\end{table}

\newpage
\subsection{Y Basis}
An alternative basis for the holoraumy is similar to that used for the algebra in~\cite{N4VTM}. In this basis, we have
\begin{align}
\holcrosstil{}\Psi\downup ck &= 2i\zeeh1\gpart\tilpsi\downup cl
+2i\zeeh2\gupcommdown\m\n ab(\g\dn\n)\sw cd\pd\m\tilpsi\downup dl\cr
&{~~\,~}+2i\zeeh3(\g\dn\n)\ddn ab\gupcommsw\n\m cd\pd\m\tilpsi\downup dl+2i\zeeh4(\gup5\gupbrack\m\n)\ddn ab(\gup5\g\dn\n)\sw cd\pd\m\tilpsi\downup dl\cr
&{~~\,~}+2i\zeeh5\ggfiupdown\m ab\gupsw5cd\pd\m\tilpsi\downup dl+2i\zeeh6(\gup5\g\dn\n)\ddn ab(\gup5\gupbrack\n\m)\sw cd\pd\m\tilpsi\downup dl\cr
&{~~\,~}+2i\zeeh7\gupdown5ab\ggfiupsw\m cd\pd\m\tilpsi\downup dl+2i\zeeh8C\ddn ab\gupsw\m cd\pd\m\tilpsi\downup dl\\
\holcrosstil{}\tilpsi\downup ck &= 2i\tilzeeh1\gpart\Psi\downup cl
+2i\tilzeeh2\gupcommdown\m\n ab(\g\dn\n)\sw cd\pd\m\Psi\downup dl\cr
&{~~\,~}+2i\tilzeeh3(\g\dn\n)\ddn ab\gupcommsw\n\m cd\pd\m\Psi\downup dl+2i\tilzeeh4(\gup5\gupbrack\m\n)\ddn ab(\gup5\g\dn\n)\sw cd\pd\m\Psi\downup dl\cr
&{~~\,~}+2i\tilzeeh5\ggfiupdown\m ab\gupsw5cd\pd\m\Psi\downup dl+2i\tilzeeh6(\gup5\g\dn\n)\ddn ab(\gup5\gupbrack\n\m)\sw cd\pd\m\til\Psi\downup dl\cr
&{~~\,~}+2i\tilzeeh7\gupdown5ab\ggfiupsw\m cd\pd\m\til\Psi\downup dl+2i\tilzeeh8C\ddn ab\gupsw\m cd\pd\m\Psi\downup dl\\\nonumber\\
\zeeh x&= \a_x\sig1ij\sig3kl+\b_x\sig1ik\sig3jl+\d_x\sig1jk\sig3il+\k_x\sig1il\sig3jk
\end{align}
\begin{table}[!h]
\centering
\begin{tabular}{|c|c|c|c|c|}
\hline
 & $\a$ & $\b$ & $\d$ & $\k$ \\ \hline
$\zeeh1$ & 0 & (1/4) &(-1/4) & 0 \\ \hline
$\zeeh2$ &0 &(-1/8)  & (1/8) & 0 \\ \hline
$\zeeh3$ & (-1/4) & (1/8) & (1/8) & 0 \\ \hline
$\zeeh4$ & (1/4) & (-1/8) & (-1/8) & 0 \\ \hline
$\zeeh5$ & 0 & (-1/4) & (-1/4) & (1/2) \\ \hline
$\zeeh6$ & (1/4) & (-1/8) & (1/8) &(-1/4)  \\ \hline
$\zeeh7$ & (1/2) & (3/4) & (-3/4) & (-1/2) \\ \hline
$\zeeh8$ & 0 & (-1/4) & (-1/4) & (1/2) \\ \hline
\end{tabular}
\quad
\begin{tabular}{|c|c|c|c|c|}
\hline
 & $\a$ & $\b$ & $\d$ & $\k$ \\ \hline
$\tilzeeh1$ & 0 & (-1/4) & (1/4) & 0 \\ \hline
$\tilzeeh2$ & 0 & (1/8) & (-1/8) &0  \\ \hline
$\tilzeeh3$ & (1/4) & (1/8) & (-1/8) & (-1/4) \\ \hline
$\tilzeeh4$ & (-1/4) & (-1/8) & (1/8) & (1/4) \\ \hline
$\tilzeeh5$ & 0 & (3/4) & (-1/4) & (-1/2)\\ \hline
$\tilzeeh6$ & (1/4) & (-3/8) & (1/8) & 0 \\ \hline
$\tilzeeh7$ & (1/2) & (1/4) & (-3/4) & 0 \\ \hline
$\tilzeeh8$ & 0 & (3/4) & (-1/4) &(-1/2)  \\ \hline
\end{tabular}
\end{table}
$ ~$


$ ~$\newpage
\section{Explicit Form of L-Matrices and \texorpdfstring{$\tilde{\rm V}$}{V}-Matrices For (4,0) Formulations}\label{appen:LRmatrices}

\begin{align}\label{e:LSMI}
	{  {\rm L}}_{1}^{(\rm SM-I)} = &\left(
\begin{array}{cccc}
 1 & 0 & 0 & 0 \\
 0 & 0 & 0 & -1 \\
 0 & 1 & 0 & 0 \\
 0 & 0 & -1 & 0 \\
\end{array}
\right)
	~~~,~~~
{  {\rm L}}_{2}^{(\rm SM-I)} =\left(
\begin{array}{cccc}
 0 & 1 & 0 & 0 \\
 0 & 0 & 1 & 0 \\
 -1 & 0 & 0 & 0 \\
 0 & 0 & 0 & -1 \\
\end{array}
\right)
~~~,~~~\cr
{  {\rm L}}_{3}^{(\rm SM-I)} =&\left(
\begin{array}{cccc}
 0 & 0 & 1 & 0 \\
 0 & -1 & 0 & 0 \\
 0 & 0 & 0 & -1 \\
 1 & 0 & 0 & 0 \\
\end{array}
\right)
~~~,~~~{  {\rm L}}_{4}^{(\rm SM-I)} =\left(
\begin{array}{cccc}
 0 & 0 & 0 & 1 \\
 1 & 0 & 0 & 0 \\
 0 & 0 & 1 & 0 \\
 0 & 1 & 0 & 0 \\
\end{array}
\right)~~~.
\end{align}

\begin{align}\label{e:LSMII}
	{  {\rm L}}_{1}^{(\rm SM-II)} = &\left(
\begin{array}{cccc}
 0 & 1 & 0 & 0 \\
 0 & 0 & 0 & -1 \\
 1 & 0 & 0 & 0 \\
 0 & 0 & -1 & 0 \\
\end{array}
\right)
	~~~,~~~
{  {\rm L}}_{2}^{(\rm SM-II)} =\left(
\begin{array}{cccc}
 1 & 0 & 0 & 0 \\
 0 & 0 & 1 & 0 \\
 0 & -1 & 0 & 0 \\
 0 & 0 & 0 & -1 \\
\end{array}
\right)
~~~,~~~\cr
{  {\rm L}}_{3}^{(\rm SM-II)} =&\left(
\begin{array}{cccc}
 0 & 0 & 0 & 1 \\
 0 & 1 & 0 & 0 \\
 0 & 0 & 1 & 0 \\
 1 & 0 & 0 & 0 \\
\end{array}
\right)
~~~,~~~{  {\rm L}}_{4}^{(\rm SM-II)} =\left(
\begin{array}{cccc}
 0 & 0 & 1 & 0 \\
 -1 & 0 & 0 & 0 \\
 0 & 0 & 0 & -1 \\
 0 & 1 & 0 & 0 \\
\end{array}
\right)~~~.
\end{align}

\begin{align}\label{e:LSMIII}
	{  {\rm L}}_{1}^{(\rm SM-III)} = &\left(
\begin{array}{cccc}
 0 & -1 & 0 & 0 \\
 0 & 0 & 0 & 1 \\
 -1 & 0 & 0 & 0 \\
 0 & 0 & -1 & 0 \\
\end{array}
\right)
	~~~,~~~
{  {\rm L}}_{2}^{(\rm SM-III)} =\left(
\begin{array}{cccc}
 0 & 0 & 0 & -1 \\
 0 & -1 & 0 & 0 \\
 0 & 0 & -1 & 0 \\
 1 & 0 & 0 & 0 \\
\end{array}
\right)
~~~,~~~\cr
{  {\rm L}}_{3}^{(\rm SM-III)} =&\left(
\begin{array}{cccc}
 1 & 0 & 0 & 0 \\
 0 & 0 & 1 & 0 \\
 0 & -1 & 0 & 0 \\
 0 & 0 & 0 & 1 \\
\end{array}
\right)
~~~,~~~{  {\rm L}}_{4}^{(\rm SM-III)} =\left(
\begin{array}{cccc}
 0 & 0 & -1 & 0 \\
 1 & 0 & 0 & 0 \\
 0 & 0 & 0 & 1 \\
 0 & 1 & 0 & 0 \\
\end{array}
\right)~~~.
\end{align}

\begin{align}\label{e:LSMIV}
	{  {\rm L}}_{1}^{(\rm SM-IV)} = &\left(
\begin{array}{cccc}
 -1 & 0 & 0 & 0 \\
 0 & 0 & 1 & 0 \\
 0 & 1 & 0 & 0 \\
 0 & 0 & 0 & -1 \\
\end{array}
\right)
	~~~,~~~
{  {\rm L}}_{2}^{(\rm SM-IV)} =\left(
\begin{array}{cccc}
 0 & 0 & -1 & 0 \\
 -1 & 0 & 0 & 0 \\
 0 & 0 & 0 & 1 \\
 0 & 1 & 0 & 0 \\
\end{array}
\right)
~~~,~~~\cr
{  {\rm L}}_{3}^{(\rm SM-IV)} =&\left(
\begin{array}{cccc}
 0 & 1 & 0 & 0 \\
 0 & 0 & 0 & 1 \\
 1 & 0 & 0 & 0 \\
 0 & 0 & 1 & 0 \\
\end{array}
\right)
~~~,~~~{  {\rm L}}_{4}^{(\rm SM-IV)} =\left(
\begin{array}{cccc}
 0 & 0 & 0 & 1 \\
 0 & -1 & 0 & 0 \\
 0 & 0 & 1 & 0 \\
 -1 & 0 & 0 & 0 \\
\end{array}
\right)~~~.
\end{align}

\begin{align}
  {\rm \tilde{V}}^{(\rm SM-I)}_{12} =&  \left(
\begin{array}{cccc}
 0 & -i & 0 & 0 \\
 i & 0 & 0 & 0 \\
 0 & 0 & 0 & -i \\
 0 & 0 & i & 0 \\
\end{array}
\right)~~~,~~~
{\rm \tilde{V}}^{(\rm SM-I)}_{13} = \left(
\begin{array}{cccc}
 0 & 0 & -i & 0 \\
 0 & 0 & 0 & i \\
 i & 0 & 0 & 0 \\
 0 & -i & 0 & 0 \\
\end{array}
\right)~~~, \cr
{\rm \tilde{V}}^{(\rm SM-I)}_{14} =& \left(
\begin{array}{cccc}
 0 & 0 & 0 & -i \\
 0 & 0 & -i & 0 \\
 0 & i & 0 & 0 \\
 i & 0 & 0 & 0 \\
\end{array}
\right)~~~,~~~
{\rm \tilde{V}}^{(\rm SM-I)}_{23} = \left(
\begin{array}{cccc}
 0 & 0 & 0 & -i \\
 0 & 0 & -i & 0 \\
 0 & i & 0 & 0 \\
 i & 0 & 0 & 0 \\
\end{array}
\right)~~~,~~~\cr
{\rm \tilde{V}}^{(\rm SM-I)}_{24} =& \left(
\begin{array}{cccc}
 0 & 0 & i & 0 \\
 0 & 0 & 0 & -i \\
 -i & 0 & 0 & 0 \\
 0 & i & 0 & 0 \\
\end{array}
\right)~~~,~~~
{\rm \tilde{V}}^{(\rm SM-I)}_{34} = \left(
\begin{array}{cccc}
 0 & -i & 0 & 0 \\
 i & 0 & 0 & 0 \\
 0 & 0 & 0 & -i \\
 0 & 0 & i & 0 \\
\end{array}
\right)~~~.
\end{align}

\begin{align}
  {\rm \tilde{V}}^{(\rm SM-II)}_{12} =&\left(
\begin{array}{cccc}
 0 & i & 0 & 0 \\
 -i & 0 & 0 & 0 \\
 0 & 0 & 0 & -i \\
 0 & 0 & i & 0 \\
\end{array}
\right)  ~~~,~~~
{\rm \tilde{V}}^{(\rm SM-II)}_{13} = \left(
\begin{array}{cccc}
 0 & 0 & -i & 0 \\
 0 & 0 & 0 & -i \\
 i & 0 & 0 & 0 \\
 0 & i & 0 & 0 \\
\end{array}
\right) ~~~, \cr
{\rm \tilde{V}}^{(\rm SM-II)}_{14} =& \left(
\begin{array}{cccc}
 0 & 0 & 0 & i \\
 0 & 0 & -i & 0 \\
 0 & i & 0 & 0 \\
 -i & 0 & 0 & 0 \\
\end{array}
\right)~~~,~~~
{\rm \tilde{V}}^{(\rm SM-II)}_{23} = \left(
\begin{array}{cccc}
 0 & 0 & 0 & -i \\
 0 & 0 & i & 0 \\
 0 & -i & 0 & 0 \\
 i & 0 & 0 & 0 \\
\end{array}
\right)~~~,~~~\cr
{\rm \tilde{V}}^{(\rm SM-II)}_{24} =& \left(
\begin{array}{cccc}
 0 & 0 & -i & 0 \\
 0 & 0 & 0 & -i \\
 i & 0 & 0 & 0 \\
 0 & i & 0 & 0 \\
\end{array}
\right)~~~,~~~
{\rm \tilde{V}}^{(\rm SM-II)}_{34} = \left(
\begin{array}{cccc}
 0 & -i & 0 & 0 \\
 i & 0 & 0 & 0 \\
 0 & 0 & 0 & i \\
 0 & 0 & -i & 0 \\
\end{array}
\right)~~~.
\end{align}

\begin{align}
  {\rm \tilde{V}}^{(\rm SM-III)}_{12} =& \left(
\begin{array}{cccc}
 0 & 0 & -i & 0 \\
 0 & 0 & 0 & -i \\
 i & 0 & 0 & 0 \\
 0 & i & 0 & 0 \\
\end{array}
\right) ~~~,~~~
{\rm \tilde{V}}^{(\rm SM-III)}_{13} = \left(
\begin{array}{cccc}
 0 & -i & 0 & 0 \\
 i & 0 & 0 & 0 \\
 0 & 0 & 0 & i \\
 0 & 0 & -i & 0 \\
\end{array}
\right) ~~~, \cr
{\rm \tilde{V}}^{(\rm SM-III)}_{14} =&  \left(
\begin{array}{cccc}
 0 & 0 & 0 & i \\
 0 & 0 & -i & 0 \\
 0 & i & 0 & 0 \\
 -i & 0 & 0 & 0 \\
\end{array}
\right)~~~,~~~
{\rm \tilde{V}}^{(\rm SM-III)}_{23} = \left(
\begin{array}{cccc}
 0 & 0 & 0 & -i \\
 0 & 0 & i & 0 \\
 0 & -i & 0 & 0 \\
 i & 0 & 0 & 0 \\
\end{array}
\right)~~~,~~~\cr
{\rm \tilde{V}}^{(\rm SM-III)}_{24} =& \left(
\begin{array}{cccc}
 0 & -i & 0 & 0 \\
 i & 0 & 0 & 0 \\
 0 & 0 & 0 & i \\
 0 & 0 & -i & 0 \\
\end{array}
\right)~~~,~~~
{\rm \tilde{V}}^{(\rm SM-III)}_{34} = \left(
\begin{array}{cccc}
 0 & 0 & i & 0 \\
 0 & 0 & 0 & i \\
 -i & 0 & 0 & 0 \\
 0 & -i & 0 & 0 \\
\end{array}
\right)~~~.
\end{align}

\begin{align}
  {\rm \tilde{V}}^{(\rm SM-IV)}_{12} =&  \left(
\begin{array}{cccc}
 0 & 0 & -i & 0 \\
 0 & 0 & 0 & -i \\
 i & 0 & 0 & 0 \\
 0 & i & 0 & 0 \\
\end{array}
\right)~~~,~~~
{\rm \tilde{V}}^{(\rm SM-IV)}_{13} = \left(
\begin{array}{cccc}
 0 & i & 0 & 0 \\
 -i & 0 & 0 & 0 \\
 0 & 0 & 0 & -i \\
 0 & 0 & i & 0 \\
\end{array}
\right)~~~, \cr
{\rm \tilde{V}}^{(\rm SM-IV)}_{14} =& \left(
\begin{array}{cccc}
 0 & 0 & 0 & i \\
 0 & 0 & -i & 0 \\
 0 & i & 0 & 0 \\
 -i & 0 & 0 & 0 \\
\end{array}
\right)~~~,~~~
{\rm \tilde{V}}^{(\rm SM-IV)}_{23} = \left(
\begin{array}{cccc}
 0 & 0 & 0 & i \\
 0 & 0 & -i & 0 \\
 0 & i & 0 & 0 \\
 -i & 0 & 0 & 0 \\
\end{array}
\right)~~~,~~~\cr
{\rm \tilde{V}}^{(\rm SM-IV)}_{24} =& \left(
\begin{array}{cccc}
 0 & -i & 0 & 0 \\
 i & 0 & 0 & 0 \\
 0 & 0 & 0 & i \\
 0 & 0 & -i & 0 \\
\end{array}
\right)~~~,~~~
{\rm \tilde{V}}^{(\rm SM-IV)}_{34} = \left(
\begin{array}{cccc}
 0 & 0 & -i & 0 \\
 0 & 0 & 0 & -i \\
 i & 0 & 0 & 0 \\
 0 & i & 0 & 0 \\
\end{array}
\right)~~~.
\end{align}

\section{Explicit Form of \texorpdfstring{$\rL$}{L}- matrices for the 4D, \texorpdfstring{$\mathcal{N}=4$}{N=4} vector-tensor multiplet and the 4D, \texorpdfstring{$\mathcal{N}=4$}{N=4} vector-chiral multiplet }\label{a:L16}
To succinctly and efficiently write these matrices in tensor product notation, we will define a new symbol $\mathscr{V}^A_{(i)}$. The definition of this symbol is to begin with the Klein Vierergruppe element $\mathscr{V}^A$ and then set all entries to zero except for the  $i$-th row. A couple examples are
\begin{align}
    \mathscr{V}^2_{(3)} =& \left(\begin{array}{cccc}
        0 & 0 & 0 & 0  \\
        0 & 0 & 0 & 0 \\
        0 & 0 & 0 & 1 \\
        0 & 0 & 0 & 0 
    \end{array} \right)
    ~~~,~~~
    \mathscr{V}^3_{(1)} =\left(\begin{array}{cccc}
        0 & 0 & 1 & 0  \\
        0 & 0 & 0 & 0 \\
        0 & 0 & 0 & 0 \\
        0 & 0 & 0 & 0 
    \end{array} \right)
\end{align}
as the $\mathscr{V}^A$ are in matrix form
\begin{align}
    \mathscr{V}^1 =& \left(\begin{array}{cccc}
        1 & 0 & 0 & 0  \\
        0 & 1 & 0 & 0 \\
        0 & 0 & 1 & 0 \\
        0 & 0 & 0 & 1 
    \end{array}\right)
    ~~~,~~~ \mathscr{V}^2 = \left(\begin{array}{cccc}
        0 & 1 & 0 & 0  \\
        1 & 0 & 0 & 0  \\
        0 & 0 & 0 & 1  \\
        0 & 0 & 1 & 0 
    \end{array}\right)
   \cr
    \mathscr{V}^3 =& \left(\begin{array}{cccc}
        0 & 0 & 1 & 0  \\
        0 & 0 & 0 & 1 \\
        1 & 0 & 0 & 0 \\
        0 & 1 & 0 & 0 
    \end{array}\right)
    ~~~,~~~ \mathscr{V}^4 = \left(\begin{array}{cccc}
        0 & 0 & 0 & 1  \\
        0 & 0 & 1 & 0  \\
        0 & 1 & 0 & 0  \\
        1 & 0 & 0 & 0 
    \end{array}\right)
\end{align}
We also introduce the Boolean notation of~\cite{Chappell:2012ms} to multiply the $\rL_\rI^{(SM-i)}$ from appendix~\ref{appen:LRmatrices} for embedding into the $\rL_\rI^{(VT)}$
\begin{align}
    (p_1 2^0 + p_2 2^1 + p_3 2^2 + p^4 2^3)_b =& \left(\begin{array}{cccc}
        (-1)^{p_1} & 0 & 0 & 0  \\
        0 & (-1)^{p_2} & 0 & 0 \\
        0 & 0 & (-1)^{p_3} & 0 \\
        0 & 0 & 0 & (-1)^{p_4}
    \end{array}\right)~~~,~~~p_i = 0,1
\end{align}
So for example
\begin{align}
    (5)_b \rL_2^{(SM-I)} =& \left(\begin{array}{cccc}
        -1 & 0 & 0 & 0  \\
        0 & 1 & 0 & 0 \\
        0 & 0 & -1 & 0 \\
        0 & 0 & 0 & 1
    \end{array}\right) \left(
\begin{array}{cccc}
 0 & 1 & 0 & 0 \\
 0 & 0 & 1 & 0 \\
 -1 & 0 & 0 & 0 \\
 0 & 0 & 0 & -1 \\
\end{array}
\right) = \left(
\begin{array}{cccc}
 0 & -1 & 0 & 0 \\
 0 & 0 & 1 & 0 \\
 1 & 0 & 0 & 0 \\
 0 & 0 & 0 & -1 \\
\end{array}
\right)
\end{align}
With this, we can succinctly write the $\rL_\rI^{(VC)}$ matrices as
\begin{align}
    \rL_\rI^{(VC)} =& \mathscr{V}^{1}_{(1)}\otimes \rL_\rI^{(SM-I)} + \mathscr{V}^{1}_{(2)}\otimes \rL_\rI^{(SM-I)}+\mathscr{V}^{1}_{(3)}\otimes \rL_\rI^{(SM-I)}+\mathscr{V}^{1}_{(4)}\otimes \rL_\rI^{(SM-II)} \\
    \rL_{\rI+4}^{(VC)} =& \mathscr{V}^{2}_{(1)}\otimes \left[(5)_b\rL_\rI^{(SM-I)}\right] + \mathscr{V}^{2}_{(2)}\otimes \left[(10)_b \rL_\rI^{(SM-I)}\right] \cr
    & +\mathscr{V}^{2}_{(3)}\otimes \left[(8)_b \rL_\rI^{(SM-I)}\right]+\mathscr{V}^{2}_{(4)}\otimes \left[(7)_b \rL_\rI^{(SM-II)}\right] \\
    \rL_{\rI+8}^{(VC)} =& \mathscr{V}^{4}_{(1)}\otimes \left[(8)_b\rL_\rI^{(SM-I)}\right] + \mathscr{V}^{4}_{(2)}\otimes \left[(5)_b \rL_\rI^{(SM-I)}\right] \cr
    & +\mathscr{V}^{4}_{(3)}\otimes \left[(10)_b \rL_\rI^{(SM-I)}\right]+\mathscr{V}^{4}_{(4)}\otimes \left[(7)_b \rL_\rI^{(SM-II)}\right] \\
    \rL_{\rI+12}^{(VC)} =& \mathscr{V}^{3}_{(1)}\otimes \left[(10)_b\rL_\rI^{(SM-I)}\right] + \mathscr{V}^{3}_{(2)}\otimes \left[(8)_b \rL_\rI^{(SM-I)}\right] \cr
    & +\mathscr{V}^{3}_{(3)}\otimes \left[(5)_b \rL_\rI^{(SM-I)}\right]+\mathscr{V}^{3}_{(4)}\otimes \left[(7)_b \rL_\rI^{(SM-II)}\right]
\end{align}
With this same notation, the $\rL_\rI^{(VT)}$ matrices are
\begin{align}
    \rL_\rI^{(VT)} =& \mathscr{V}^{1}_{(1)}\otimes \rL_\rI^{(SM-I)} + \mathscr{V}^{1}_{(2)}\otimes \rL_\rI^{(TM)}+\mathscr{V}^{1}_{(3)}\otimes \rL_\rI^{(SM-I)}+\mathscr{V}^{1}_{(4)}\otimes \rL_\rI^{(SM-II)} \\
    \rL_{\rI+4}^{(VT)} =& \mathscr{V}^{2}_{(1)}\otimes \left[(1)_b\rL_\rI^{(SM-I)}\right] + \mathscr{V}^{2}_{(2)}\otimes \left[(14)_b \rL_\rI^{(TM)}\right] \cr
    & +\mathscr{V}^{2}_{(3)}\otimes \left[(8)_b \rL_\rI^{(SM-I)}\right]+\mathscr{V}^{2}_{(4)}\otimes \left[(7)_b \rL_\rI^{(SM-II)}\right] \\
    \rL_{\rI+8}^{(VT)} =& \mathscr{V}^{4}_{(1)}\otimes \left[(14)_b\rL_\rI^{(SM-I)}\right] + \mathscr{V}^{4}_{(2)}\otimes \left[(15)_b \rL_\rI^{(TM)}\right] \cr
    & +\mathscr{V}^{4}_{(3)}\otimes \left[(15)_b \rL_\rI^{(SM-I)}\right]+\mathscr{V}^{4}_{(4)}\otimes \left[(7)_b \rL_\rI^{(SM-II)}\right]\\
    \rL_{\rI+12}^{(VT)} =& \mathscr{V}^{3}_{(1)}\otimes \left[(0)_b\rL_\rI^{(SM-I)}\right] + \mathscr{V}^{3}_{(2)}\otimes \left[(14)_b \rL_\rI^{(TM)}\right] \cr
    & +\mathscr{V}^{3}_{(3)}\otimes \left[(8)_b \rL_\rI^{(SM-I)}\right]+\mathscr{V}^{3}_{(4)}\otimes \left[(15)_b \rL_\rI^{(SM-II)}\right]
\end{align}
where the $\rL_\rI^{(TM)}$ are
\begin{align}
   \rL_1^{(TM)} =&  \left(
\begin{array}{cccc}
 1 & 0 & 0 & 0 \\
 0 & 0 & -1 & 0 \\
 0 & 0 & 0 & -1 \\
 0 & -1 & 0 & 0 \\
\end{array}
\right)
~~~,~~~
\rL_2^{(TM)} =\left(
\begin{array}{cccc}
 0 & 1 & 0 & 0 \\
 0 & 0 & 0 & 1 \\
 0 & 0 & -1 & 0 \\
 1 & 0 & 0 & 0 \\
\end{array}
\right) \cr
\rL_3^{(TM)} =&\left(
\begin{array}{cccc}
 0 & 0 & 1 & 0 \\
 1 & 0 & 0 & 0 \\
 0 & 1 & 0 & 0 \\
 0 & 0 & 0 & -1 \\
\end{array}
\right)
~~~,~~~
\rL_4^{(TM)} =\left(
\begin{array}{cccc}
 0 & 0 & 0 & 1 \\
 0 & -1 & 0 & 0 \\
 1 & 0 & 0 & 0 \\
 0 & 0 & 1 & 0 \\
\end{array}
\right)
\end{align}
The $\rR_\rI^{(VC)}$ and $\rR_\rI^{(VT)}$ matrices satisfy the trace-orthogonality relation in Eq.~\eqref{eq:ortho} for all $\rI = 1,2,3, \dots ,16$.

We note that the $\rL_\rI^{(VC)}$ are identified with the $\rL_\rI^{[0]}$ and $\rL_\rI^{[\mathcal{I}]}$ from~\cite{Calkins:2014exa} with $\rI = 1,2,3,4$ as~\footnote{\textcolor{black}{Here we correct two typos in~\cite{Calkins:2014exa}: $\rL_2^{[0]}$ should have $(12)_b(23)$ instead of $(4)_b(23)$ for its $\mathscr{V}^{1}_{(4)}$ term and $\rL_1^{[2]}$ should have $(13)_b(1243)$ instead of $(13)_b(1234)$ in its $\mathscr{V}^3_{(4)}$ term in the conventions of~\cite{Calkins:2014exa}.}}
\begin{align}
    \rL^{(VC)}_\rI =& \rL_\rI^{[0]}~~~,~~~\rL^{(VC)}_{\rI+4} = \rL^{[3]}_\rI~~~,~~~\rL^{(VC)}_{\rI+8} = \rL^{[1]}_\rI~~~,~~~\rL^{(VC)}_{\rI+12} = \rL^{[2]}_\rI
\end{align}
per the identification of the supercharges as in Eq.~\eqref{e:VCSupercharges}.

\newpage

\end{document}